\documentclass[pdftex,a4paper,11pt]{book}
\usepackage{hyperref}
\usepackage[version=3]{mhchem}
\usepackage[rightcaption]{sidecap}
\def\d{\mathrm{d}}

\usepackage [latin1] {inputenc}
\usepackage [english] {babel}
\usepackage {amsmath}
\usepackage [pdftex]{graphicx}
\usepackage {subfigure}
\usepackage{fancyhdr}                                               
    \fancypagestyle{plain}{\fancyhead{}}
\begin{document}

\begin{titlepage}

\begin{center}
\includegraphics {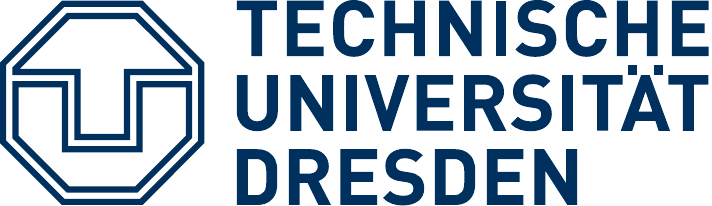}\\[3cm]

{ \LARGE \bfseries Dynamics of heterogeneous clusters\\
 under intense laser fields}\\[2cm]

A dissertation submitted to the Technical University of Dresden,\\
 Faculty of Science, Department of Physics\\
 in partial fulfillment of the requirements for the academic degree of\\
Doctor rerum naturalium (Dr. rer. nat.)\\[1cm]

by\\[1cm]
{ \LARGE \bfseries Pierfrancesco Di Cintio }\\[.5cm]

\end{center}
\vfill

\begin{minipage}{.8\textwidth}
Thesis supervisor: Prof. Dr. Jan-Michael Rost\\[.5cm]
Second Thesis supervisor: Prof. Dr. Ulf Saalmann\\[.5cm]
Dresden, 7 August 2014
\vfill
\end{minipage}
\begin{minipage}{.18\textwidth}
{\footnotesize\sffamily supported by}\\[.4cm]
\includegraphics[width=\textwidth]{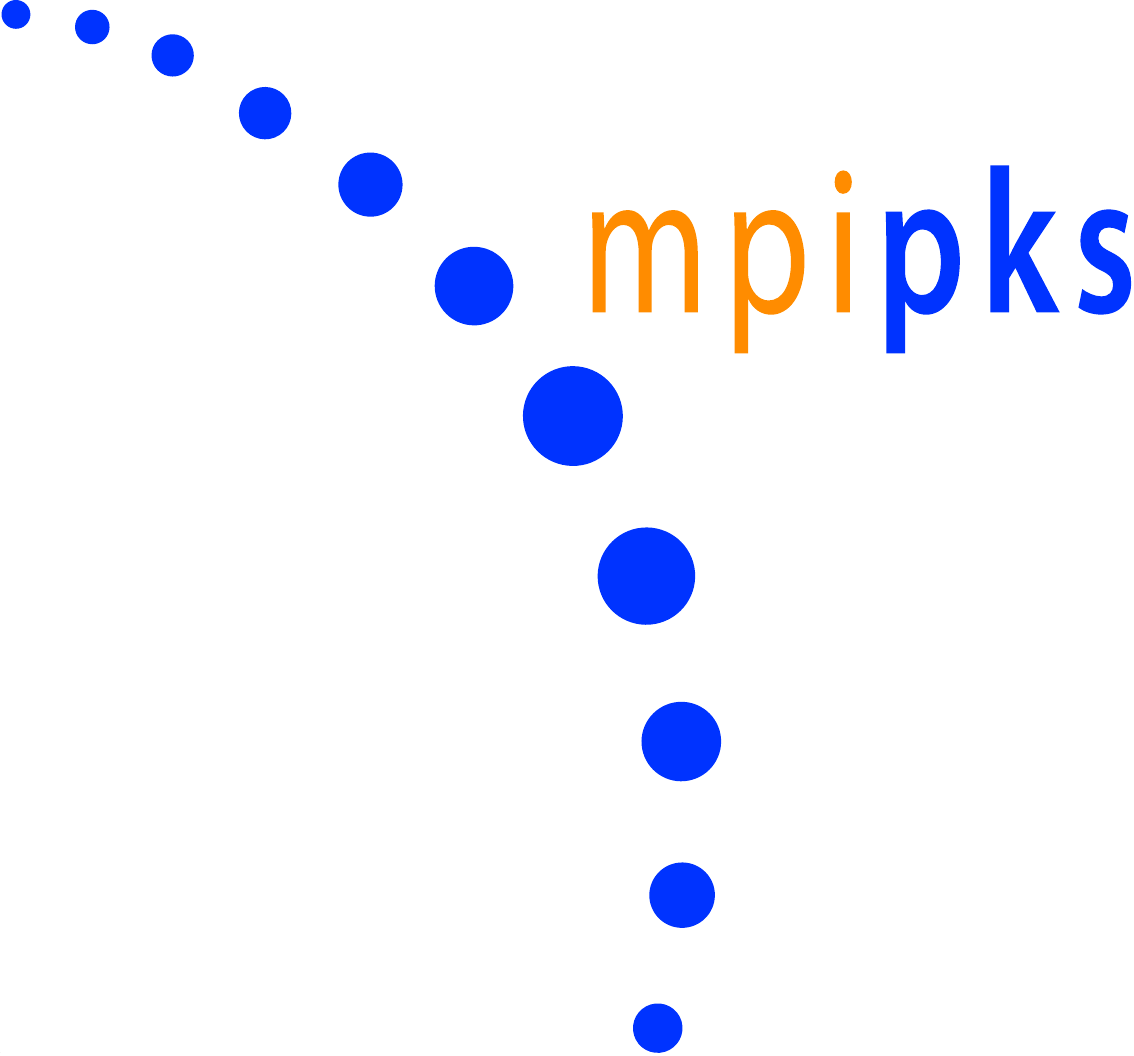}
\end{minipage}

\clearpage

\end{titlepage}

\newcommand\de{{\rm d}}
\newcommand\ri{r_i}
\newcommand\rj{r_j}
\newcommand\vi{v_i}
\newcommand\gi{g_i}
\newcommand\mi{m_i}
\newcommand\mj{m_j}
\newcommand\Ji{J_i}
\newcommand\aji{g_{ji}}
\newcommand\aii{g_{ii}}
\newcommand\tdyn{t_{\rm dyn}}
\newcommand\rz{r_0}
\newcommand\Mt{M_{\rm tot}}
\newcommand\vtyp{v_{\rm typ}}
\newcommand\Ei{E_i}
\newcommand\Ki{K_i}
\newcommand\phii{\phi_{ii}}
\newcommand\phji{\phi_{ji}}
\newcommand\xv{{\bf x}}
\newcommand\yv{{\bf y}}
\tableofcontents
\chapter{Introduction}\label{intro}
\begin{figure} [h!t]
        \centering 
         \includegraphics[width=0.74\textwidth]{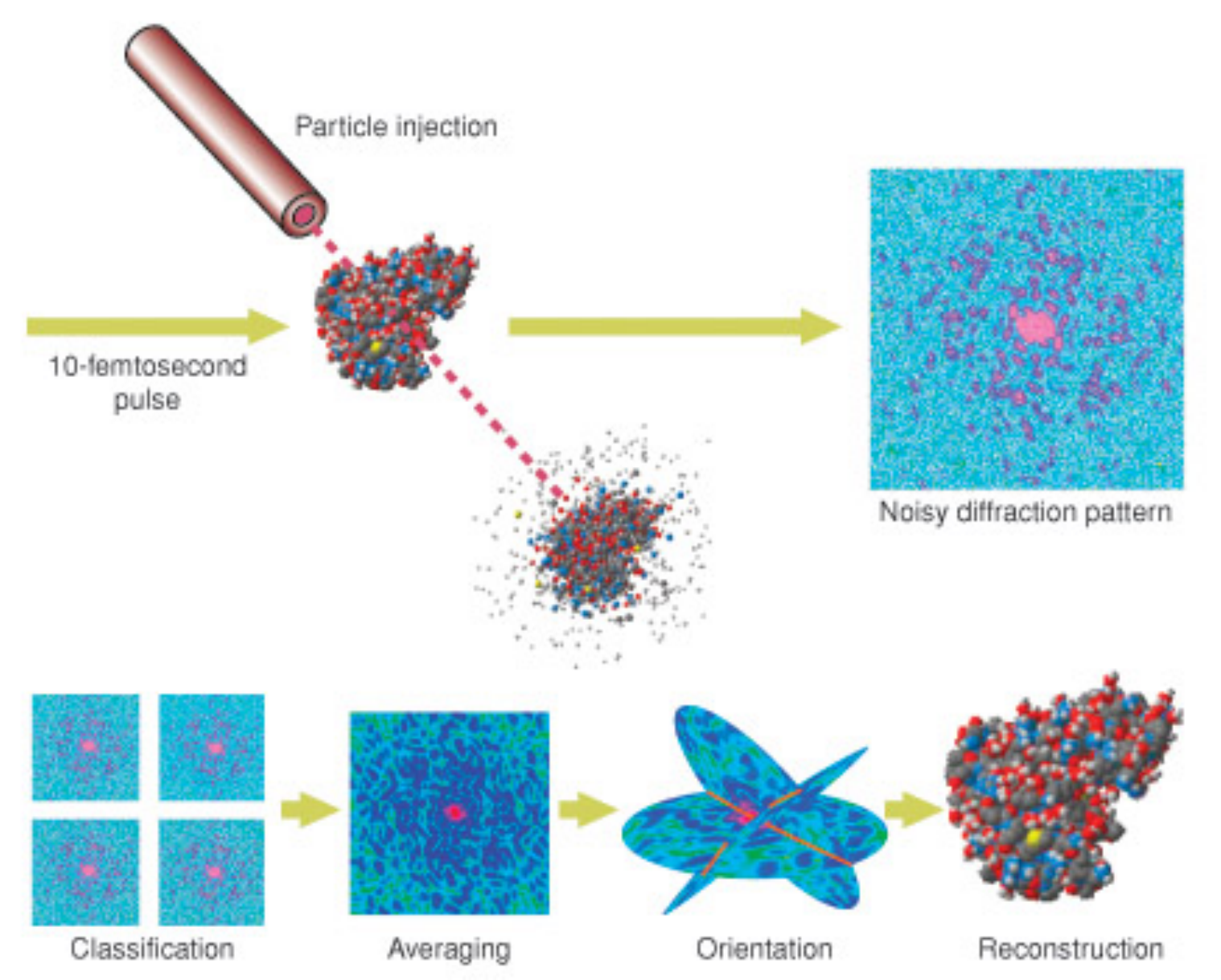}
         \caption{The steps of x-ray biomolecular imaging are illustrated.}
\label{image}
\end{figure}

\begin{figure} [h!t]
        \centering 
         \includegraphics[width=0.8\textwidth]{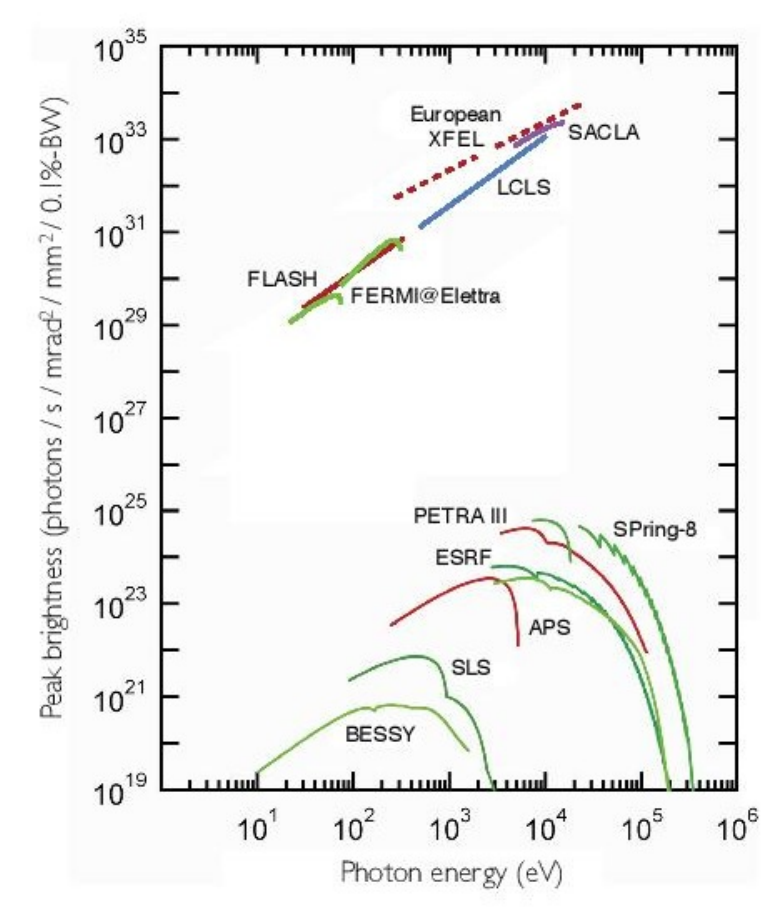}
         \caption{Peak brightness as function of the photon energy for different modern light sources. Note the enormous differences in the brightness. Figure taken from Ref. \cite{lcls1}.}
\label{lcls}
\end{figure}

\begin{figure} [h!t]
        \centering 
         \includegraphics[width=0.9\textwidth]{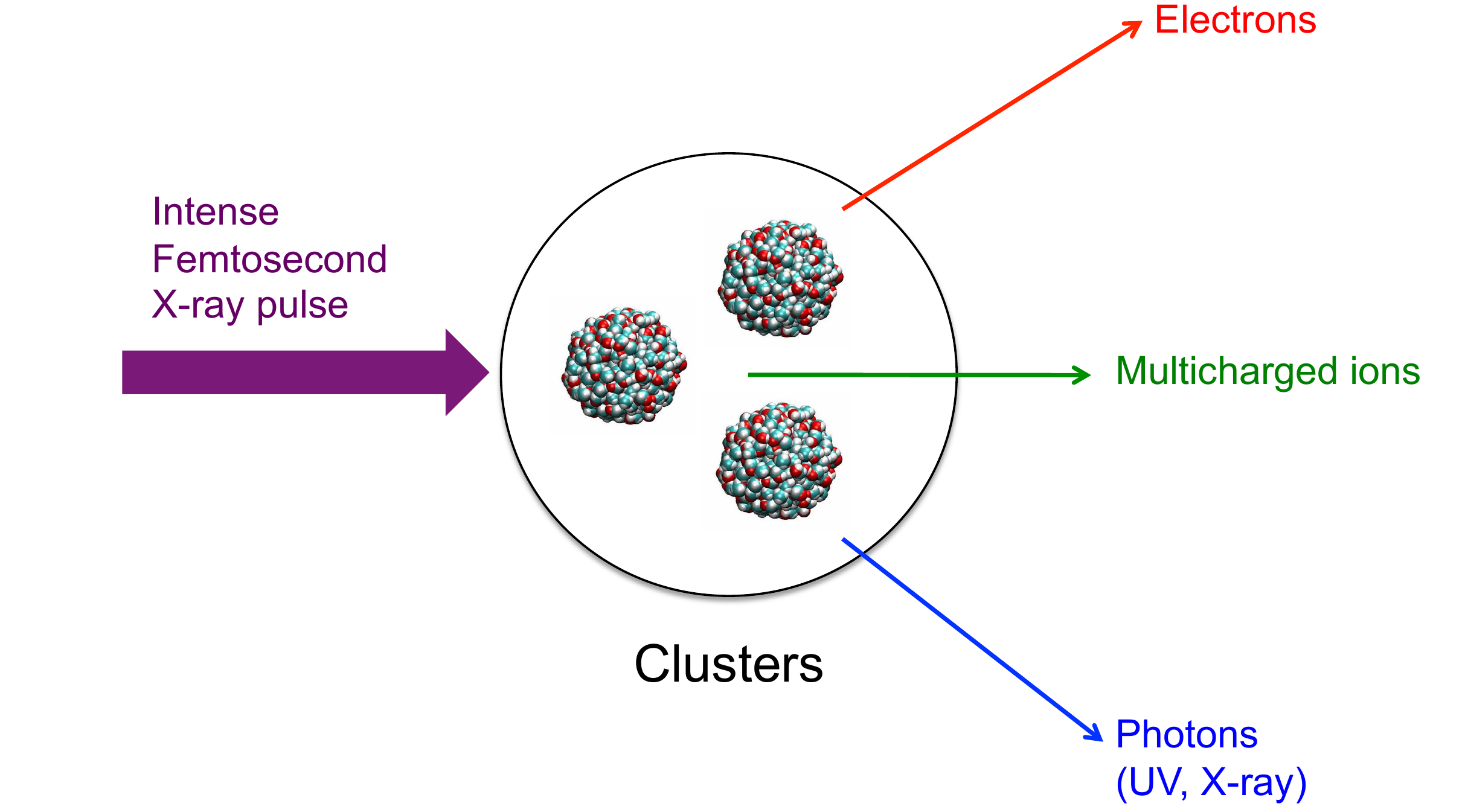}
         \caption{The different products of laser irradiation of clusters.}
\label{schemacluster}
\end{figure}

Understanding the response of matter when exposed to intense radiation is relevant to many areas of modern research, such as for example particle acceleration \cite{2013RvMP...85..751M}, 
astrophysics of active galactic nuclei \cite{2012ApJ...744...21P} and biomolecular imaging with x-rays pulses \cite{2008NanoL...8..310B}. 
The latter has been considered in recent years, as a promising application for the modern x-ray free electron lasers (XFEL) like the Linear Coherent Light Source (LCLS) at Stanford \cite{weblcls}, The Japanese XFEL SACLA at Kouto \cite{websacla}, or the European XFEL currently under construction in Hamburg \cite{webxfel}.\\
\indent Since the  pioneering paper by Neutze et al. \cite{neuze}, where single-molecule diffractive imaging was suggested for the first time, many theoretical and experimental studies have been aimed at finding an optimal parameter range for the imaging of large biological molecules using x-ray pulses (see e.g. Refs \cite {2006NatPh...2..839C}, \cite{2013PhRvA..88f1402B}). 
In a nutshell, as sketched in Fig. \ref{image}, a jet of ``different copies" of the molecule to be imaged is irradiated by a pulsated x-ray laser field and the diffraction pattern resulting from the interaction of the radiation with matter is collected for every illuminated sample \cite{jurek}. 
Due to the fact that every molecule of the jet comes with a different orientation with respect to the laser's  direction of propagation, a complicated reconstruction procedure is then needed to extract the three dimensional structure of the molecule from many different two dimensional patterns with unknown orientation, see \cite{2004AcCrA..60..294H} and \cite{2013SPIE.8845E..02Y}.\\
\indent However, this method is additionally plagued by the fact that at photon energies of a few keV (soft x-rays), the cross section for $K-$shell photoionization of carbon and oxygen (that are the most common elements in biological molecules that one is interested in imaging) 
are considerably larger -roughly one order of magnitude- than the electronic elastic scattering cross sections. Due to this, the sample is likely to be destroyed by radiation damage within the duration of the pulse \cite{2007PhRvL..98s8302H}. 
Obtaining pulse lengths of a few femtoseconds ($1 {\rm fs} =10^{-15}{\rm s}$) and keeping large intensities (of the order of $10^{17}{\rm W/cm^2}$) in order to have enough photons hitting the targets 
has encountered several theoretical and technical 
difficulties through the years. Modern x-ray laser sources, such as the aforementioned LCLS \cite{lcls1}, are now able to reach pulse lengths of the order of one femtosecond for photon energies of up to 10-12 keV (see e.g. \cite{2010NJPh...12c5024B}, \cite{2013EPJWC..5704001B}, \cite{2013OExpr..2124647L}) reaching 
peak brightness up to $10^{33}$ photons per ${\rm s~mrad^2~mm^2}$, as shown in Fig. \ref{lcls}. 
Therefore, such machines also represent perfect candidates to attempt single-molecule imaging, as well as to probe the properties of matter under extreme conditions of irradiation \cite{2009OExpr..1718271N}.\\
\indent As optimal ``test systems'' with respect to the molecular imaging, clusters (i.e. droplets of atoms or molecules containing from a few up 
to $\sim 10^6$ particles and with typical number densities of $10^{-2}{\rm \AA}^{-3}$ ) have been suggested. They have particularly simple structure and can be tailored in 
size from 1 up to $10^3$ nm to match that of more complex organic molecules that one currently aims at imaging.\\ 
\indent It must be pointed out that clusters, when exposed to laser pulses with focusing of the order of few micrometers, feel the same laser intensity on their whole volume contrary to what happens for larger solid-state targets. Thus clusters can absorb a larger fraction of the energy ``pumped in'' by 
the laser field with respect to solids with similar particle density. The interaction of strong lasers with clusters leads to different fragmentation products (as sketched in Fig. \ref{schemacluster}), such as energetic electrons \cite{1996PhRvL..77.3343S}, ions \cite{1996PhRvA..53.3379D}, \cite{1997Natur.386...54D}, \cite{2000PhRvA..61f3201S}, 
as well as 
photons \cite{1994Natur.370..631M}, \cite{1998JPhB...31.2825D}, \cite{2001PhRvE..64c6404T}, emitted in the decay of inner shell vacancies or due to {\it bremsstrahlung}.\\
\indent The strong x-ray pulses generated by contemporary machines, with their high intensities (up to $10^{19}{\rm W/cm^2}$) and lengths from 1 up to 100 femtoseconds, efficiently depose large amounts of energy in the target. Once charged, due to multiple almost simultaneous photoionzation events, the cluster 
becomes what we will hereafter refer as {\it nanoplasma} (i.e. a plasma of ions and electrons with typical size of a few up to thousand nm, \cite{2012PhRvL.108x5005G}). 
Note that, with respect to long wave length lasers (for instance infrared or VUV light), the charging of the target happens via different processes when exposed to x-ray pulses. In particular, photons with energies of the order of one keV 
ionize mainly the inner electronic shells of elements such as carbon, nitrogen and oxygen that are among the principal constituents of biological molecules. The photoelectrons released in this way have typical kinetic energies of the order of 500-700 eV allowing them to leave the charged 
cluster on a time scale of few femtoseconds, and for electrons absorbing 10 keV photons, such times are even of the order of attoseconds ($1{\rm as}=10^{-18}$ s). The inner shell vacancies, with life times comparable if not shorter than the pulse lengths considered here, decay via Auger processes thus forming a secondary population of electrons 
with kinetic energies of roughly 200 eV, implying that the absorption of one photon results in the emission of two electrons.  
We stress the fact that this picture is radically different from that which one has when longer wavelength are employed. In that case, the cluster is charged by the laser removing the electrons on time scales, typically of the order of one picosecond ($1{\rm ps}=10^{-12}$ s), and in the meantime 
ions have started moving under the combined action of their repulsive Coulomb forces and the laser electric field. The fast and at the same time massive charging, obtainable with the contemporary x-ray sources, was until a decade ago unreachable when employing that time's conventional laser machines. 
These extreme ionization conditions (i.e. high ion charge states produced in short time) thus open the door on previously unexperimented regimes of non-neutral plasmas, deserving therefore the interest of the theorist.\\
\indent When photoelectrons (and the faster Auger electrons) leave the cluster having kinetic energies $K_e$ larger than their potential energy in the cluster's electrostatic potential, the system is rapidly torn apart by mutual 
repulsive Coulomb forces among the ions. The latter have suffered initially little to no displacement due to the pulse. 
This regime of expansion is called Coulomb explosion, and for a cluster of initial radius $R_0$ and number density $n_0$, is obtained when
\begin{figure} [h!t]
        \centering 
         \includegraphics[width=0.9\textwidth]{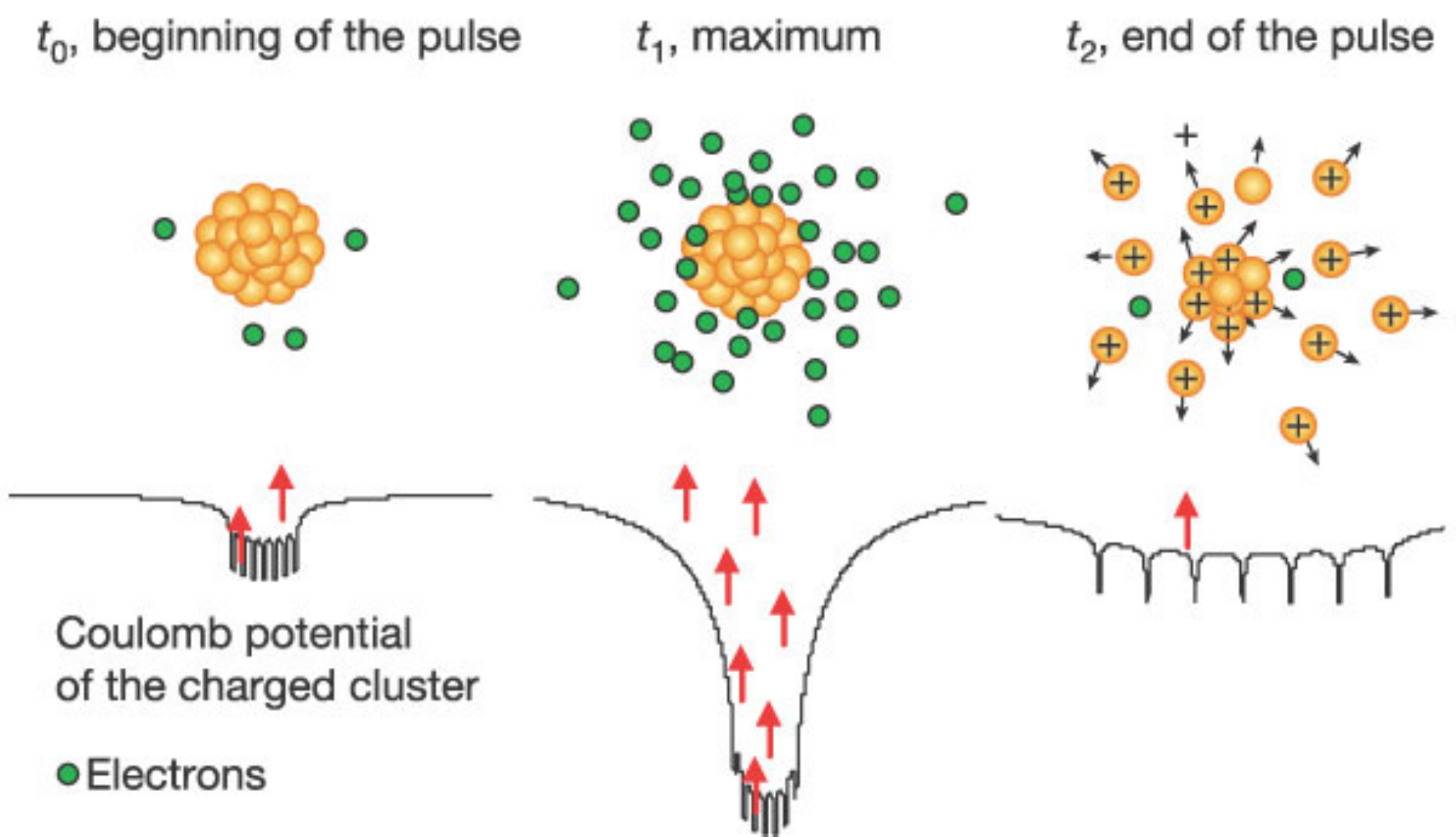}
         \caption{Upper row: schematic view of the explosion of a charged cluster within the laser pulse. Lower row: evolution of the cluster potential as the photoelectrons leave the system (red arrows). Figure taken from Ref. \cite{2004AcCrA..60..294H}.}
\label{schemace}
\end{figure} 
\begin{equation}\label{cece}
\frac{K_e}{|e\bar q4 \pi n_0R_0^2/3|}> 1
\end{equation}
where $\bar q$ is the average charge state of the ions and $e$ the charge of the electron. Figure \ref{schemace} sketches the formation of the cluster potential and the system's expansion. At the beginning of the pulse ($t_0$) the first photoelectrons leave. 
Then, as the overall charge increases the cluster potential deepens reaching its maximum ``depth'' ($t_1$). 
At this stage, some of the less energetic electrons, produced via secondary processes such as Auger decay of inner shell holes or ionization of valence electrons due to the clusters electrostatic field, are trapped. The cluster expansion however rapidly, makes the potential well shallower ($t_2$) letting them evaporate.\\
\indent Albeit being structurally simpler than biological molecules, clusters still present several ``degrees of freedom'' that make their dynamics after strong ionization non trivial \cite{2010RvMP...82.1793F}, and therefore interesting to study. 
To mention only a few, in molecular clusters the different species of ions can be accelerated differently in the cluster potential leading to dynamics with different time scales. Moreover, 
the kinetic energy spectrum of the ions is strongly influenced by the initial structure of the cluster and its shape. Systems starting with flattened or elongated geometries are expected to behave qualitatively very differently. 
In addition, charge migration due to rapid electron motion after an almost homogeneous spatial charging, also, influences the energetics of the cluster fragments and it is therefore important to have a clear picture also of the electronic component.\\   
\indent In this thesis, with the focus of shedding some light on the points mentioned above, in a regime of laser parameters relevant for molecular imaging with x-ray pulses, we have carried out a quantitative study of the dynamics of clusters irradiated by femtosecond x-ray pulses modelling those produced by contemporary laser sources such as LCLS or the European X-FEL. 
Our study is based on classical $N-$body simulations for the dynamics of particles, coupled with rate equations and Monte Carlo samplings to treat photoionization processes.\\     
\indent To prepare the field, the pure Coulomb explosion (i.e. no electrons are considered) has been treated for different systems. We started by reviewing the simple continuum model idealizing ionized clusters as uniformly charged spherical systems. 
Since clusters are in reality constituted by particles, we discussed the discrepancy with the approach based on particles. As mentioned before, the initial shape of the cluster, or in general of the ionized target, determines the energies of the products of its explosion.\\ 
\indent To this purpose, we have studied the expansion of non-neutral cluster plasmas with non spherical symmetry for different families of ellipsoidal system, both by means of a (semi-)analytic continuum model and numerical simulations. We have also analyzed the effects of initial conditions characterized by non uniform charging and 
non negligible ion temperature, as they are of some relevance with respect to regimes of laser irradiation where some ion motion is possible within the laser pulse.
Finally, we have implemented a detailed model of laser cluster interaction incorporating Auger transitions and electron recombination 
and used it to study the response of molecular hydride clusters (i.e. \ce{H2O}, \ce{NH3}, \ce{CH4} clusters) to femtosecond x-ray pulses. This was motivated by the puzzling experimental findings for methane clusters exposed to short and intense x-ray irradiation \cite{nirmala1}.\\
\indent The thesis is structured as follows: First of all, in Chapter \ref{chapmultice}, the physics of the Coulomb explosion is introduced and discussed for the case of spherical systems with homogeneous and heterogeneous composition. 
In the latter case, particular attention is devoted to the effect of the charge to mass ratio on the dynamics of the explosion, the results here reported will be finally contained in a forthcoming publication.\\
\indent In Chapter \ref{chapspheroid} we extend the discussion to (homogeneous) systems significantly departing from the spherical symmetry (i.e. ellipsoidal geometry), focusing on the structure of the final energy spectrum. 
Part of the work contained here, appears in publication \cite{grech2011}.\\
\indent Chapter \ref{chapLCLS} is devoted to the theoretical study of molecular hydrides clusters irradiated by extreme XFEL pulses. Its content has been published in \cite{nirmala1} and \cite{dicintio2013}.\\
\indent The numerical methods used throughout this work are discussed in Chapter \ref{numerica}. In Chapter \ref{sommario} the main results of this study are summarized and the future application and aims are discussed.\\
\indent Finally, the thesis is completed by three appendixes treating respectively the structure and production of clusters, the effect of multiple binary collisions on a charge travelling through a background of particles, 
and Coulomb explosion modelled with kinetic theory.  
\chapter{Coulomb explosion of spherical cluster plasmas}\label{chapmultice}
In this chapter we discuss the physics of the Coulomb explosion of small spherically symmetric targets (i.e. atomic or molecular clusters) irradiated by intense laser pulses. We start by reviewing the simple case of single-component homogeneously dense systems in the non-relativistic regime, where exact analytical results are available. We compare these results to calculations, where the spherical cluster is composed of particles. The characteristic differences between the two cases are discussed in detail. We then treat the case of non uniform density profiles and the formation of shock shells. Finally, we discuss the case of systems with heterogeneous composition.
\section{Mono-component systems}
As outlined before in the introduction, when an initially neutral cluster is irradiated by an intense laser pulse it gets charged and therefore starts to expand. According to the duration of the pulse, its intensity, photon energy and the structural and atomic properties of the target itself, different types of expansion can take place.\\
\indent When almost all the electrons stripped from the atoms in the cluster remain trapped by the space charge of the overall cluster, we speak of quasi-neutral or hydrodynamical expansion (see Refs. \cite{crow75}, \cite{mora79} and \cite{sack87}). On the contrary, when all the electrons are taken away, leaving a non-neutral plasma (see e.g. \cite{davidson01}, cfr. also Eq. \ref{cece}) whose dynamics is governed only by the repulsive inter-particle Coulomb forces, one has instead the so called Coulomb explosion (see Refs. \cite{kovalev05}, \cite{2008PlPhR..34..920N}). Note that intermediate regimes are also possible, for instance, due to non homogeneous charging of the cluster (see e.g. \cite{2010PhPl...17b3110A}).\\
\indent In this thesis, we concentrate on the case of Coulomb explosion, since we deal with x-ray pulses with intensities and photon energies for which the majority of the electrons escape the system.\\
\indent The dynamics of an expanding non-neutral plasma can be modelled using different approaches, depending on what quantity or observable one is interested in. First we present the main analytical results in the continuum model and then we discuss the results of numerical calculations in a particle-based approach.
\subsection*{Continuum model}\label{contmod}
Let us consider an isolated single component cluster that has been stripped of all its electrons composed by $N$ ions of the same charge and mass $q$ and $m$. In the so called continuum approximation\footnote{It must be pointed out that {\it continuum approximation} or continuum model is not the same concept as {\it continuum limit} (i.e. $N\rightarrow\infty$; $q_i\rightarrow0$). In the latter case one refers to a system in which each particle essentially behaves as a test particle in the field produced by the others. In our case, we are substituting a discrete density with a continuum distribution, regardless of the number of discrete particles $N$ of the original system. For an extensive discussion see Refs. \cite{2003LNP...626..154K} and \cite{kandrup04} and references therein} the cluster is replaced with a smooth number density $n(r)$, so that its radial mass and charge densities are given by $\rho_m(r)=mn(r)$ and $\rho_c=qn(r)$, where $m$ and $q$ are the unit mass and charge respectively and $r$ is the radial coordinate. We assume here no angular dependence. Therefore, the spherical symmetry of the system is preserved during the explosion.\\
\indent Under the assumption of incompressibility, one can treat this model with the equations of (non relativistic) fluid dynamics, see e.g. \cite{2007PhPl...14j3110K}, namely the continuity and momentum equations, that are respectively
\begin{equation}\label{continuity}
\frac{\partial n(r,t)}{\partial t}+\frac{1}{r^2}\left[r^2n(r,t)v(r,t)\right]=0
\end{equation} 
and
\begin{equation}\label{momentum}
n(r,t)\left[\frac{\partial v(r,t)}{\partial t}+v(r,t)\frac{\partial v(r,t)}{\partial r}\right]=\frac{Q}{M}\nabla\Phi(r,t),
\end{equation} 
where $Q$ and $M$ are the total mass and charge of the system, whose ratio equals the ratio $q/m$ between unit mass and charge, and $v(r,t)$ is the velocity field. The electrostatic potential $\Phi$ at position $r$ is given by
\begin{equation}\label{potential1}
\Phi(r)=\frac{4\pi}{r}\int_0^r\rho_c(r^\prime)r^{\prime 2}\d r^\prime+4\pi\int_r^{+\infty}\rho_c(r^\prime)r^\prime\d r^\prime.
\end{equation}
Equation (\ref{potential1}) is obtained from the radial Poisson equation
\begin{equation}\label{poissonspher}
\Delta\Phi(r)=\frac{1}{r}\frac{\partial}{\partial r}\left(r\frac{\partial \Phi(r)}{\partial r}\right)=4\pi\rho(r).
\end{equation}
Here $\Delta$ is the radial part of the Laplace operator which we give explicitly. Note that we are using here the atomic units for which the constant $\kappa_0=1/4\pi\epsilon_0$ is set to 1.\\
\indent The derivative of the potential $\Phi(r)$ needed in Eq. (\ref{momentum}) can be obtained from Eq. (\ref{potential1}) and reads  
\begin{equation}\label{efieldint}
\nabla\Phi(r)=\frac{4\pi}{r^2}\int_0^r \rho_c(r^{\prime})r^{\prime 2}\d r^\prime.
\end{equation}
The system of partial differential equations (PDEs) given by Eqs. (\ref{continuity}), (\ref{momentum}) and (\ref{efieldint}) formally describes in closed form the Coulomb explosion of a mono-component cluster plasma and could be easily generalized to heterogeneous systems, as well as to cases where an electron density is present (see Refs. \cite{murakami06} and \cite{beck09}).\\
\indent We will now discuss a case which represents the continuum version of a homogeneously charged cluster of total charge $Q$ and mass $M$, initial radius $R_0$ and with initial kinetic energy $K_{0}=0$. Therefore the initial charge density and velocity distribution read
\begin{eqnarray}\label{steplikerho}
\rho(r,0)&=&\rho_0n(r,0)=\rho_0\theta(R_0-r)\nonumber\\
v(r,0)&=&0
\end{eqnarray}
where $\theta(x)$ is the Heaviside's step function and $\rho_0=3Q/4\pi R_0^3$. It turns out that in this case the full dynamics can be modelled simply by a second order ordinary differential equation (ODE) with a class of self similar solutions, instead of a system of PDEs.\\
\indent For such initial conditions the electrostatic field at a generic $r_0<R_0$ is given by Eq. (\ref{efieldint}) and corresponds to the linear function of the radial coordinate $r_0$ itself
\begin{equation}
E(r_0)=\nabla\Phi(r_0)=\frac{4\pi}{3}\rho_{0}r_0.
\end{equation}
In an infinitesimal time $\delta t$, an infinitesimal volume element of mass $\delta m$ and charge $\delta q$ (so that $\delta q/\delta m=Q/M$) placed initially at radius $r_0$ will reach the velocity
\begin{equation}
\delta v(r_0,0+\delta t)=\frac{\delta q}{\delta m}E(r_0)\delta t=\frac{\delta q}{\delta m}\frac{4\pi}{3}\rho_{0}r_0\delta t.
\end{equation}
Due to the linearity with $r_0$, matter initially ``sitting" at a given radius can {\it not} overtake matter initially placed at larger radii, always attaining larger velocities.\\ 
\indent The differential equation for the dynamics of such element of volume reads
\begin{equation}
\frac{\d^2r(t)}{\d t^2}=\frac{\delta q}{\delta m}\frac{4\pi\int_0^{r(t)} \rho_{t} r^{\prime 2}\d r^\prime}{r^2(t)}=\frac{\delta q}{\delta m}\frac{Q(r,t)}{r^2(t)}.
\end{equation}
Since no overtaking is taking place, $Q(r,t)=Q(r_0)=4\pi\rho_{0}r_0^3/3$, and the equation above can be rewritten as
\begin{equation}\label{dinamique}
\frac{\d^2r(t)}{\d t^2}=\frac{C}{r^2(t)},
\end{equation} 
where $C=\delta q4\pi\rho_0r_0^3/3\delta m$, and its first integral reads
\begin{equation}
\left[\frac{\d r(t)}{\d t}\right]^2=2C\left[\frac{1}{r_0}-\frac{1}{r(t)}\right].
\end{equation}
\begin{figure} [h!t]
        \centering 
         \includegraphics[width=0.7\textwidth]{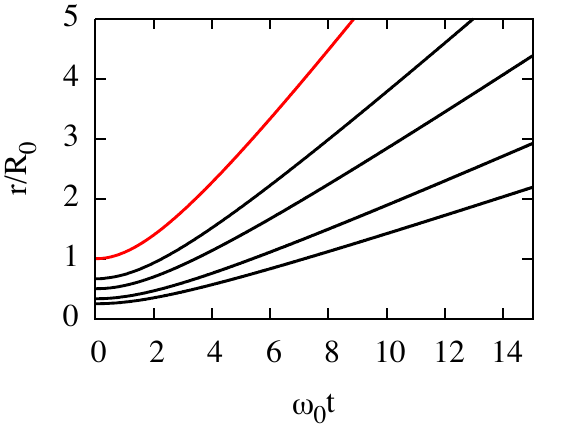}
         \caption{Trajectories of different volume elements initially placed at different radii expressed in units of the initial cluster radius $R_0$. The red line marks the expansion of the surface.}
\label{neomo1}
\end{figure} 
By further integrating the latter equation, one gets the time $t$ that takes to increase from radius $r_0$ to radius $r(t)$ as
\begin{equation}\label{sss}
t=\sqrt{\frac{3}{2}}\omega_0^{-1}\left[\sqrt{\xi(\xi-1)}+\ln\left(\sqrt{\xi}+\sqrt{\xi-1}\right)\right],
\end{equation}
where $\xi=r(t)/r_0$ and $\omega_0=\sqrt{4\pi\rho_0\delta q/\delta m}=Q\sqrt{3/MR_0^3}$.\\
\indent When $\omega_0t\ll1$ the trajectory of matter initially placed at $r_0$ is given asymptotically by
\begin{equation}\label{increase2}
r(t)\simeq r_0\left(1+\frac{\omega_0^2t^2}{6}\right),
\end{equation}
while in the limit of $\omega_0t\gg1$, one has
\begin{equation}\label{increase1}
r(t)\simeq \sqrt{\frac{2}{3}}r_0\omega_0t.
\end{equation}
Finally, the asymptotic velocity reads
\begin{equation}\label{asvel}
v_{\infty}=\sqrt{\frac{2}{3}}\omega_0r_0.
\end{equation}
In Fig. \ref{neomo1}, the trajectories $r(t)$ of different volume elements are shown. They show a quadratic increase for early times cfr. Eq. (\ref{increase2}), and become linear at large times, cfr. Eq. (\ref{increase1}). It is clearly evident that they do not intersect.\\
\indent The absence of overtaking and the fact that every choice of $r_0$ in the interval (0; $R_0$) initial condition of Eq. (\ref{sss}) leads to identical (rescaled) dynamics, means that {\it an initially cold homogeneously charged sphere expands retaining as spatially uniform density profile}. In other words Eq. (\ref{sss}) represents a {\it self-similar} solution.\\ 
\indent Knowing that a homogeneously charged sphere expands self-similarly it remains to determine its asymptotic number energy distribution 
\begin{equation}
n(\mathcal{E})=\frac{\d P}{\d \mathcal{E}}
\end{equation}
defined as the fraction of the system with energy $\mathcal{E}$. Whereas the total energy $\mathcal{E}_{\rm tot}$ is initially given by the potential energy $U_0$ if the initial kinetic energy $K_0$ equals 0, in the limit $t\rightarrow\infty$ it is $\mathcal{E}_{\rm tot}=K$.\\
\indent For the homogeneously charged sphere considered here, one has
\begin{equation} 
\mathcal{E}_{\rm tot}=U_0=2\pi\int_{0}^{+\infty}\rho(r_0)\Phi(r_0)r_0^2\d r_0=\frac{3}{5}\frac{Q^2}{R_0}.
\end {equation}
Using the fact that shells of matter starting from different radii in an uniformly charged sphere do not overtake each other and the first Newton theorem\footnote{A uniformly charged shell of charge $q$, exerts at its exterior the same force as due to a point-like particle of the same charge sitting at its centre, cfr. Ref. \cite{kellogg}. See also Eq. (\ref{efieldint})} (see e.g. \cite{2001NIMPA.464...98N}, \cite{2006PhRvA..73d1201I} and \cite{mika2013}),
one can extract the asymptotic $n(\mathcal{E})$, from the system's initial configuration as
\begin{equation}\label{derivdif1}
n(\mathcal{E})=\frac{\d P}{\d \mathcal{E}}=\frac{\d P}{\d r_0}/\frac{\d\mathcal{E}}{\d r_0}.
\end{equation}
Here we have defined the probability at $t=0$ to find an infinitesimal element of volume at radius $r_0$ as
\begin{equation}\label{neinteg}
\frac{\d P}{\d r_0}=\frac{3r_0^2}{R_0^3}\theta(R_0-r_0).
\end{equation}
The asymptotic kinetic energy (per unit charge) of a volume element is proportional to its initial radius through Eq. (\ref{asvel}) and reads
\begin{equation}\label{asie}
\mathcal{E}(r_0)=\frac{Qr_0^2}{R_0^3},
\end{equation}
therefore
\begin{equation}\label{derivdif2}
\frac{\d\mathcal{E}}{\d r_0}=\frac{2Qr_0}{R_0^3}.
\end{equation}
Substituting Eqs. (\ref{derivdif2}) and (\ref{neinteg}) into Eq. (\ref{derivdif1}) and expressing $r_0=\sqrt{\mathcal{E}R_0^3/Q}$, which follows from Eq. (\ref{asie}), leads to a square root distribution
\begin{equation}\label{nesqrt}
n(\mathcal{E})=\frac{3}{2}\sqrt{\frac{\mathcal{E}R_0^3}{Q^3}}\theta({\mathcal{E}(R_0)-\mathcal{E}})=\frac{3}{2}\sqrt{\frac{\mathcal{E}}{\mathcal{E}^3_{\rm max}}}\theta({\mathcal{E}_{\rm max}-\mathcal{E}}),
\end{equation}
\begin{figure} [h!t]
        \centering 
         \includegraphics[width=\textwidth]{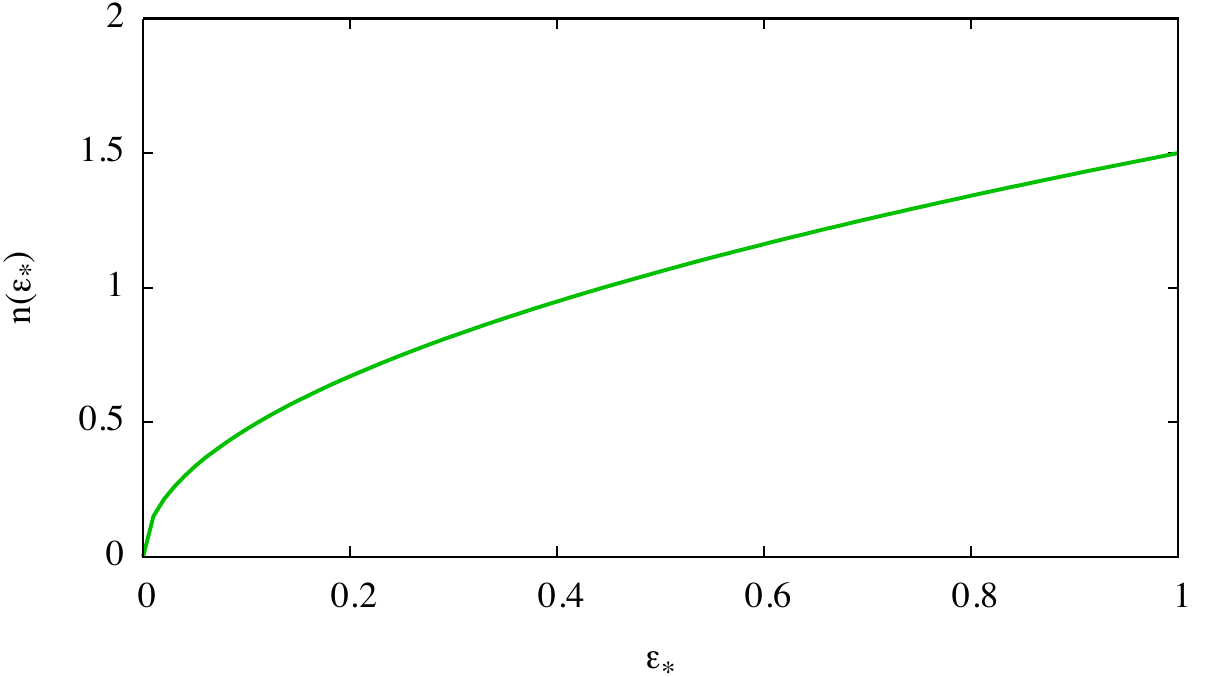}
         \caption{Asymptotic number energy distribution given by Eq. (\ref{nesqrt}). The scaled energy is defined as $\mathcal{E}_*=\mathcal{E}/\mathcal{E}_{\rm max}$.}
\label{neomo2}
\end{figure} 
where $\mathcal{E}_{\rm max}=Q/R_0$, is the maximal energy per unit of charge, reached by the fraction of the system initially sitting at its surface, see Fig. \ref{neomo2}. Alternatively, the energy distribution can be also derived by substituting Eq. (\ref{asie}) into
\begin{equation}\label{derivne1}
n(\mathcal{E})=\frac{\d P}{\d\mathcal{E}}=\frac{3}{R_0^3}\int_0^{R_0}\delta(\mathcal{E}-\mathcal{E}(r_0)) r_0^2\d r_0,
\end{equation}
and performing the integration in $r_0$.\\
\indent The total (conserved) energy $\mathcal{E}_{\rm tot}$ is recovered by
\begin{equation}
\mathcal{E}_{\rm tot}=\int_0^{\mathcal{E}_{\rm max}}n(\mathcal{E})\mathcal{E}\d\mathcal{E}.
\end{equation}
Remarkably, the expression for the asymptotic $n(\mathcal{E})$ given in Eq. (\ref{nesqrt}) holds true also in case of relativistic velocities, as it has been proved in \cite{2011JETPL..94...97B}.\\
\indent The simple case discussed here, albeit bearing a high level of abstraction, serves as a reference system for problems involving Coulomb explosion.
\begin{figure} [h!t]
        \centering 
         \includegraphics[width=\textwidth]{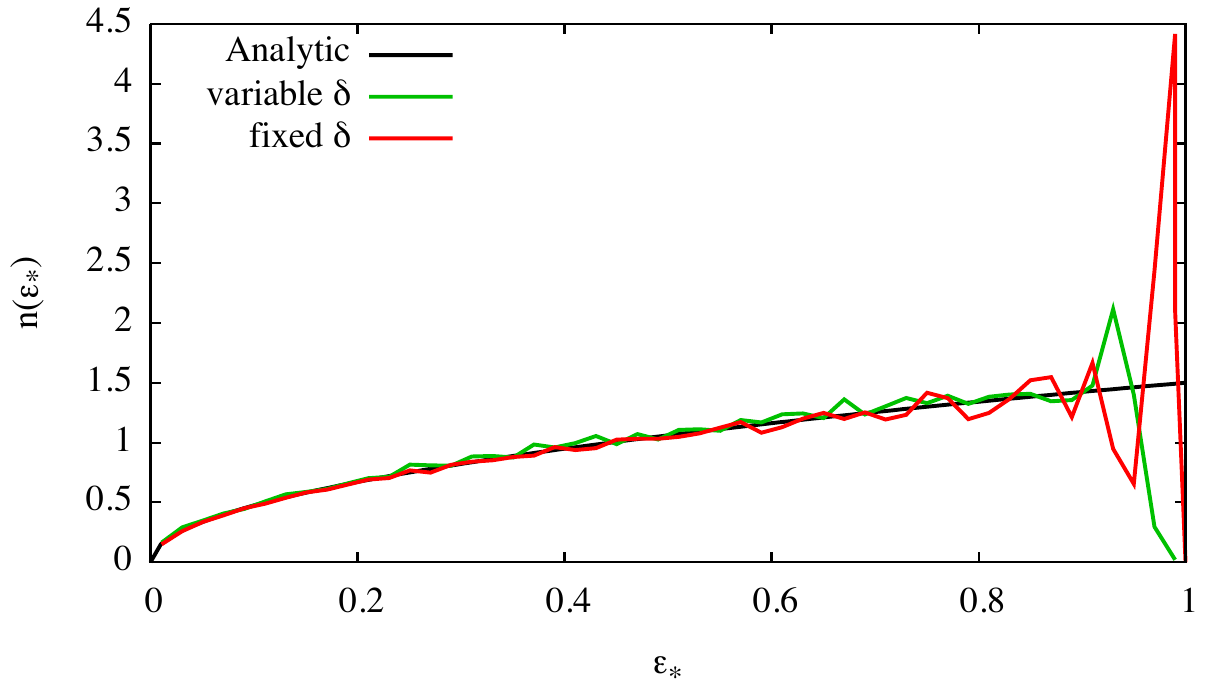}
         \caption{Normalized differential energy distributions at $t=\tau_{99\%}$ so that $\mathcal{E}=99\%U_0$, for two simulated systems of $N=4\times10^4$ particles starting with homogeneous density profiles and $K_0=0$. The red curve corresponds to the case where a minimum nearest neighbour distance $\delta$ is enforced in the initial condition at every radius. The green curve instead, to the case of initially randomly displaced particles (i.e. an arbitrary small $\delta$ may occur). The two systems have the same initial radius and total charge, and therefore the same {\it averaged density}. As comparison, the black line marks the theoretical expression given by Eq. (\ref{nesqrt}). Energies are normalized in units of the final cutoff energy $\mathcal{E}_{\rm max}$.}
\label{testsquareroot}
\end{figure} 
\subsection*{Numerical simulations using particles}\label{simulationomo}
\begin{figure} [h!t]
        \centering 
         \includegraphics[width=0.9\textwidth]{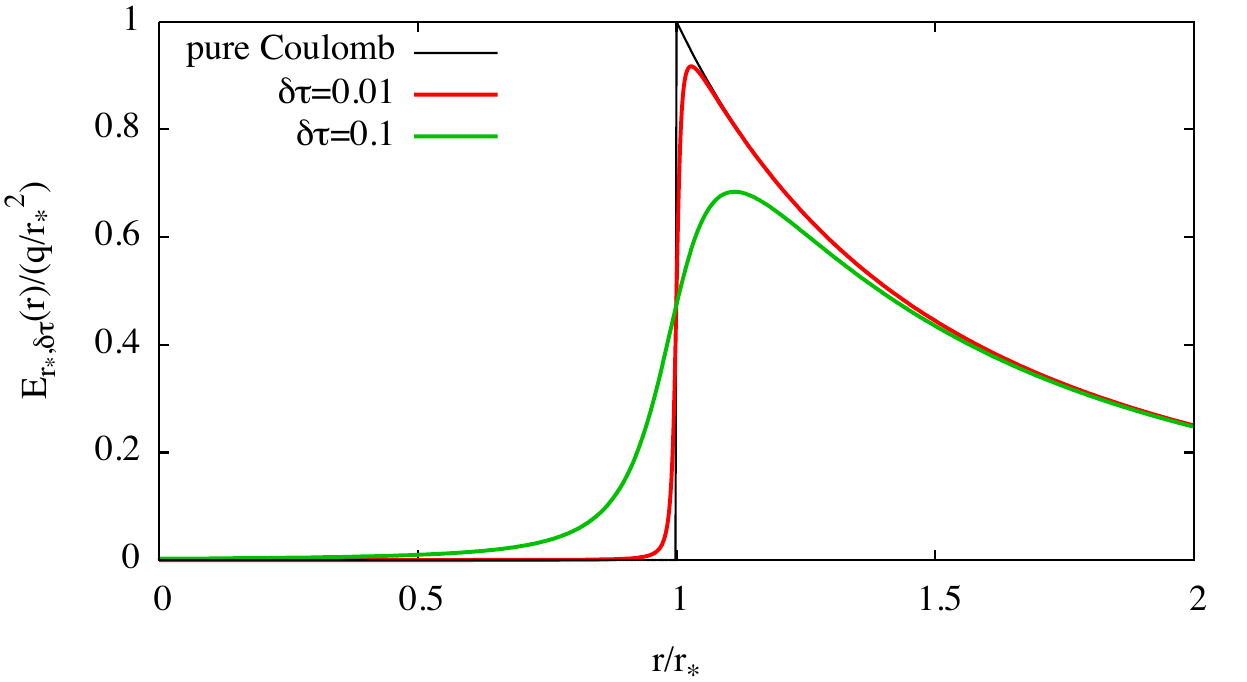}
         \caption{Electric field $E(r)$ produced by a shell with a hole with $\delta\tau=0.1$ (green curve) and 0.01 (red curve) compared to the field due to a continuum shell (black curve).}
\label{errorea}
\end{figure}

\begin{figure} [h!t]
        \centering 
         \includegraphics[width=0.7\textwidth]{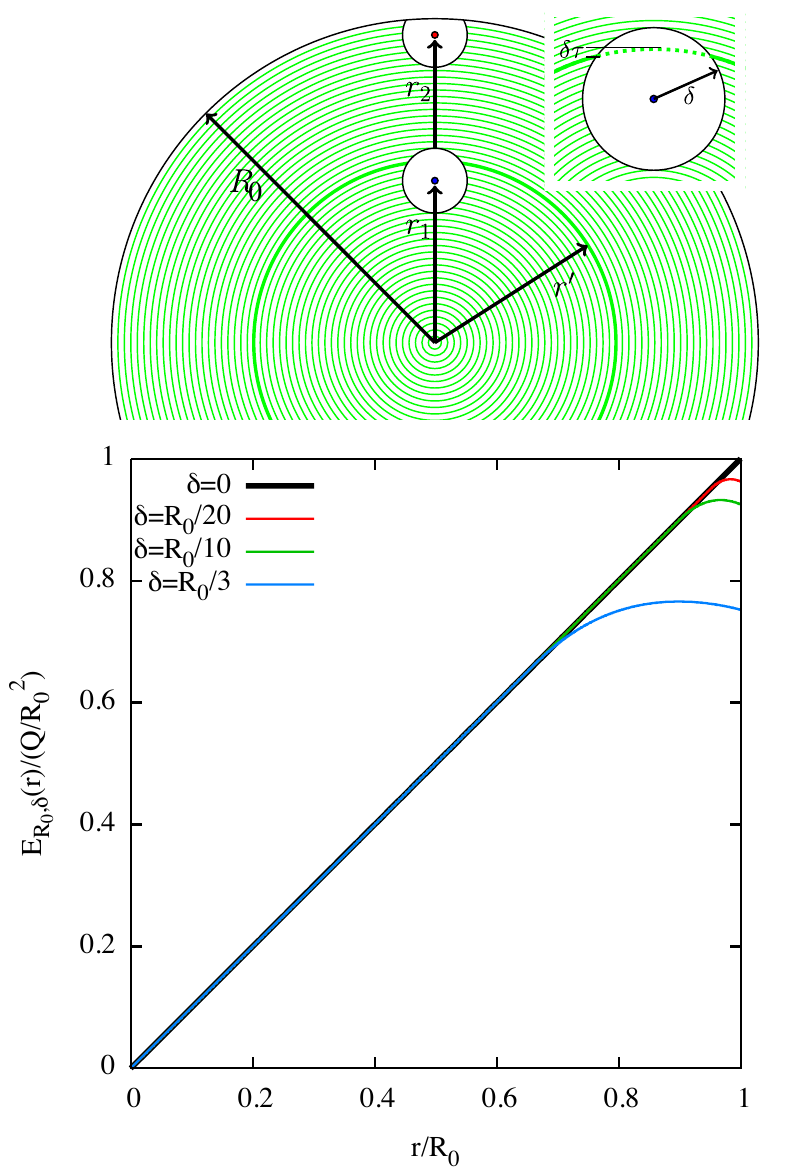}
         \caption{Top panel: Sketch of two cases when the correlation hole of radius $\delta$ is around a particle placed in the bulk at radius $r_1$, or close to the surface at radius $r_2$ inside a system of total radius $R_0$. Bottom panel: For different values of $\delta$, electric field inside a homogeneous, albeit granular, spherically symmetric distribution of total charge $Q$ and radius $R_0$ according to Eq. (\ref{attenuated}). The heavy solid line marks the ideal case of a perfectly continuos distribution ($\delta=0$).}
\label{corhole}
\end{figure}

If one takes into account its particle nature, the system discussed in Sect. \ref{contmod}, and in general every mono-component plasma, can be described by the Hamiltonian 
\begin{equation}\label{nbodyhamcl}
H=\sum_{i=1}^N\left( \frac{\mathbf{p}_i^2}{2m}+\frac{q^2}{2}\sum_{j\neq i=1}^{N}\frac{1}{|\mathbf{r}_j-\mathbf{r}_i|}\right)
\end{equation}
where $\mathbf{r}_i$ and $\mathbf{p}_i$ are position and momentum of particle $i$ and $m$ and $q$ its charge and mass. Particle's trajectories are obtained by integrating the system of $6N$ Hamilton equations
\begin{eqnarray}\label{ham1}
 \frac{{\rm d}\mathbf{r}_i}{{\rm d} t}=\frac{\partial H}{\partial\mathbf{p}_i};\quad \frac{{\rm d}\mathbf{p}_i}{{\rm d} t}=-\frac{\partial H}{\partial\mathbf{r}_i};\quad i=1,N.
\end{eqnarray}
It is possible to study the dynamics of a charged particles system, accounting for its particulate nature, only with the aid of $N$-body numerical simulations, also known as molecular dynamics simulations (MD). Several numerical approaches do exist in order to compute the forces between particles and we redirect the reader to Chap. \ref{numerica} for a description of the most widely used ones.\\
\indent Here we discuss the simulations of homogeneous systems performed with direct force calculations (see Sect. \ref{metodinbody}) and the origin of discrepancies with the continuum model.\\  
\indent Figure \ref{testsquareroot} shows the number energy distribution $n(\mathcal{E})$, when the initial potential energy $U_0$ has been essentially completely converted into kinetic energy, for two systems of $N=4\times10^4$ identically charged particles homogeneously distributed at rest in a spherical volume of initial radius $R_0$. The only difference between the two, is the way particles have been distributed in the initial condition. In one case (red curve) a minimum inter-particle distance $\delta$ is fixed, while in the other (green curve) the positions are randomly generated with no limit on the minimum distance between nearest neighbours.\\
\indent One notices immediately that both curves reproduce the theoretical square root trend up to $\mathcal{E}_*\sim0.9$. For larger energies they are instead characterized by a sharp peak which makes the numerical curve departing considerably from its analytical counterpart (black curve in Fig. \ref{testsquareroot}). For the system with arbitrarily close particle at $t=0$, the peak is ``milder'' and broader.\\
\indent The origin of such feature in the case of a homogeneous, albeit made by particles, density profile is not straightforward. It has been shown recently (see \cite{mika2013}), that due to its granular nature, a homogeneously charged sphere of radius made by particles exerts on a test particle placed close to its surface a force that is considerably lower than that due to a continuous distribution with the same total charge.\\
\indent For an ion placed at position $\mathbf{r}$ inside the cluster, the probability to find another ion at radius $\mathbf{r}^\prime$ vanishes if $|\mathbf{r}-\mathbf{r}^\prime|<\delta$, where $\delta$ is the radius of the so called correlation hole only in one of the two cases shown in Fig. \ref{testsquareroot}.\\
\indent It is possible to account analytically for the effect of a correlation hole in a distribution of charge. Let us first consider the electrostatic field at position $\mathbf{r}$ due to an infinitesimally thin shell of radius $r_*$ and charge $q$ that is given by angularly integrating
\begin{equation}\label{fullhole}
\mathbf{E}_{r_*}(\mathbf{r})=\frac{q}{4\pi}\int_0^{2\pi}\int_0^{\pi}\frac{\partial}{\partial\mathbf{r}}\frac{1}{|\mathbf{r}-\mathbf{r}_*|}\sin\theta\d \theta\d \phi,
\end{equation}
where $\mathbf{r}_*=r_*(\sin\theta\cos\phi,\sin\phi\cos\theta)$ is a vector on the shell.\\
\indent  Without loss of generality, due to the spherical symmetry of the problem, we take into account only the the radial component of  $\mathbf{E}_{r_*}(\mathbf{r})$. Setting $\tau=\cos\theta$ and performing the integration over $\phi$, the latter reads
\begin{eqnarray}\label{discr}
E_{r_*,\delta\tau}(r)=\frac{q}{2}\int_{-1}^{+1-\delta\tau}\frac{\d}{\d r}\frac{1}{\sqrt{r^2+r_*^2-2rr_*\tau}}\d\tau\nonumber\\
=\frac{q}{r^2}\left[\frac{1}{2}+\frac{(1-\delta\tau)r-r_*}{2\sqrt{r_*^2+r^2-2(1-\delta\tau)r_*r}}\right],
\end{eqnarray}
where restricting the upper boundary of integration to $+1-\delta\tau$ accounts for the presence of the correlation hole.\\
\indent  Note that in the limit of $\delta\tau\rightarrow0$ one recovers the first and the second Newton Theorems for which an homogeneously charged shell exerts no field at its interior and is proportional to $q/r^2$ outside, cfr. Ref. \cite{kellogg}.\\
\indent In Fig. \ref{errorea} the field produced by a shell with a hole is shown for two values of $\delta\tau$. Note how the field is non zero at the interior of the shell. It is expected that, 
for particles belonging to shells placed in the bulk of the system, the modification of the electric field due to the  hole is on average compensated by the contribution of outer shells. For particles placed at the surface such compensation should be in principle not possible.\\
\indent Let us now evaluate the field on a particle with a hole of radius $\delta$ placed close to the surface (particle radius $r_2$ in the sketch in Fig. \ref{corhole}), due to an extended distribution with homogeneous charge density. Setting $\delta\tau=(\delta^2-(r-r_*)^2)/(2r_*r)$ for the shells intersecting the hole (i.e. $|r-r_*|\leq\delta$) and zero otherwise in Eq. (\ref{discr}), and integrating over $r_*$ gives
\begin{align}\label{attenuated}
E_{R_0,\delta}(r)=
\begin{cases}
Qr/R_0^3;\quad\quad\quad &{\rm for}\quad r<R_0-\delta \\
A_{R_0,\delta}(r)Qr/R_0^3;\quad &{\rm for}\quad R_0-\delta<r<R_0,
\end{cases}
\end{align}
where the ``attenuation factor" $A_{R_0,\delta}(r)$ is defined by
\begin{equation}
A_{R_0,\delta}(r)=\frac{(r+R_0-\delta)^2(2(r+R_0)\delta+\delta^2-3(r-R_0)^2)}{16\delta r^3}.
\end{equation}
As seen in Fig. \ref{corhole} (bottom panel), the radial electric field $E_{R_0,\delta}(r)$, does not increase linearly with $r$ up to the surface, but starts abruptly to increase markedly sublinearly\footnote{In other words, the closer a particle is to the surface, the smaller is the compensation on the underestimated radial electrostatic field generated by particles at lower radii, due to the spurious (with respect to the continuum picture) internal field of charged discrete shell placed at larger radii.} for $r>R_0-\delta$.\\
\indent Since the asymptotic energy distribution is entirely determined by the initial state of the system, if we assume a self similar expansion with uniform scaling factor $\eta$ also in presence of correlation holes, and therefore
\begin{equation}
E_{\eta R,\eta\delta}(\eta r)=E_{\eta R_0,\delta}(r_0)/\eta^2, 
\end{equation}
\begin{figure} [h!t]
        \centering 
         \includegraphics[width=0.85\textwidth]{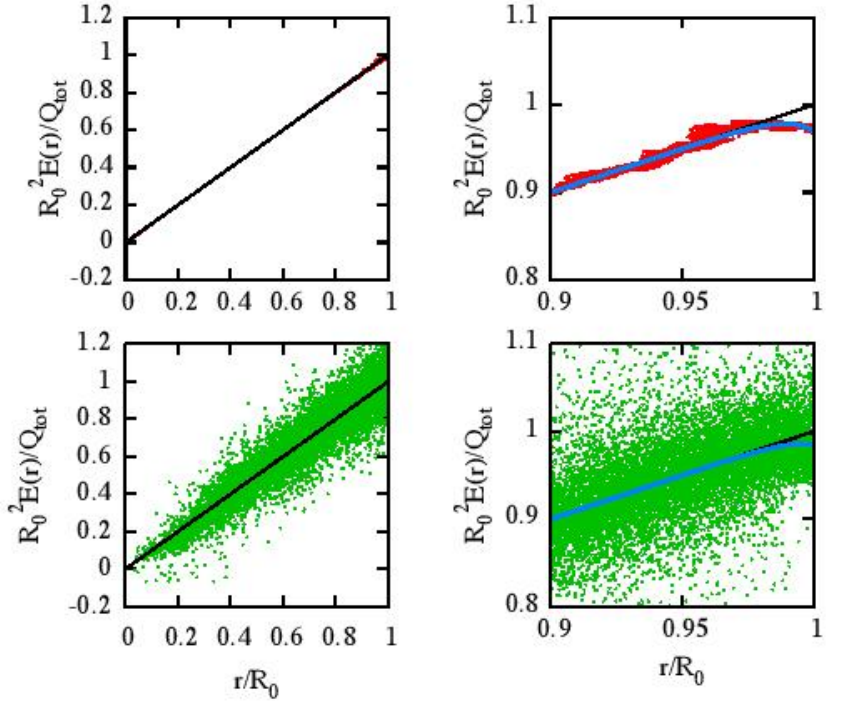}
         \caption{Left panels: radial component of the electric field acting on particle at radial coordinate $r$ for the initial condition relative to Fig. \ref{testsquareroot}, (points) and radial electric field inside an homogeneously charged sphere of radius $R_0$, (solid line) for the two cases where a minimum inter-particle distance is fixed (top) and where the particles are randomly displaced (bottom). Right panels: same as left but in the radial interval $0.9\leq r/R_0\leq 1$, the heavy blue line marks the averaged radial electric field $\langle E(r)\rangle$ in the particle distribution, note how in both cases it becomes markedly sublinear as $r$ approaches the surface.}
\label{erroreb}
\end{figure} 
we can compute the asymptotic (kinetic) energy as function of the position $r_0$ in the initial state.
The asymptotic kinetic energy as a function of $r_0$ reads
\begin{equation}
\mathcal{E}(r_0)=r_0\int_1^{+\infty}E_{\eta R_0,\eta\delta}(\eta r_0)\d\eta=rE_{R_0,\delta}(r_0).
\end{equation}
By plugging the latter into Eq. (\ref{derivne1}), and having assumed a finite energy resolution $\delta\mathcal{E}$ so that the Dirac $\delta(x)$ is replaced by 
\begin{equation}
D_{\delta\mathcal{E}}(x)=\exp((-x/\delta\mathcal{E})^2)/\sqrt{\pi}\delta\mathcal{E},
\end{equation}
one obtains an asymptotic $n(\mathcal{E})$ following the square root behavior for a broad interval of energies and peaking close to the cutoff energy in a similar fashion of the numerical curves shown in Fig. \ref{testsquareroot}.\\
\indent The high energy behavior of $n(\mathcal{E})$, when a correlation hole of radius $\delta$ is present, is to be interpreted as the fact that the kinetic energies reached by the particles initially placed at a distance from the surface smaller than $\delta$ are lower than those attained by an element of volume in an idealized continuous distribution starting from the same radius. This causes the ``bunching" of such energies that is seen as a peak in the differential energy distribution.\\
\indent We shall call this feature a {\it discreteness peak} to distinguish it from the similarly looking feature of systems with non-uniform initial density profile that we will discuss in the next section.\\ 
\indent It must be stressed out that the discrete nature of the system implies that, the discrepancy between $E(r)$ computed in the continuum model and for particles, depends strongly on how the positions of the particles in the initial state are selected.\\
\indent In Fig. \ref{testsquareroot} the discreteness peak is sharper for the system where particles are initially placed enforcing a minimal nearest neighbours distance (blue curve). In the case without a correlation hole, i.e. fully random positions, (green curve), this feature is broadened by the combined effect of a more randomized contribution to force due to the near neighbours and position specific size of the correlation hole. However, the discreteness peak is not entirely removed since the local arrangement of the particles in proximity of the surface acts like a correlation hole.\\
\indent In Fig. \ref{erroreb} the radial component of the electrostatic field $E(r)$ acting on the particles is shown for the initial states relative to Fig. \ref{testsquareroot}. In both cases, where a minimum inter-particle distance is or is not imposed, $E(r)$ presents large deviations from its theoretical value (indicated by the thin black line) due to the randomization of the contribution of the nearby particles. The averaged radial electric field $\langle E(r)\rangle$ (blue lines in right panels of same figure) clearly drops close to the surface with respect to the linear trend of $E(r)$ predicted by the continuum model, in the same fashion of Fig. \ref{corhole} (lower panel).
For the model with enforced minimum inter-particle distance $\delta$ (red dots), $\langle E(r)\rangle$ departs more from the linear trend although overall $E(r)$ is less noisy than in the model with arbitrarily small $\delta$ (green dots). Nevertheless, in the latter case the noisiness of $E(r)$ partially compensates the effect of discreteness related to the correlation hole which results also in a less pronounced peak 
in the final $n(\mathcal{E})$.
\subsubsection*{Smoothed Coulomb interaction}
Curiously, in direct $N$-body simulations, the modification of the pair Coulomb potential and force introduced in order to avoid its divergence for vanishing separation (see Chap. \ref{numerica}), introduces an effect that lowers in a similar way the electrostatic field at the surface of an homogeneous sphere made by particles.\\
\indent If the Coulombian $1/r$ potential is substituted with 
\begin{equation}\label{pluplu}
V(r)=\frac{1}{\sqrt{r^2+\epsilon^2}},
\end{equation}
\begin{figure} [h!t]
        \centering 		
         \includegraphics[width=\textwidth]{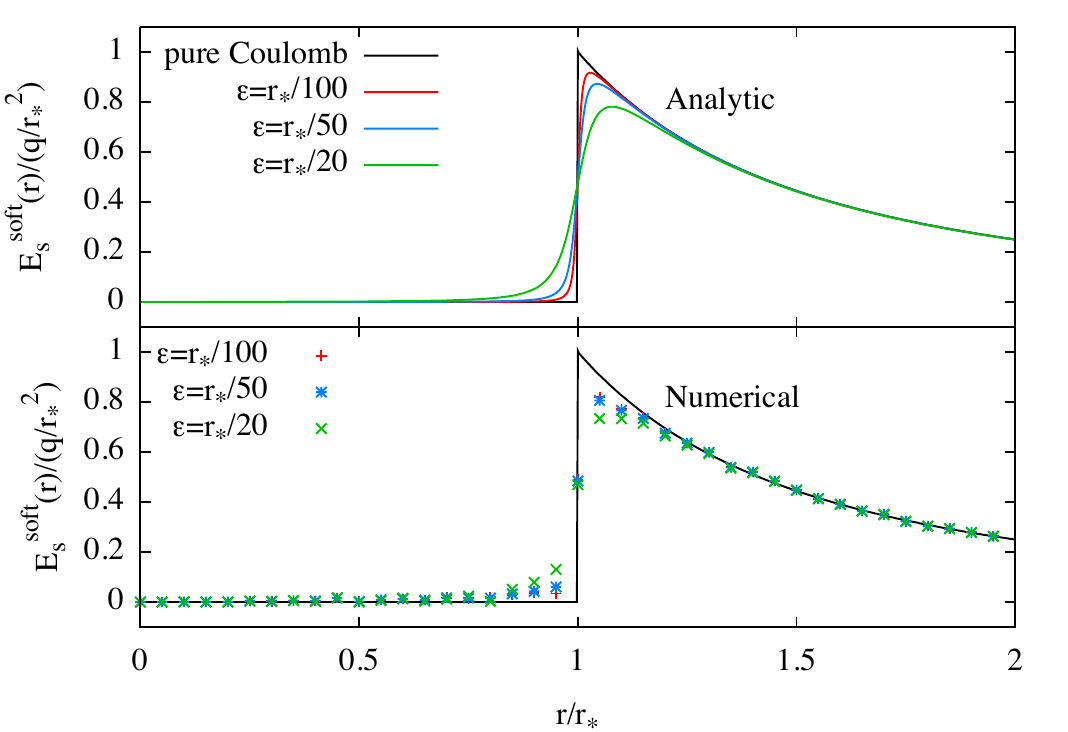}
         \caption{Top panel: for different values of $\epsilon$ in units of $r_*$, softened field $E_{\rm s}^{\rm soft}(r)$ at radius $r$ evaluated analytically for a shell of density $\sigma$. Bottom panel: same as above but computed for a discrete distribution of $N=10^{5}$ particles placed on a shell, computed with direct summation and averaged over 20 realizations.}
\label{figsofterr}
\end{figure} 
where $\epsilon$ is the so called softening length, the electrostatic field produced by a point charge is also non Coulombian.\\
\indent Since $1/\sqrt{r^2+\epsilon^2}$ is not a valid Green function for the Laplace operator, also the field produced by an extended distribution of particles (or continuum) is expected to differ from that calculated using the real Coulomb interaction.\\
\indent Let us consider an infinitesimally thin shell of radius $r_*$ and surface density $\sigma=q/(4\pi r_*^2)$, where $q$ is its total charge. With the modification of the Coulomb interaction given in Eq. (\ref{pluplu}), the radial component of the electric field exerted by the surface element $\delta s$ on a point placed at distance $r$ from the centre of the shell, reads
\begin{equation}\label{softeq}
\delta E_{\rm s}^{\rm soft}(r)=\frac{\sigma\delta s(r_*\cos\theta-r)}{(a^2+\epsilon^2)^{3/2}},
\end{equation} 
where $a=\sqrt{r_*^2+r^2-2r_*r\cos\theta}$ is the distance of the shell element from the point where the field is evaluated and $\theta$ the angle between the vectors of length $a$ and $r$, with origin set in that point. 
Setting $\delta s=r_*^2{\rm d}\psi\sin\theta{\rm d}\theta$ in Eq. (\ref{softeq}) and performing the integration over the two angular variables $\psi$ and $\theta$, the field at 
\begin{figure} [h!t]
        \centering 		
         \includegraphics[width=0.7\textwidth]{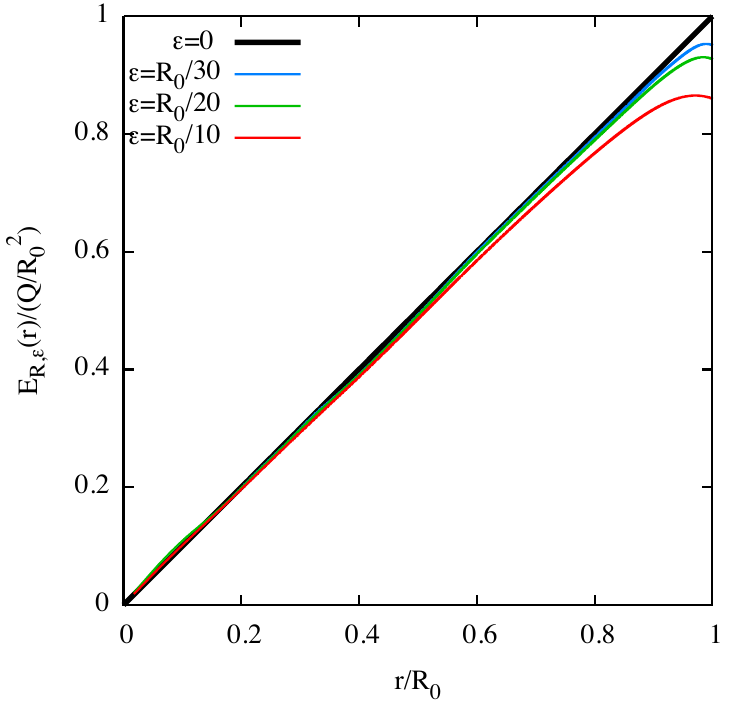}
         \caption{For different values of $\epsilon$ in units of the system's $R_0$, softened electric field at $E_{R_0,\epsilon}(r)$ inside an homogeneous sphere with softened interaction.}
\label{figsofthomo}
\end{figure} 
distance $r$ due to the whole shell is
\begin{equation}\label{esoft}
E_{\rm s}^{\rm soft}(r)=2\pi\sigma r_*\left[\frac{r_*r+r_*^2+\epsilon^2}{r^2\sqrt{\epsilon^2+(r_*+r)^2}}+\frac{r_*r-r_*^2-\epsilon^2}{r^2\sqrt{\epsilon^2+(r_*-r)^2}}\right],
\end{equation}
from which in the limit of $\epsilon\rightarrow0$ one obtains the known result that $E_{\rm s}^{\rm soft}(r)=q/r^2$ for $r\geq r_*$ and $E_{\rm s}^{\rm soft}(r)=0$ for $r<r_*$.\\
\indent In Fig. \ref{figsofterr} (top panel) we show the analytical estimation of the softened $E_{\rm s}^{\rm soft}(r)$ for different choices of $\epsilon$, and its numerical computation averaged over 20 realizations, with a direct summation code for a system of $N=10^5$ particles distributed on a spherical shell  (bottom panel).\\
\indent It is clearly evident that with this modification of the Coulomb interaction, a test particle placed inside a spherical charged shell would be prone to a radial force, directed towards the centre if the particle and the shell have charges of opposite sign, or directed outside in the opposite case. For $r>r_*$ instead, the field $E_{\rm s}^{\rm soft}(r)$ results underestimated respect to the field $E_{\rm s}(r)$ calculated with the real coulomb interaction. The effect is more pronounced the larger $\epsilon$ is in units of the shell's radius $r_*$.\\
\indent For a homogeneous charge distribution of density $\rho$ and radius $R_0$, integrating the contribution of every infinitesimal shell given by Eq. (\ref{esoft}), gives a trend of the softened electrostatic field $E_{R,\epsilon}(r)$ similar to that obtained in Eq. (\ref{attenuated}) which takes into account the effect of the correlation hole.
Figure \ref{figsofthomo} shows the softened electric field produced by a uniformly charged sphere acting on a particle of unit charge placed at its interior.\\ 
\indent It is evident how $E_{R_0,\epsilon}(r)$ substantially departs from the linear trend of the real force as $r$ approaches the system's edge $R_0$. The larger $\epsilon$ is the more important such deviation is.\\
\indent In conclusion, this means that using a softened interaction in numerical calculations may spuriously increase the effect of the correlation hole between particles. This is overcome by choosing $\epsilon<\delta$.
\section{Shock shells in non-uniform density profiles}\label{modellinonuni}
To this point we have studied the Coulomb explosion of systems having a uniform initial density profile. In reality, clusters ionized by strong laser pulses may have radially non uniform charge densities, due to their intrinsic initial structure or to the details of the ionization process.\\
\indent In this case, the dynamics of the expansion might be highly non trivial (see e.g. \cite{2003PhRvL..91n3401K} and \cite{kovalev05}). For instance, when the radial component of the electric field $E(r)$ is non monotonic with $r$ and has its maximum inside the cluster, inner ions may eventually move faster than those initially placed at larger radii. This implies that sooner or later they will reach and then overtake them. Hereafter, we will refer to overtaking as ``shell crossing".  While inner ions rapidly catch up the outer ones, the charge density depletes in the central region of the cluster and starts increasing near its edge leading to the formation of what is called a shock.\\
\indent In the continuum or fluid picture, the shock wave, or shock front, is defined as a propagating discontinuity in the systems properties (i.e. velocity field, density, pressure), \cite{meyer}. In our case it corresponds to a divergence in the cluster's density profile.\\
\indent The Coulomb explosion of non-uniform systems and the control of the shock wave's dynamics have received much attention (see \cite{2005PhRvL..94c3401P}, \cite{peano2007}) due to their possible application in the context of intra-cluster fusion, \cite{2001PhRvL..87c3401L}, \cite{2013PhRvE..87b3106B}. We are now going to sketch the problem in the continuum picture and then we will discuss the results of our particle based numerical calculations. 
\subsection*{Continuum model}
Under the assumption of non collisionality (i.e. the effects of particles encounters are negligible), when passing to the continuum approximation, the Coulomb explosion can be treated in principle with the equations of pressureless ($P=0$) hydrodynamics written in Sect. \ref{contmod}. Note that, in correspondence of a shell crossing, the solution of the latter becomes multivalued since at the same radius elements matter with different velocities are found.\\
\indent  The problem of an exploding cluster with non uniform initial density profile has been studied in Ref. \cite{2003PhRvL..91n3401K}, for the one parameter family of initial density profiles 
\begin{figure} [h!t]
        \centering 
         \includegraphics[width=\textwidth]{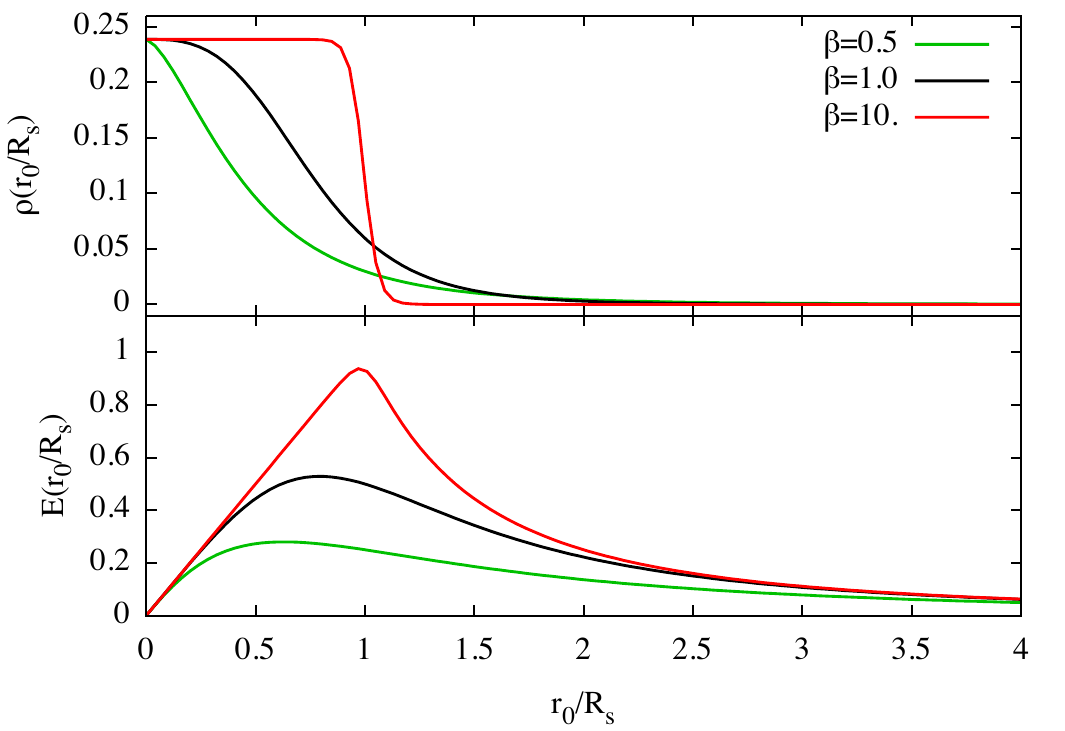}
         \caption{Top panel: density profiles given by Eq. (\ref{proflekaplan}) for $\beta=0.5,1$ and 10. Bottom panel: corresponding radial electric field. Radii are given in units of the scale radius $R_s$.}
\label{nonuniform}
\end{figure} 
given by
\begin{equation}\label{proflekaplan}
\rho(r_0)=\frac{3Q}{4\pi R_s^3}{\left[1+\left(\frac{r_0}{R_s}\right)^{3\beta}\right]^{-1-1/\beta}};\quad 1/3<\beta<\infty,
\end{equation}
where $Q$ is the total charge and $R_s$ a scale radius. The charge enclosed by radius $r_0$ then reads
\begin{equation}
Q(r_0)=\int_0^{r_0}\rho(r_0^\prime)r_0^{\prime 2}\d r^\prime=\frac{Qr_0^3}{R_s^3\left[1+\left(\frac{r_0}{R_s}\right)^{3\beta}\right]^{1/\beta}}.
\end{equation}
Note that for finite $\beta$ in Eq. (\ref{proflekaplan}), the density falls off to zero at infinity, while in the limit of $\beta\rightarrow\infty$, $\rho(r_0)$ tends to the uniform profile of Eq. (\ref{steplikerho}), where $R_s=R_0$. In Fig. \ref{nonuniform} we show for some values of $\beta$ the density profile $\rho(r_0)$ and the radial component of the electric field it generates.\\
\indent Formally, the dynamics of the expansion is given by the solutions of the system of ODEs
\begin{eqnarray}\label{cenonuinisistem}
\rho(r)&=&\frac{1}{4\pi r^2}\frac{\d Q(r)}{\d r};\nonumber\\
\frac{\d^2r}{\d t^2}&=&\frac{Q}{M}\frac{Q(r)}{r^2}, 
\end{eqnarray}
that are nothing but the characteristics of the equations of hydrodynamics.\\  
\indent For reasons of clarity, let us now use normalized variables with respect to the system's initial parameters, $s=r/R_s$ and $\tau=t/t_s$ where
\begin{equation}
t_s=\sqrt{MR_s^3/Q^2}.
\end{equation}
With this choice, the normalized density is
\begin{equation}  
\rho(s)=\rho(r)R_s^3/Q
\end{equation}
and the system (\ref{cenonuinisistem}) becomes
\begin{eqnarray}\label{cenonuinisistemnorm}
\rho(s)&=&\frac{1}{4\pi s^2}\frac{\d Q(s)}{\d s};\nonumber\\
\frac{\d^2s}{\d \tau^2}&=&\frac{Q(s)}{s^2}.
\end{eqnarray}
Up to the critical time $\tau_{\rm c}$ when the shell crossing generates the shock, the first integral of the second equation is simply
\begin{equation}\label{firstint}
\frac{1}{2}\left(\frac{\d s}{\d\tau}\right)^2=Q(s_0)\left(\frac{1}{s_0}-\frac{1}{s}\right),
\end{equation}
where $Q(s_0)$ is the charge enclosed by the radius $s_0$ at $\tau=0$. Trajectories $s(\tau)$ of elements of volume initially placed at $s_0$ are the implicit solutions of
\begin{equation}
\sqrt{\xi(\xi-1)}+\ln(\sqrt{\xi}+\sqrt{\xi-1})=\tau\sqrt{2Q(s_0)/s_0^3};\quad \xi=s/s_0,
\end{equation}
in the same fashion of the case of a homogeneous sphere, cfr. Eq. (\ref{sss}).
The charge density then reads
\begin{equation}\label{ultimarho}
\rho(s)=\left(\frac{s_0}{s}\right)^2\frac{\d s_0}{\d s}.
\end{equation}
For $\tau\geq\tau_{\rm cr}$, the solutions above are invalid since instead of Eq. (\ref{firstint}), one has now
\begin{equation}
\frac{1}{2}\left(\frac{\d s}{\d\tau}\right)^2=\int\frac{Q(s)}{s^2}\d s.
\end{equation}
At $\tau=\tau_{\rm cr}$ the derivative of the velocity profile
\begin{equation}
\frac{\d v(s)}{\d s}=\frac{\d}{\d s}\left(\frac{\d s}{\d \tau}\right)
\end{equation}
diverges at the critical coordinate $s_{\rm cr}$ where the crossing takes place. Such divergence in the velocity profile corresponds to diverging density in $s_{\rm cr}$, as one can still\footnote{Once $\rho(s)$ becomes multivalued, 
Eq. (\ref{ultimarho}) is invalid and it is substituted by $\rho(s)=(4\pi s^2)^{-1}\d Q(s)/\d s$.} obtain from Eq. (\ref{ultimarho}).\\
\begin{figure} [h!t]
        \centering 		
         \includegraphics[width=0.95\textwidth]{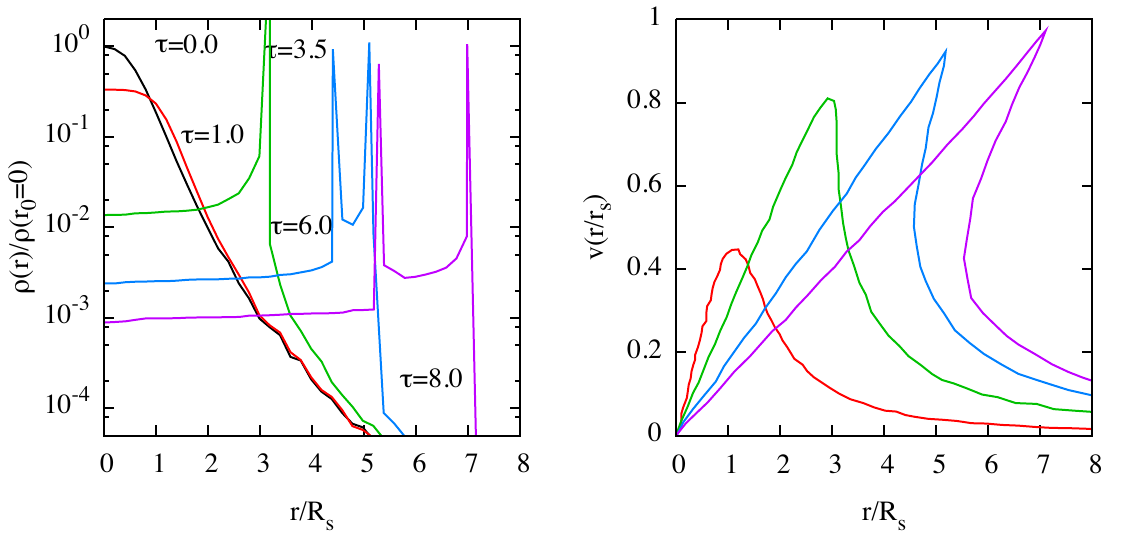}
         \caption{Left panel: formation and evolution of the shock shell in a cluster starting with density profile given by Eq. (\ref{proflekaplan}) where $\beta=1$. Right panel: velocity field at the same times. To improve visualization, the densities are normalized in units of the initial central density $\rho(r=0)$ and radii in units of the initial scale radius $r_s$. Times $\tau$ are given in units of $\sqrt{Mr_s^3/Q^2}$.}
\label{hydro}
\end{figure} 
\indent At later times, $v(s)$ stays multivalued in the region (shock shell) $s_{\rm 2}<s<s_{\rm 1}$ where $s_{\rm 1}$ and $s_{\rm 2}$ are the radii of the so called leading and trailing shocks respectively.\\ 
\indent Numerical integration of (\ref{cenonuinisistemnorm}) for various initial density profiles\footnote{Actually this is the case for all densities smoothly going to 0 at infinity.} given by Eq. (\ref{proflekaplan}), revealed that the initial width of the shock shell is narrower in the limit of large $\beta$. In each case, however, it always broadens with time, spanning almost the entire system.\\  
\indent Interestingly, even in the limit of large $\beta$, the coordinate and the critical time of the shock formation do not tend respectively to the systems' edge and to 0 as one would expect. Instead one has $s_{\rm cr}\simeq0.635$ and $\tau_{\rm cr}\simeq1.237$ as roots of  
\begin{eqnarray}
{\sqrt{s(s-1)}+\ln\left(\sqrt{s}+\sqrt{s-1}\right)}=\frac{2s^{3/2}}{3\sqrt{s-1}}
\end{eqnarray}
and
\begin{eqnarray}
\sqrt{s(s-1)}+\ln\left(\sqrt{s}+\sqrt{s-1}\right)=\tau\sqrt{2},
\end{eqnarray}
see e.g. \cite{2003PhRvL..91n3401K}.\\
\indent This implies in other words, that even an infinitesimal initial perturbation near the edge of the homogeneous sphere treated in Sect. \ref{contmod}, is enough to give rise to shocks, thus breaking the self similarity of the expansion.\\
\indent Remarkably, if one considers truncated initial density profiles, defined as
\begin{equation}
 \rho(r_0)=\varrho(r_0)\theta(R_0-r_0)
\end{equation} 
where $\varrho(r_0)$ is a monotonically decreasing function of $r_0$ and $R_0$ the cutoff radius, the leading shock front disappears as it rapidly meets the discontinuity of the density gradient $\d\rho/\d r$ at the edge of the system. This is qualitatively similar to what happens to a shock front moving through a non homogeneous medium, (see e.g. \cite{1976ApJ...209..424C}, \cite{bykov}).\\
\indent If a singularity arises in the density profile, associated with a singularity in the velocity profile, one may expect that also the kinetic energy distribution presents a divergence in correspondence of the kinetic energy of the shock front.\\
\indent As example, in Fig. \ref{hydro}, we show at different times the density for a cluster with 
initial profile given by Equation $\ref{proflekaplan}$ for $\beta=1$, as well as the associated velocity profile. The curves are obtained by numerically integrating the hydrodynamics equations. The formation of 
a ``singularity" (due to the finite resolution it is a sharp peak) in the density profile, happens already at early stages of the explosion ($\tau\sim3.5$) when the velocity profile has become multivalued (see right panel and analogous plots in Ref. \cite{2003PhRvL..91n3401K}).\\
\indent Unfortunately, to derive the asymptotic $n(\mathcal{E})$ from the potential energy of the initial state is in principle not possible for an arbitrary initial density profile, using the method described in Sect. \ref{contmod}. One has to use instead an approach based on kinetic theory, we redirect the interested reader to Appendix \ref{kt} and the literature therein referenced.
\subsection*{Numerical simulations using particles}\label{singlece}
\begin{figure} [h!t]
        \centering 
         \includegraphics[width=\textwidth]{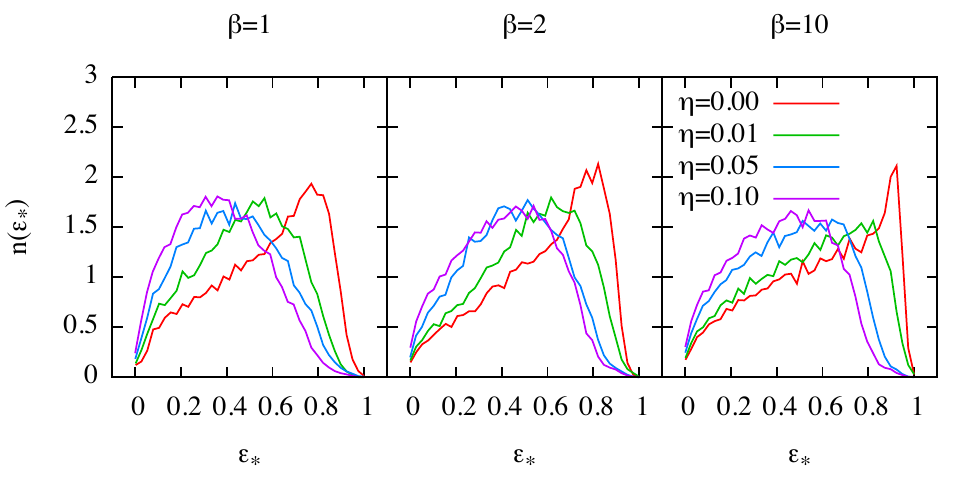}
         \caption{From left to right, scaled final differential energy distribution for $\beta=1,2$ and 10 in the initial condition given by Eq. (\ref{proflekaplan}), and different values of the ratio $\eta$.}
\label{betane}
\end{figure}

Numerical simulations based on particles, aiming at studying intra-cluster nuclear fusion,  have been performed in \cite{2005PhRvL..94c3401P}, \cite{peano2007} and \cite{2006PhRvA..73e3202P} with more realistic initial conditions where a non negligible electron density were considered. This has revealed that radial shocks in the ion density profile still occur, even when the residual electron density screens part of the ions.\\ 
\indent In this work we are interested mainly in the structure of the asymptotic number energy distribution $n(\mathcal{E})$, since it is an experimentally deducible quantity (from the time of flight spectrum), and how it is affected by an initial velocity distribution. To this scope, we have performed $N$-body simulations of pure Coulomb explosion (i.e. no electron contribution) for different non uniform initial density profiles, with a single kind of particles of mass $m$ and charge $q$.\\
\indent We discuss here the properties of the final states of two families of initial density profiles. 
In the first case, to model systems characterized by a flat core (i.e. almost homogeneous central region) and a smoothly decaying density in the outer layer, we use the expression for $\rho(r_0)$ given by Eq. (\ref{proflekaplan}) where $\beta$ controls the steepness of the system's edge. 
In the second case, where we model instead systems with a highly dense central region and an outer layer decaying with a fixed slope, we use the family of $\gamma-$models 
\begin{equation}\label{denhen1}
\rho(r_0)=\frac{Q(3-\gamma)}{4\pi }\frac{R_s}{r_0^{\gamma}(R_s+r_0)^{4-\gamma}};\quad 0\leq\gamma<3,
\end{equation}
where $R_s$ is a scale radius and $\gamma={\rm d}{\log}(\rho)/{\rm d}s$ is the so called logarithmic density slope. The charge enclosed by radius $r_0$ is given by
\begin{equation}
Q(r_0)=Q\left(\frac{r_0}{r_0+R_s}\right)^{3-\gamma}. 
\end{equation}
Note that, all the profiles given by Eq. (\ref{denhen1}) fall of as $r_0^{-4}$ for $r_0\gg R_s$ and for $\gamma\neq 0$ diverge for $r_0\rightarrow 0$; the latter is not really an issue since we consider discrete particles in the simulations. 
The expression (\ref{denhen1}) was introduced by Dehnen \cite{1993MNRAS.265..250D} in the context of galactic dynamics to model spherical galaxies with prominent central density cusp. We use it here to generate our initial conditions only for the fact that varying $\gamma$ allows one to control the ``importance" of the central density cusp, from a flat core ($\gamma=0$) to an extremely steep cusp ($\gamma\rightarrow3$).\\
\indent In both families, the density profiles are formally extended to infinity where $\rho$ falls to zero. In particles simulations they are obviously intrinsically truncated due to the finite number of particles, however the initial conditions are not characterized by a sharp edge.\\
\indent The initial ion velocities $v_{0,i}$ are sampled from a position independent Maxwellian distribution  
\begin{equation}
n(v)=4\pi\left(\frac{1}{2\pi}\right)^{3/2}v^2\exp(-v^2/2).
\end{equation}
and then renormalized to obtain the wanted value of the total initial kinetic energy 
\begin{equation}
K_0=\sum_{i=1}^N \frac{mv_{0, i}^2}{2}.
\end{equation}
For a system of $N$ particles sampled from a given density profile, the initial potential energy is given by
\begin{equation}
U_0=\frac{q^2}{2}\sum_{j\neq i=1}^N \frac{1}{|\mathbf{r}_{0,i}-\mathbf{r}_{0,j}|},
\end{equation}
where $\mathbf{r}_{0,i}$ are the particle positions at time 0. Hereafter, we refer to the ratio $\eta={K_0}/{U_0}$ as ``initial ion temperature".\\
\indent We define the dynamical time scale of the simulated system as
\begin{equation}\label{dyntime}
t_{\rm dyn}\equiv2\pi\sqrt\frac{m}{q^2\tilde n_0},
\end{equation}
where $\tilde n_0$ is the average particle density inside the radius $r_h$ enclosing half of the particles at initial time. As a rule, the calculations are run up to the time when the system's kinetic energy equals $99\%$ of the total energy $\mathcal{E}_{\rm tot}=K_0+U_0$.\\ 
\indent Figure \ref{betane} shows $n(\mathcal{E})$ at final time for systems starting from the smoothed-step density profile of Eq. (\ref{proflekaplan}). In order to compare the curves on the same scale, energies are rescaled with respect to the maximal energy $\mathcal{E}_{\rm max}$ attained by 
\begin{figure} [h!t]
        \centering 
         \includegraphics[width=\textwidth]{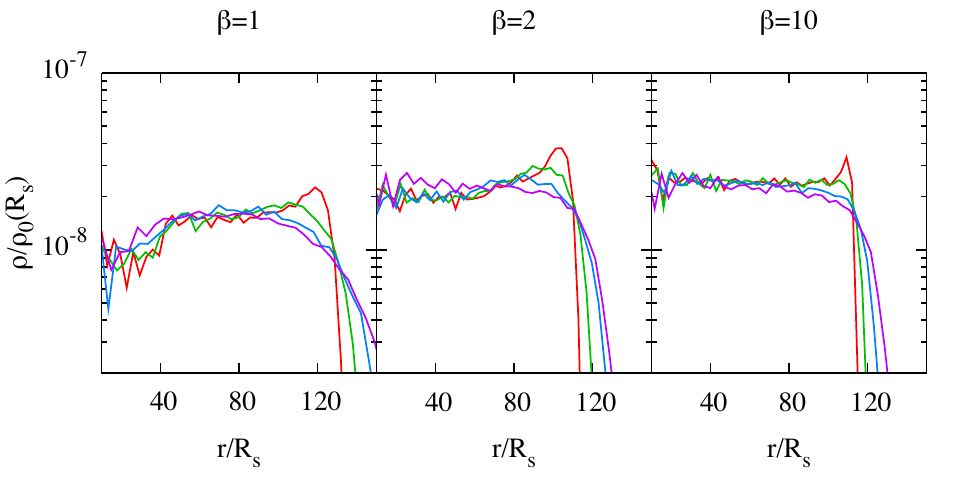}
         \caption{Density profiles for the systems of Fig. \ref{betane}.}
\label{betarho}
\end{figure} 
the particles (averaged over the 20 fastest ones). We observe that when $\eta=0$ (red curves), the final $n(\mathcal{E})$ is characterized by a peak close to $\mathcal{E}_{\rm max}$ that gets sharper as $\beta$ increases (i.e. the initial density profile tends to the uniform), and a long tail at low energies.\\
\indent If the particles have already some kinetic energy at $t=0$, as for instance in system where some energy has been transferred from the laser pulse also to the ions, the peak in the final energy is ``smeared'' and $n(\mathcal{E})$ has its maximum value at 
\begin{figure} [h!t]
        \centering 
         \includegraphics[width=\textwidth]{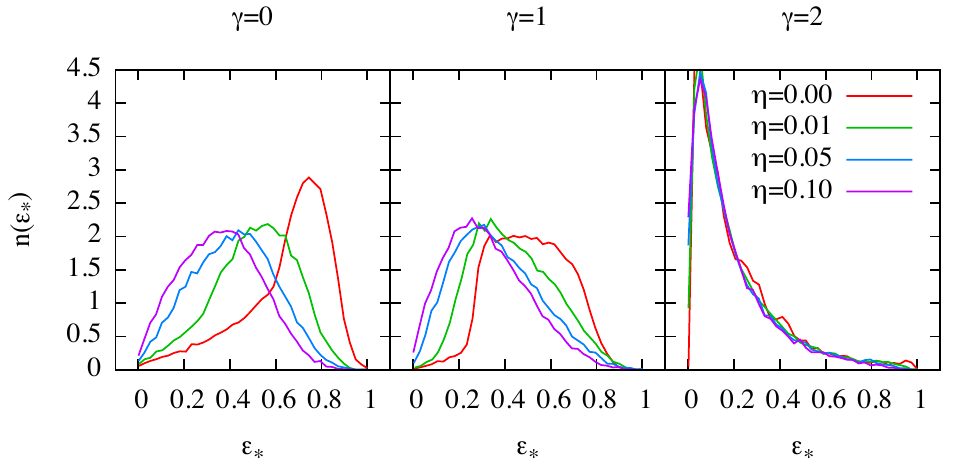}
         \caption{From left to right, scaled final differential energy distribution, for $\gamma=0,1$ and 2 in the initial condition given by Eq. (\ref{denhen1}), and different values of the ratio $\eta$.}
\label{gammane2}
\end{figure} 
\begin{figure} [h!t]
        \centering 
         \includegraphics[width=\textwidth]{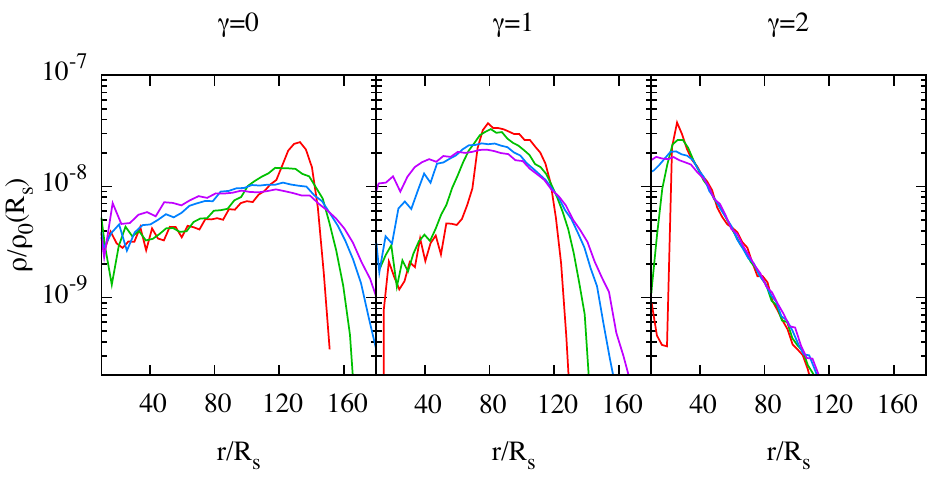}
         \caption{Density profiles for the systems of Fig. \ref{gammane2}.}
\label{gammarho2}
\end{figure} 
lower values of $\mathcal{E}$.\\
\indent The corresponding density profiles $\rho(r)$ are shown in Fig. \ref{betarho}. A density peak in proximity of the cluster's edge is evident for all the systems starting form cold initial conditions (red curves). Such feature is the ``remnant'' of the trailing front of the shock shell and appears to be narrower for larger values of $\beta$, consistently with what predicted in \cite{2003PhRvL..91n3401K}.\\
\indent When ions are starting with a non zero temperature, the density peak becomes 
\begin{figure} [h!t]
        \centering 
         \includegraphics[width=\textwidth]{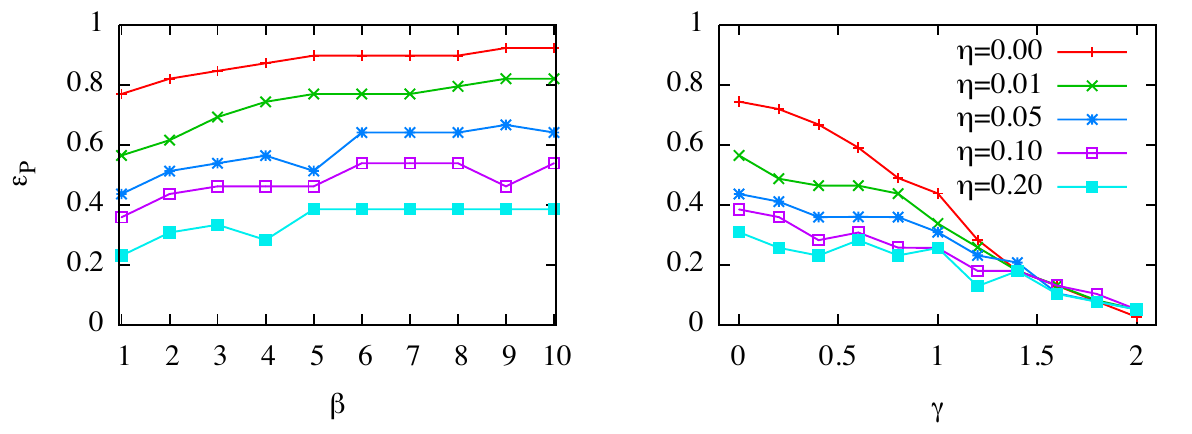}
         \caption{Left panel: as a function of $\beta$, position of the maximum of the final $\mathcal{E}$ for different values of the initial energy ratio $\eta$ in the interval (0; 0.2). Right panel: same as left but in this case the initial distribution of the simulation particles is extracted from a truncated $\gamma-$model.}
\label{displaced}
\end{figure} 
less prominent and disappears entirely for $\eta>0.02$. Generally, different initial temperature produce qualitatively different final density profiles with several slope changes. This is less evident for large values of $\beta$, see the case $\beta=10$ in Fig. \ref{betarho}.\\ 
\indent Simulations of clusters with initial density profiles with central cusp, show a completely different picture. In Figure \ref{gammane2} the final differential energy distribution is shown for $\gamma=0$ (no central density cusp), $\gamma=1$ (mild cusp) and $\gamma=2$ (extreme cusp).
The cases starting with $\eta=0$ have, as expected, three different behaviors, with $n(\mathcal{E})$ peaking at high energy for $\gamma=0$, and at low energy for $\gamma=2$. For $\gamma=1$ the distribution is instead flat for a broad range of energy values.\\    
\indent  From the values of $\gamma$ presented here, it appears that introducing a non zero initial temperature has no effect on the final $n(\mathcal{E})$ for large values of $\gamma$. In fact, (see Fig. \ref{gammane2}), in the largest $\gamma=2$ case, the normalized $n(\mathcal{E})$ for different initial values of $\eta$ are practically indistinguishable.
For $\gamma=0$ the situation is qualitatively similar to that of the above discussed flat cored models, with the peak energy smoothly shifting towards lower values. Curiously, the final $n(\mathcal{E})$ for systems with intermediate values of $\gamma$ (in this case $\gamma=1$) has a much more abrupt transition from the perfectly cold case ($\eta=0$) to gradually initially hotter systems, 
loosing its ``plateau structure'' even for very small values of $\eta$.\\
\indent The density profiles shown in Fig. \ref{gammarho2} display essentially the same picture, with no significant effect (except at low radii) due to the initial temperature for systems with large $\gamma$, and a similar behavior to that of the flat-cored systems for the low $\gamma$
cases.\\
\indent As a general remark, we observe from both $n(\mathcal{E})$ and $\rho(r)$, that even a small amount of initial kinetic energy can significantly affect the explosion dynamics, leading to end states that considerably depart from those of the initially cold model, whether the initial density belongs to one or the other family.\\
\indent Figure \ref{displaced} summarizes for the two families of initial density profiles, the effect of the initial temperature in shifting the maximum of the energy distribution. For flat cored systems, Eq. (\ref{proflekaplan}) (left panel), the maximum of the energy distribution of the final states at fixed $\eta$ falls to roughly the same energy $\mathcal{E}_p$ (in units of the cutoff energy), for $\beta\geq5.5$. This implies that the energy 
spectrum tends to differ less and less as the initial density profile approaches the perfect step.\\ 
\indent For systems with central density cusp, Eq. (\ref{denhen1}) (right panel), it is found that for $\gamma\geq1.4$, the normalized distributions peak at the same $\mathcal{E}_p$ independently of the ratio $\eta$, while at lower slopes of the initial density cusp, it is shifted to lower values the greater $\eta$ is.
\section{Multi-component systems}\label{multisystem}
To study the Coulomb explosion of systems composed of two or more different species of particles is particularly relevant with respect to the modelling of ionized clusters of heteronuclear molecules, as well as core-shell clusters, \cite{2012PhRvA..86c3203S}, \cite{2007JChPh.126u4706L} 
or atomic clusters embedded in helium droplets \cite{krishna11}.\\
\indent From a theoretical point of view, the problem of multi-component (also called multi-species) Coulomb explosion has been studied in \cite{kov} by means of kinetic theory, see also Appendix \ref{kt}, and in \cite{murakamimima} and \cite{2010PhPl...17b3110A}
 using a simple two fluid model. The effect of a residual electron density has been treated in \cite{2009LPB....27..321P}.\\
\indent In this part of the work we have performed simulations of pure Coulomb explosions of clusters containing two species of particles, aiming at studying their final differential energy distributions. More detailed calculations involving 
ionization end treating dynamics of the electrons are presented in the next chapter. 
\subsection*{Continuum model}
As we have seen, the dynamics of mono-component clusters is already quite complex if the initial density profile is non-uniform. Adding another degree of freedom, by making the system heterogeneous in composition, makes the problem even more complicated. 
However, assuming a continuum picture and restricting ourselves to the uniform initial density still allows, under certain assumptions, to derive the asymptotic $n(\mathcal{E})$ for one of the two components analytically.\\
\indent Following the approach of \cite{murakamimima}, let us consider a heterogeneous cluster of initial radius $R_0$, whose initial total charge density is given by
\begin{equation}
\rho(r)=\rho_1\theta(R_0-r)+\rho_2\theta(R_0-r),
\end{equation}
as the sum of the individual (constant) densities of the two components $\rho_1$ and $\rho_2$. The total charge $Q$ is given in this case by
\begin{equation}\label{multiq}
Q=Q_1+Q_2=\frac{4\pi\rho_1R_0^3}{3}+\frac{4\pi\rho_2R_0^3}{3},
\end{equation}
and the radial component of the electric field is inside $R_0$ reads
\begin{equation}
E(r)=\frac{4\pi}{3}(\rho_1+\rho_2)r. 
\end{equation}
The infinitesimal element of volume occupied by the component with $\rho_1$ has unit mass $\delta m_1$ and unit charge $\delta q_1$, for the other component we have instead $\delta m_2$ and $\delta q_2$. Hereafter we assume $\delta m_1/\delta m_2\geq 1$, regardless of the ratio 
of their unit charges. Therefore we refer to component 1 as {\it heavy component} and to component 2 as {\it light component}.\\
\indent In order to characterize the system, we define now the two parameters $a$ and $b$ as
\begin{equation}\label{qsum}
a=\frac{\delta q_2/\delta m_2}{\delta q_1/\delta m_1};\quad b=\frac{\rho_2}{\rho_1+\rho_2}.
\end{equation}
Note that in the limit of $a\rightarrow\infty$ the heavy component does not move, while for $b=1$ (or $b=0$) one retrieves the single component case treated in Sect \ref{contmod}.\\
\indent For large values of $a$, during the explosion the light component overtakes {\it entirely} the heavy one. Assuming that no shell crossing happens among elements of the same component, one can write the asymptotic energy (per unit charge) of an element of light component as function of the initial coordinate $r_0$ as
\begin{equation}\label{multie}
\mathcal{E}(r_0)=U_{0,1}+U_{0,2}=\frac{Q_1}{2R_0^3}(3R_0^2-r_0^2)+\frac{Q_2}{R_0^3}r_0^2.
\end{equation}
From Eq. (\ref{multiq}) and the definition of $b$, Equation (\ref{multie}) can be rewritten as 
\begin{equation}\label{multie2}
\mathcal{E}(r_0)=\frac{Q}{2R_0^3}\left[3(1-b)R_0^2+(3b-1)r_0^2\right].
\end{equation}
With these assumptions and making the same steps as in the case of a single component uniform system (cfr. Sect. \ref{contmod} and Ref. \cite{mika2013}), the asymptotic differential energy distribution for the light component is obtained as
\begin{equation}
n_2(\mathcal{E})=\frac{3}{(3b-1)^{3/2}}\sqrt{\frac{2\mathcal{E}}{\mathcal{E}_{\rm max}^3}-\frac{3(1-b)}{\mathcal{E}_{\rm max}^2}}\theta(\mathcal{E}_{\rm max}-\mathcal{E}),
\end{equation}
where $\mathcal{E}_{\rm max}=Q/R_0$. For $b=1$ from the formula above one obtains Eq. (\ref{nesqrt}), while for $b=1/3$ (and $a\rightarrow\infty$) one has 
\begin{equation}
 n_2(\mathcal{E})=\delta(\mathcal{E}-\mathcal{E}_{\rm max}),
\end{equation}
as $\mathcal{E}(r_0)$ is independent on $r_0$, crf. Eq. (\ref{multie2}).\\
\indent Note also, that independently on $b$, in case of large $a$, the asymptotic distribution for the heavy component $n_1(\mathcal{E})$ is expected to resemble qualitatively that for the one component model, since the dynamics of the latter is barely influenced by the fast explosion of the lighter component.    
\subsection*{Numerical simulations using particles}
\begin{figure} [h!t]
        \centering 
         \includegraphics[width=0.8\textwidth]{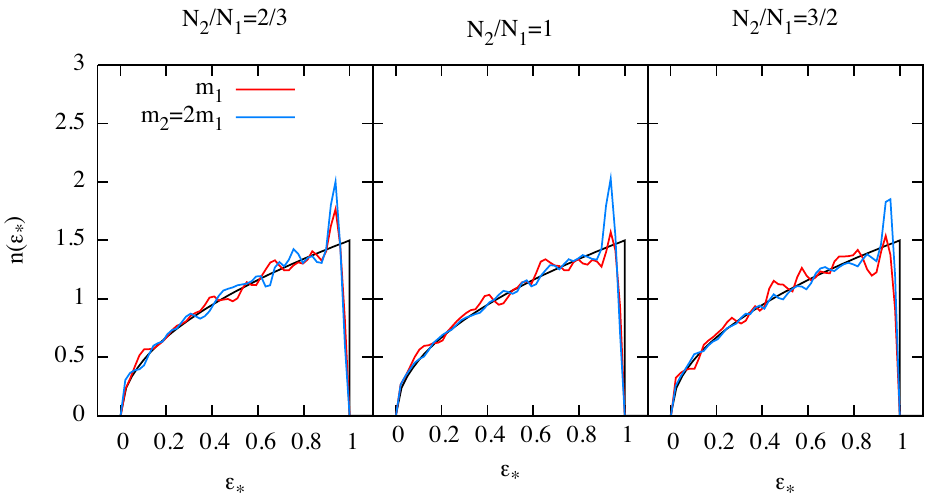}
         \caption{Final normalized energy distributions of the two components. For the three combinations of $N_2/N_1$ indicated above, for systems with total number of particles $N_1+N_2=N=2\times10^4$, $a=1$ and $m_2=2m_1$.}
\label{multine1qm}
\end{figure}

\begin{figure} [h!t]
        \centering 
         \includegraphics[width=0.8\textwidth]{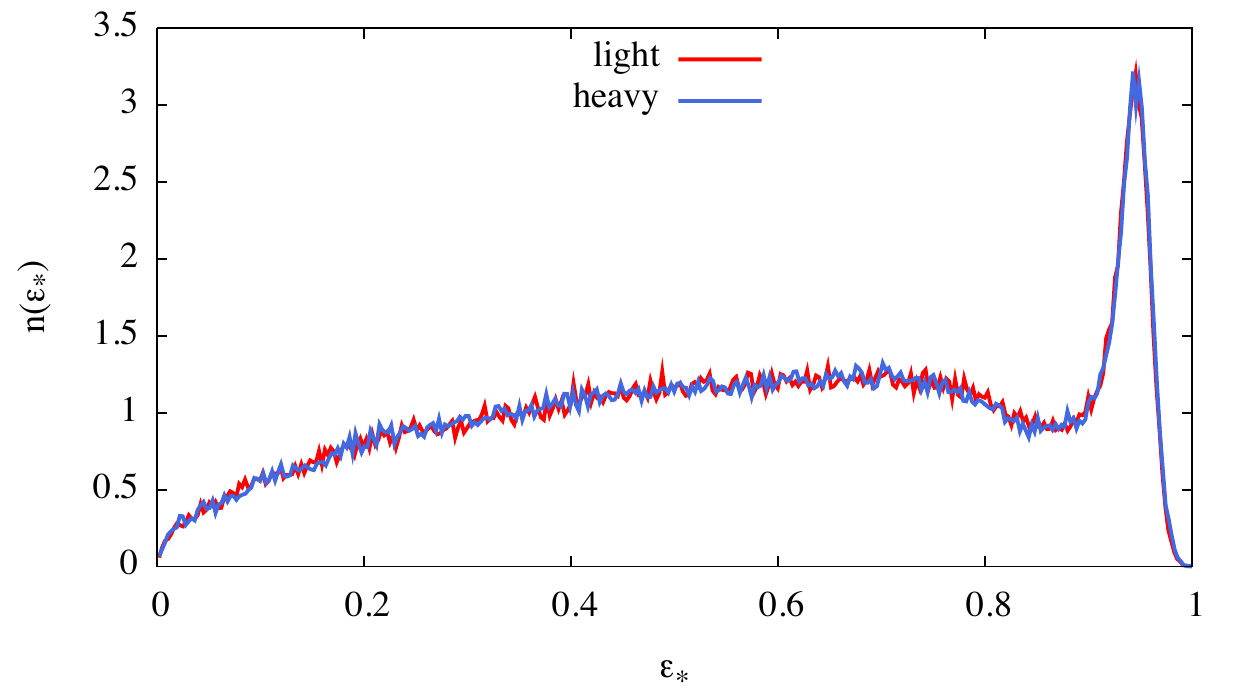}
         \caption{Final normalized $n(\mathcal{E})$ for the two components of an $N_1+N_2=N=8.5\times10^5$ multi-species spherical and homogeneous cluster, from a collisionless simulation with a particle-mesh code. Each curve is normalized respect to its cutoff $\mathcal{E}$. For $\mathcal{E}>0.8$ the distribution departs considerably from the $\sqrt{\mathcal{E}_*}$ trend showing a dip followed sharp cusp, signatures of an energy bunching, due to the worse representation of $\rho_c$ and the potential for a spherical system on a cartesian grid. See also Chap. \ref{numerica}.}
\label{neheavylightpm}
\end{figure}

We have investigated the effect of a heterogeneous composition in the Coulomb explosion of a charged cluster by means of molecular dynamics calculations. The initial conditions are implemented as follows: $N_1$ particles of mass $m_1$ and charge $q_1$ and $N_2$ particles of mass $m_2$ and charge $q_2$, are homogeneously distributed inside a spherical volume of radius $R_0$. No minimum inter-particle distance is enforced and, in all cases, the initial velocities $v_{0,i}$ are set to 0, therefore $K_0=0$.\\
\indent The total energy of the system is given by its initial potential energy
\begin{equation}\label{umulti}
 U_0=\frac{1}{2}\sum_{i\neq j=1}^N \frac{q_iq_j}{|\mathbf{r}_{0,i}-\mathbf{r}_{0,j}|};\quad N=N_1+N_2.
\end{equation}
As usual, we define the final state of the system when the kinetic energy $K$ equals 99\% of $U_0$.\\
\indent In presence of two species, the scale time in the simulations is redefined as
\begin{equation}
t_{\rm dyn}\equiv2\pi\sqrt\frac{\bar m}{\bar q^2\tilde n_{0}},
\end{equation}
where $\bar q$ and $\bar m$ are the average charge and mass respectively. Hereafter, we characterize each system with its fractional charge to mass ratio (cfr. also Eq. (\ref{qsum})) given by
\begin{equation}
 a=\frac{q_2/m_2}{q_1/m_1},
\end{equation}
\begin{figure} [h!t]
        \centering 
         \includegraphics[width=\textwidth]{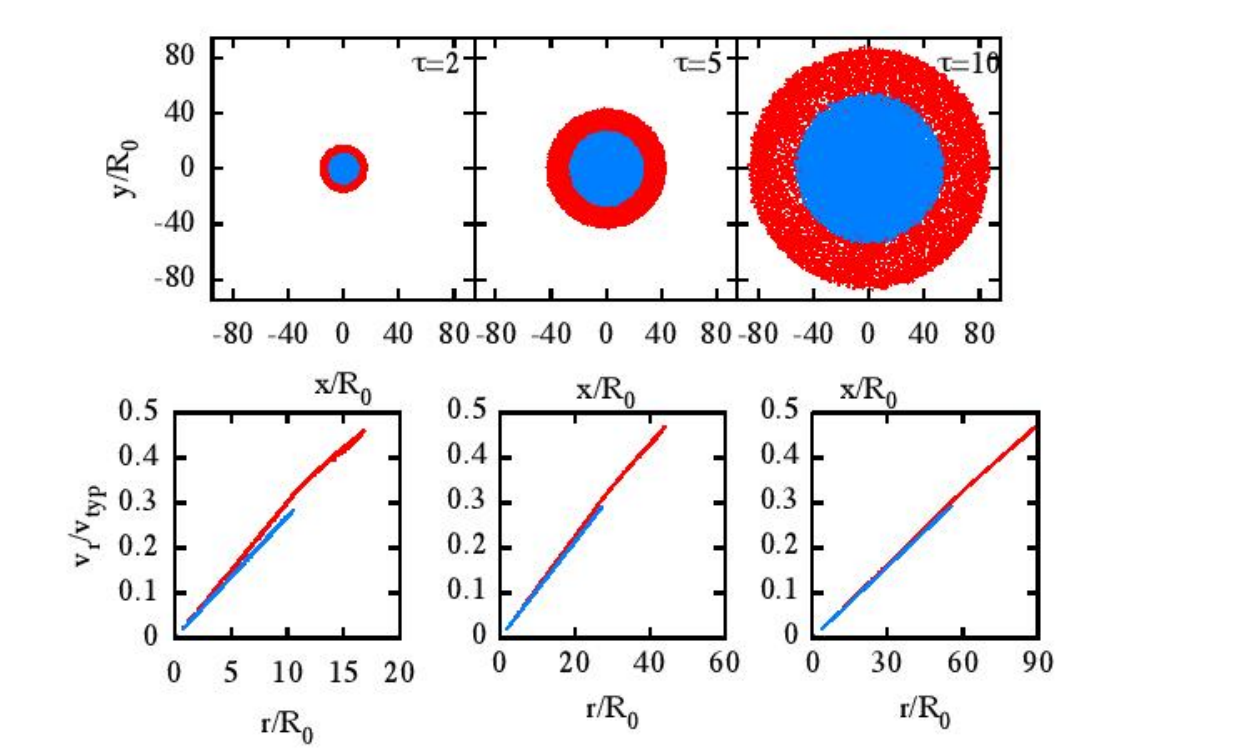}
         \caption{Upper row: particles positions in the $x-y$ plane for a multi-component cluster ($N_1=N_2=10^4$) with $a=2$ at $\tau=2t_{\rm dyn}$, $5t_{\rm dyn}$ and $10t_{\rm dyn}$. Red dots mark the particles of the component with higher charge to mass ratio and blue dots those with the lower.
 Lower row: for the same times as above, phase-space pairs $r$ and $v_{r}$. Coordinates are normalized in units of the initial cluster radius $R_0$ while the radial velocities in units of the scale velocity $v_{\rm typ}=R_0/t_{\rm dyn}$.}
\label{phasemulti}
\end{figure} 
and for reasons of convenience, we will label here with $2$ the species with the higher charge to mass ratio, so that $a\geq 1$.\\ 
\indent The acceleration felt by the $i-$th particle placed at position $\mathbf{r}_i$, due to the cluster's electrostatic field $\mathbf{E}$,  is given by
\begin{equation}
\mathbf{a}_i=q_i\mathbf{E}(\mathbf{r}_i)/m_i,
\end{equation}
therefore, as expected, particles of the component with higher charge to mass ratio $q/m$ are accelerated to higher velocities with respect to particles of the other component. 
Due to that the cluster expansion is not uniform and even if the initial density profile is homogeneous, one of the two components overtakes the other.\\
\indent This is clearly evident in Figure \ref{phasemulti} (upper row) where we show at three different times, for a cluster with $a=2$, the projection of the particle's positions. An outer shell containing the faster particles is already present at early stages of the explosion. From the phase-space sections radial velocity $v_r$ versus radial coordinate $r$ (same figure, lower row), it can be noticed how once the faster component 
\begin{figure} [h!t]
        \centering 
         \includegraphics[width=\textwidth]{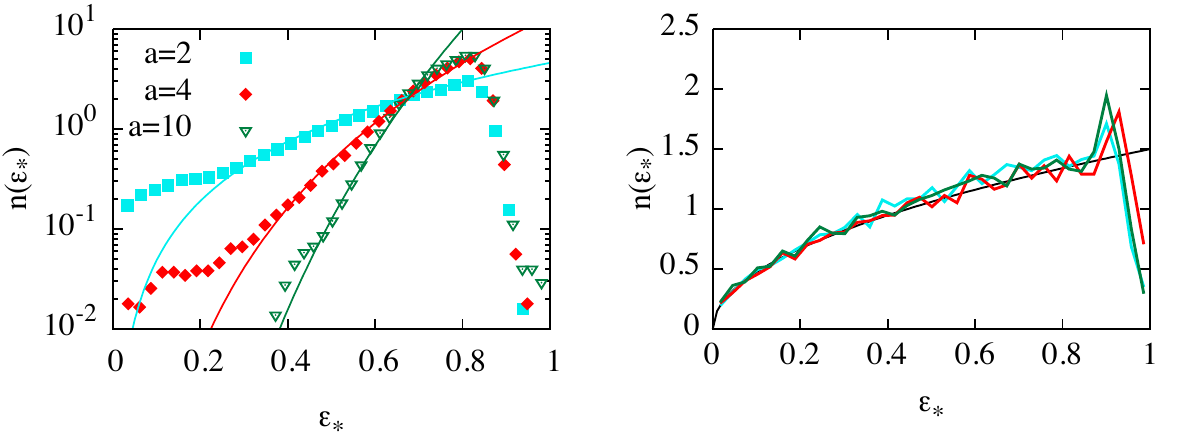}
         \caption{Left panel: asymptotic number energy distribution for the ``fast'' component of three initially cold cluster with $N_1=N_2=7500$ and $a=2$, 4 and 10 (points), the best fit curves are also shown (lines). Right panel: same quantity for the ``slow'' component. For comparison, the thin black line line marks the theoretical asymptotic $n(\mathcal{E})$ of the single species homogeneous sphere. Each curve is normalized to its individual cutoff energy.}
\label{neheavylight}
\end{figure} 
has overtaken the other, its velocity profile is not linear with $r$ (showing a little multivalued region). However, (as expected, e.g. \cite{2003PhRvL..91n3401K} and \cite{murakamimima}) at later times (here $\tau=10$) it has again a linear trend.\\
\indent Figure \ref{neheavylight} shows the final $n(\mathcal{E})$ for the two components of three clusters with different values of the parameter $a$. In all cases $N_1=N_2=7500$. The differential energy distribution of the fast component peaks in all cases at high energies, containing roughly the 70\% of the particles between $0.65\mathcal{E}_{\rm max}$ and $\mathcal{E}_{\rm max}$. For a broad range of energies is remarkably well fitted by a power law $(\mathcal{E})^{k}$, see solid lines in Fig. \ref{neheavylight}.
 For increasing values of $a$, the exponent $k$ increases making $n_2(\mathcal{E})$ steeper. For the case shown here, $k\simeq 1.98$ for $a=2$, $k\simeq4.89$ for $a=4$ and $k=9.33$ for $a=10$. In the limit of infinite $a$, the component 1 does not move, and the asymptotic distribution for component 2 is expected to have a delta-like structure, see e.g. Refs. \cite{2004JChPh.121.8329L}, \cite{murakamimima} and \cite{2009LPB....27..321P}.\\
\indent The qualitative explanation of such behavior is, if one has that for the initial charge density of the two components $\rho_{0,2}\ll\rho_{0,1}$, the cluster's electrostatic field is always dominated by the effect of component 1 that moves on a considerably larger time scale.
Due to that, when a particle of component 2 coming from the inner regions of the cluster reaches the radius $R_0$, it has already a certain amount of kinetic energy and therefore will reach a larger asymptotic energy than a particle of the same species initially sitting at $R_0$. The more the charge density of component 2 is small, the less the kinetic
 energies of particles coming from different inner radii once reaching $R_0$, differ from each other, thus steepening $n_2(\mathcal{E})$.\\ 
\indent The number energy distribution $n_1(\mathcal{E})$ for the slow component is, as expected, 
reminiscent of $n(\mathcal{E})$ observed for a single species initially uniform system. Remarkably, the cusp at high energies, 
is milder than that observed in $n(\mathcal{E})$ for single-component systems.\\ 
\indent Let us now consider the $a=1$ case. Intuitively, independently of the density profile, if the combinations of charge and mass are such that $q_1/m_1=q_2/m_2$ (for instance the situation that one would have in a mixture of deuterons $\ce{D}^+$ and carbon ions $^{12}$\ce{C}$^{6+}$) the two {\it normalized} asymptotic energy distributions $n_2(\mathcal{E})$ and $n_2(\mathcal{E})$ should in principle coincide. However, as seen in Fig. \ref{multine1qm}, for the end products of 3 direct $N-$body simulations this is not exactly true. In fact, the energy distribution for the component with larger mass (and charge), in this case $m_2=2m_1$, shows a more pronounced high energy peak than the other. This happens regardless of the ratio $N_2/N_1$ and persists using different binning for the numerical energy distribution.\\ 
\indent By contrast, this is not the case for the final states of particle-mesh simulations (see Chap. \ref{numerica}) shown in Fig. \ref{neheavylightpm} where the electric 
field acting on particles is not computed by direct summation (as in MD simulations) but solving Poisson equation on a cartesian grid. In this case the two curves perfectly coincide.\\
\indent The high energy peak is to be interpreted as an effect of the discrete grid based electric field, that close to the system's surface is slightly underestimated, leading to an energy bunching effect, analogous to that induced by the correlation hole treated in Sect. \ref{simulationomo}.\\
\indent To interpret this discrepancy, let us consider now a test particle of mass $m_k$ and charge $q_k$ moving through a homogeneous spherical background of $N_1$ particles with charges and mass $q_1$ and $q_2$, and $N_2$ particles with charge and mass $m_1$ and $m_2$, so that $q_1/m_1=q_2/m_2$. For consistency we assume $m_2>m_1$ (and obviously $q_2>q_1$).\\ 
\indent The average charge of the background particles is given by 
\begin{equation}
 \bar q = (q_1N_1+q_2N_2)/(N_1+N_2),
\end{equation}
while their total number density is simply the sum of the individual number densities of the two species $\bar n=n_1+n_2$.\\
\indent The Langevin-type equations for the radial motion of the test particle, (see e.g. \cite{murakamimima}, see also \cite{vankampen}) read
\begin{equation}\label{averagedequ}
\dot v= \frac{q_kQ(r)}{m_kr^2}-\omega_{\rm coll} v;\quad v=\dot r,
\end{equation}
where $Q(r)=(4\pi/3)r^3\bar n\bar q$ is simply the charge inside radius $r$ and 
\begin{equation}\label{omegacoll}
\omega_{\rm coll}=\frac{8\pi\bar n q_k^2\bar q^2 \ln\Lambda}{m_k\bar\mu_k v^3}
\end{equation}
is the collision frequency with the background ions, where $\ln\Lambda$ is the Coulomb logarithm\footnote{This quantity is defined as the natural logarithm of the ratio between the maximum and minimum impact parameter in the collision experienced by the test particle. See also Appendix \ref{frizione}}, (see \cite{spitz}) and 
\begin{equation}
\bar\mu_k=\frac{m_k\bar m}{(m_k+\bar m)}
\end{equation} 
is the species averaged reduced mass.\\ 
\indent If the term depending on $\omega_{\rm coll}$ in (\ref{averagedequ}) is non negligible, the test particle experiences an effective drag force along the radial direction due to the two body encounters with the background particles.\\
\indent Note that in such case, due to the different dependence of $\dot v$ on $q_k$ and $m_k$, a test particle with $m_k=m_1$ and $q_k=q_1$ is prone to a different deceleration than a test particle with $m_k=m_2$ and $q_k=q_2$ starting with the same velocity from the same initial position. 
In particular, for $m_2=2m_1$ the more massive (and more charged) test particle with mass $m_2$ suffers a deceleration of a factor roughly 1.96 larger than a particle with mass $m_1$.\\  
\indent The Coulomb explosion of a spherical heterogeneous system with one charge to mass is still to a large extent a self similar expansion without particles propagating through the others. However, at early stages of the explosion, the contribution to the force due  
to near neighbours dominates over the mean field for most of the particles, depending also on the local structure of the system. Particles of both species do not have pure radial trajectories as for $t\rightarrow\infty$, and some energy can transferred between 
radial and tangential motion via few ``two body collisions''. Since, cfr. Eq. (\ref{averagedequ}), such energy exchange depends on the (averaged) reduced mass, the heaviest component is expected to lose more kinetic energy in favour of the light ones.
Thus, in addition to the energy bunching due to the discreteness discussed before,
 the heavier component suffers an additional shift towards lower energies causing the height of the peak in the normalized $n(\mathcal{E})$ to increase.\\
\begin{figure} [h!t]
        \centering 
         \includegraphics[width=\textwidth]{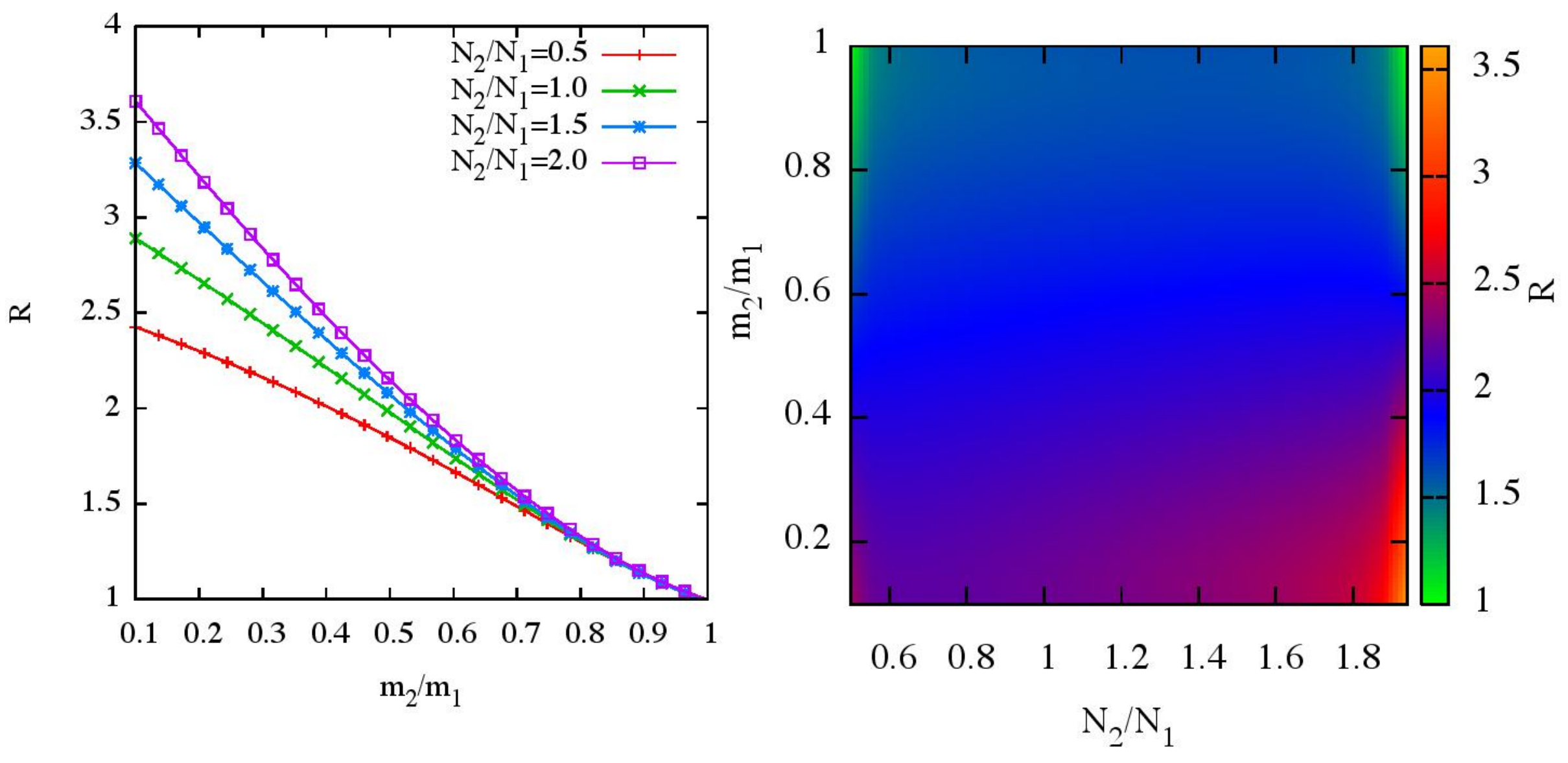}
         \caption{Left panel: for four different number ratios 0.5, 1, 1.5 and 2, ratio of the final kinetic energies of the light and heavy particles. Right panel: as a function of $0.1<m_2/m_1<1$ and $0.5<N_2/N_2<1$, the ratio of the average kinetic energies for the two species is shown.}
\label{multimass}
\end{figure} 
\indent Note that this is also influenced by the percentages of the two species $N_1$ and $N_2$ since they enter the definition of $\omega_{\rm coll}$ twice, through $\bar\mu$ and $\bar q$.\\
\indent In order to characterize the effects of different mass ratios and different percentages of the two species, we now consider test systems of fixed total charge $Q$ where $q_1=q_2=Q/(N_1+N_2)$, 
and we span the two intervals of $m_2/m_1$ and $N_2/N_1$.\\
\indent We assume, $m_2<m_1$. In this way, if in the initial conditions $K_0=0$ and the number density $\bar n$ is homogeneous inside the spherical volume of radius $R_0$, 
the total energy of the system, given by Eq. (\ref{umulti}) is always the same in all the cases considered.\\
\indent For the final states of the numerical simulations we define the quantity
\begin{equation}
\mathcal{R}=\frac{\langle \mathcal{E} \rangle_2}{\langle \mathcal{E}\rangle_1}  
\end{equation}
as a function of $m_2/m_1$ and $N_2/N_1$, where $\langle \mathcal{E} \rangle_i$ is the final average energies of the two components.\\
\indent In Fig. \ref{multimass} we show $\mathcal{R}$ as a function of $m_2/m_1$ for some some relevant number ratios as well as the quantity as function of both $N_2/N_1$ and $m_2/m_1$. Remarkably, see right panel, for mass ratios larger than roughly 0.75, the final energy ratio is not significantly influenced by different percentages of the two species in the cluster. For small values of $m_2/m_1$, $\mathcal{R}$ grows indefinitely for larger values of $N_2/N_1$ while grows sublinearly for $N_2/N_2<1$.
\section{Summary}
We have reviewed here the basic aspects of the Coulomb explosion of spherical system in order to set the stage for our study of laser irradiated atomic and molecular clusters. By means of numerical simulations we have investigated the explosion of systems starting with different density profiles 
and ion temperature as well as with heterogeneous composition.\\
\indent We have observed the discrepancy between the continuum model and systems where the particulate structure is accounted for the homogeneously charged sphere. This comes in the form of a spike in the number energy distribution $n(\mathcal{E})$ caused by the presence 
of a ``correlation'' hole in the initial particle distribution. The effect appears to be considerably reduced if no minimum inter-particle distance is enforced in the initial condition (i.e. the initial state is characterized by a non lattice-like structure). In addition, we showed that the regularization on short distances of the Coulomb (or Newton) force induces a spurious external field effect in  the interior of perfectly spherical systems of the same entity of that due of the discreteness of the system itself.\\
\indent The formation of shock shells in systems with non uniform initial charge density profile is found to be inhibited when an ion temperature $\eta>0.2$ is established in the initial condition. In our simulations, contrary to \cite{2003PhRvL..91n3401K}, the correspondent ion velocity distribution 
is position independent, thus the shock is erased for lower values of $\eta$.\\
\indent Finally, from the simulations of two component clusters we find that such systems are characterized by a mean field segregation effect induced by the different charge to mass ratios of the different species. The effect of different percentages on the redistribution of the total energy on the two species is strongly influenced by the mass ratio and tend to vanish already at $m_2/m_1\sim0.75$.  
\chapter{Coulomb explosion of ellipsoidal systems}\label{chapspheroid}
Initially spherically symmetric cluster, once irradiated by strong laser pulses, may assume non spherical (i.e. ellipsoidal) charge distributions, either due to the laser spatial polarization or to resonances between the laser frequency and the initial plasma frequency (see e.g. Refs \cite{mika08}, \cite{mika09}, \cite{skopa10} and \cite{krishna11}). Moreover, electron or ion beams 
confined by electromagnetic fields in accelerators are known to adjust to ellipsoidal shapes depending on the parameters of the confining field (see e.g. \cite{bat01}, \cite{kandrup04}, \cite{fubiani06}, \cite{gruner09} and references therein).\\
\indent It is therefore useful to extend our discussion of Coulomb explosion carried on in the previous chapter, to the case of ellipsoidal systems. limiting ourselves to single-component systems, we study the cases of homogeneously charged triaxial and axisymmetric ellipsoids (i.e. spheroids) and finally axisymmetric systems with non uniform initial density.  
In the same line of Chap. \ref{chapmultice}, we treat the problem first in the continuum picture, and then we discuss the analysis of our $N-$body calculations. 
\section{Potentials for ellipsoidal distributions of charge}
To tackle the problem of non-spherical Coulomb explosion in the continuum approach we first need the expressions for the potential due to ellipsoidal charge distributions.\\
\indent Let us consider an infinitesimally thin ellipsoidal shell of semi-axes $a^\prime$, $b^\prime$ and $c^\prime$, total charge $q$ and uniform surface density. We shall call such object a {\it homeoid}. 
The electrostatic potential exerted by the homeoid at a point $\mathbf{r}=(x,y,x)$ placed on its exterior is given (see \cite{kellogg}, \cite{chandra87}) by 
\begin{equation}\label{pothome}
\Phi^{\rm hom}(\mathbf{r})=\frac{q}{2}\int_{\lambda}^{+\infty}\frac{{\rm d}u}{\sqrt{(a^{\prime2}+u)(b^{\prime2}+u)(c^{\prime2}+u)}},
\end{equation}
where the lower boundary of integration $\lambda$ is obtained as the largest root of the algebraic equation
\begin{equation}\label{lambda}
\frac{x^2}{a^{\prime2}+\lambda}+\frac{y^2}{b^{\prime2}+\lambda}+\frac{z^2}{c^{\prime2}+\lambda}=1,
\end{equation}
and its equipotential surfaces are ellipsoids confocal to it.
The potential inside the homeoid is constant, therefore it exerts no force at its interior. Such results it is known as third Newton's theorem, see e.g. \cite{kellogg}. 
We now consider a continuum ellipsoidal charge distribution with with semi-axes $a,b,c$ and density $\rho(s)$ stratified on concentric similar homeoids, for which we have introduced the so called ellipsoidal radius
\begin{equation}
 s=\sqrt{x^2/a^2+y^2/b^2+z^2/b^2}.
\end{equation}
Using Eq. (\ref{pothome}) and the third Newton theorem, one can formally construct the expression for the potential at a point $\mathbf{r}=(x,y,z)$ generated by such a generic ellipsoidal charge distribution integrating the contribution of each homeoidal shell.\\
\indent We define the integration variable $u$ from the solution of
\begin{equation}
\frac{x^2}{a^2+u}+\frac{y^2}{b^2+u}+\frac{z^2}{c^2+u}=s^2,
\end{equation}
and the auxiliary integral function
\begin{equation}
\psi(s^2)=\int_1^{s^2(u)}\rho(s^{\prime2}){\rm d}s^{\prime2}.
\end{equation}
The potential due to the charged ellipsoid then reads
\begin{equation}\label{triaxint}
\Phi(\mathbf{r})=\pi abc\int_{l}^{+\infty}\frac{\left[\psi(1)-\psi(s^2(u))\right]}{\sqrt{(a^2+u)(b^2+u)(c^2+u)}}{\rm d}u,
\end{equation}
where $l=0$ if $\mathbf{r}$ is inside the charge distribution and $l=\lambda$, with $\lambda$ from Eq. (\ref{lambda}) otherwise.\\
\indent In the special case when the density is equal everywhere to the constant value $\rho_c$, $\psi(s^2)=s^2$ and Eq. (\ref{triaxint}) becomes
\begin{equation}
\Phi(\mathbf{r})=\pi\rho_c abc\int_{l}^{+\infty} \left(1-\frac{x^2}{a^2+u}-\frac{y^2}{b^2+u}-\frac{z^2}{c^2+u}\right)\frac{{\rm d}u}{\sqrt{(a^2+u)(b^2+u)(c^2+u)}}.
\end{equation}
Note that the equations of motion for a test particle moving under a potential of the form (\ref{triaxint}), are in general not separable in cartesian coordinates.\footnote{For particular forms of $\rho(m^2)$, the equations are indeed separable in polar ellipsoidal coordinates ($\xi,\eta,\zeta$) such that $x=(B+\zeta)\cos\xi\cos\eta$; $y=(B+\zeta)\cos\xi\cos\eta$; $z=\left[(1-e^2)B+\xi\right]\sin\eta$, where $e^2=(a^2-b^2)/a^2$ and $B=a/\sqrt{1-e^2\sin^2\xi}$, if the potential can be expressed, see \cite{dezeeuw85a}, as
\begin{equation}\label{separable3ax}
\Phi(\xi,\eta,\zeta)=\frac{{\mathcal F}(\xi)}{(\xi-\eta)(\xi-\zeta)}+\frac{{\mathcal F}(\eta)}{(\eta-\zeta)(\eta-\xi)}+\frac{{\mathcal F}(\zeta)}{(\zeta-\xi)(\zeta-\eta)}\nonumber,
\end{equation}
where ${\mathcal F}(x)$ is a generic smooth function. A potential of such form is called a St\"ackel potential \cite{Stackel}.}\\
\section{Coulomb explosion of uniform axisymmetric systems}
The first cases of interest are those of homogeneously charged rotational ellipsoids (also known as spheroids), In our discussions of both continuum and particle approaches we set $a=b=a_\perp$ and $c=a_\parallel$. In addition, we define the quantity 
\begin{equation}
\alpha\equiv\frac{a_\perp}{a_\parallel},
\end{equation}
that we will call hereafter {\it aspect ratio}. When $\alpha<1$, the spheroid has an elongated shape and it is called prolate, when $\alpha>1$ the spheroid is flattened and is called oblate. The $\alpha=1$ case corresponds to the sphere.\\
\indent Given the axial symmetry of such systems, it is convenient to introduce cylindrical coordinates:  
\begin{align}
r\quad&=\quad\sqrt{x^2+y^2}\nonumber\\
\varphi\quad&=\quad
\begin{cases}
0, & {\rm for} \quad x=y=0 \\
{\rm arcsin}(y/r), & {\rm for} \quad x\geq 0\\
\pi-{\rm arcsin}(y/r), & {\rm for} \quad x<0
\end{cases}\nonumber\\
z\quad&=\quad z,
\end{align}
where we have assumed conventionally that $a_\parallel$ is oriented along the $z$ axis.
\subsection*{Continuum model}
\begin{figure} [h!t]
        \centering 
         \includegraphics[width=\textwidth]{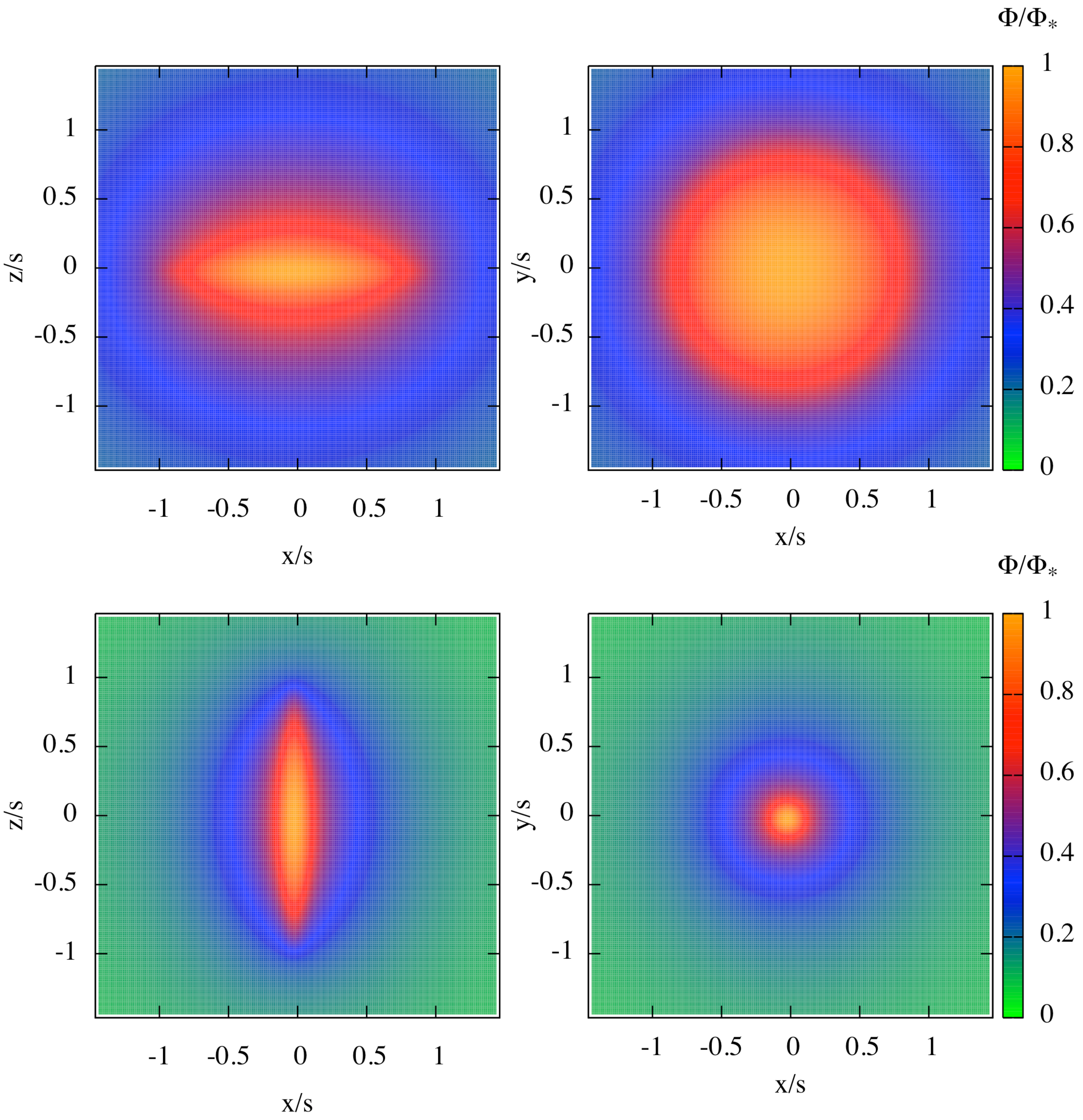}
         \caption{Upper panels: Potential in the coordinate planes $x,z$ (left) and $x,y$ (right) for a homogeneous oblate spheroid with $\alpha=10$. Lower panels: Potential in the coordinate planes $x,z$ (left) and $x,y$ (right) for a homogeneous prolate spheroid with $\alpha=0.1$. The potential is normalized to its value at the origin $\Phi_*$ and the $x,y,z$ coordinates to $s=a_\parallel$ for the prolate case and $s=a_\perp$ for the oblate case.}
\label{potspheroid}
\end{figure}
\begin{figure} [h!t]
        \centering 
         \includegraphics[width=\textwidth]{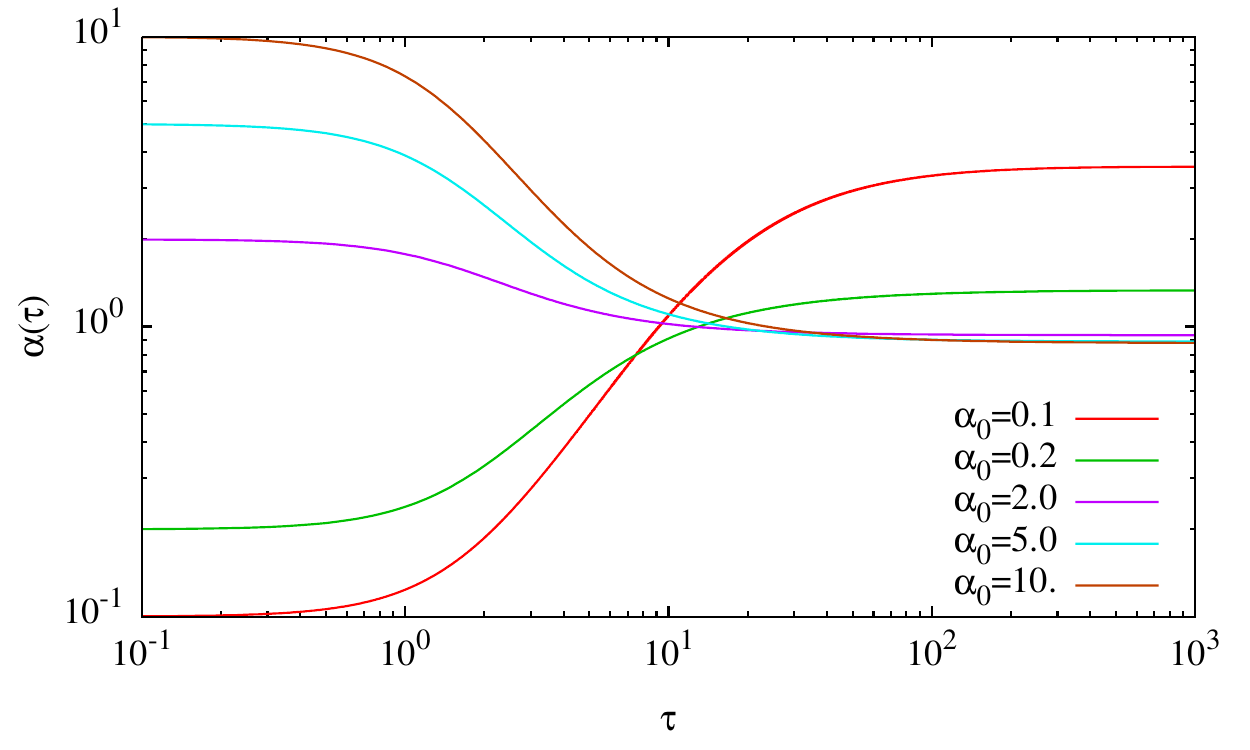}
         \caption{Evolution of the aspect ratio starting from different values in the interval (0.1-10) obtained via the finite difference integration of Eqs. (\ref{cespheronorm}).}
\label{limita}
\end{figure}

\begin{figure} [h!t]
        \centering 
         \includegraphics[width=\textwidth]{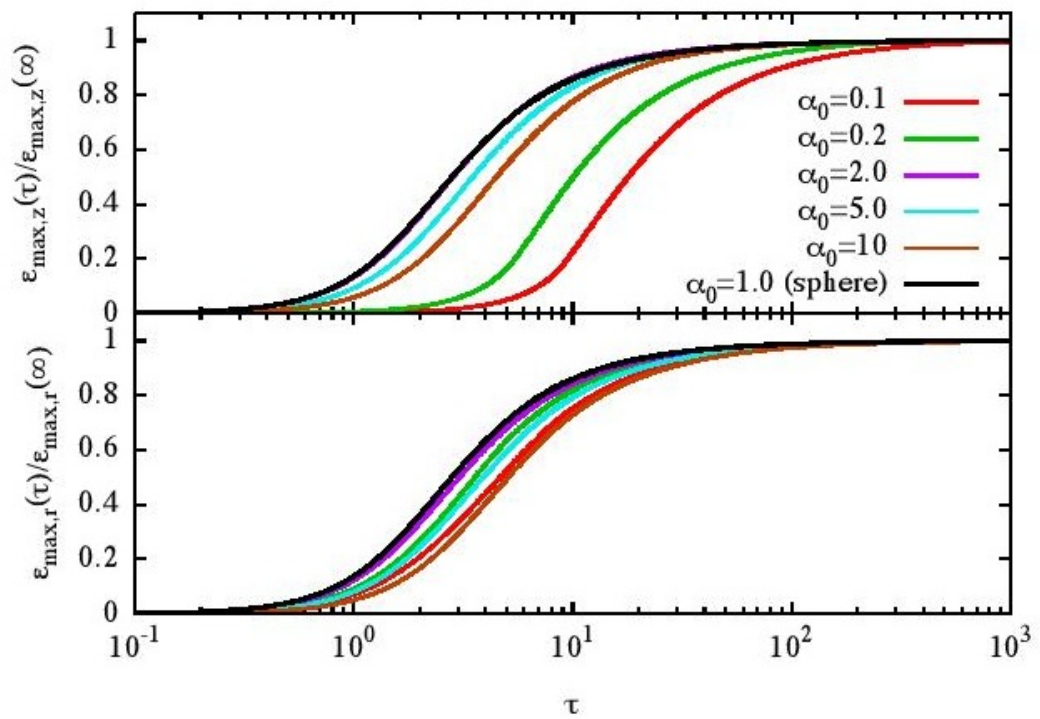}
         \caption{Upper panel: Maximum energy along the symmetry axis in units of its asymptotic value for the same values of $\alpha_0$ of Fig. \ref{limita}. Lower panel: Maximum energy in the equatorial plane in units of its asymptotic value.}
\label{limlim}
\end{figure}

\begin{figure} [h!t]
        \centering 
         \includegraphics[width=0.86\textwidth]{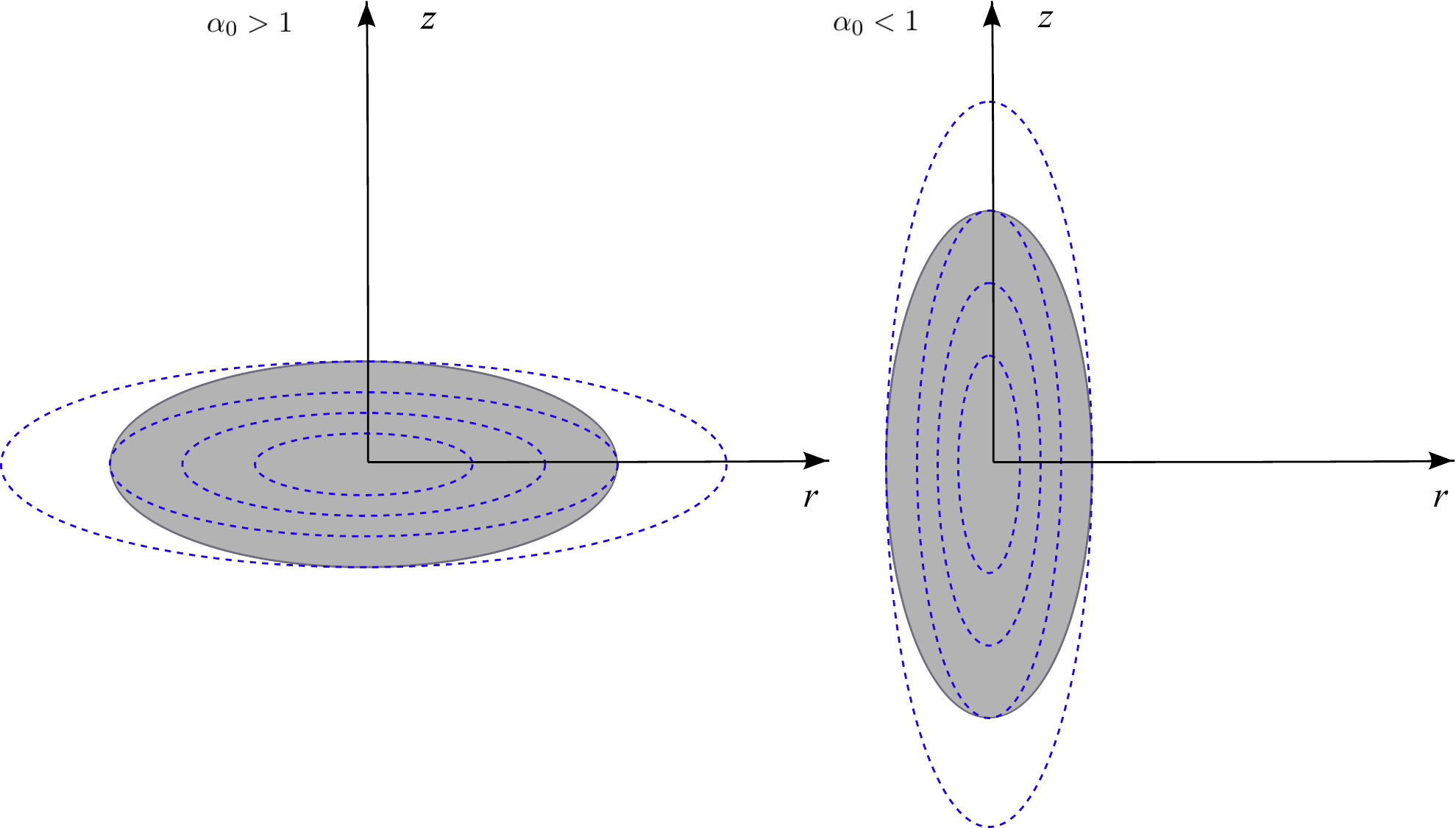}
         \caption{Section along the symmetry axis of the equivelocity surfaces (blue dashed lines) for a prolate (left) and an oblate (right) spheroid (gray shaded area).}
\label{aspetto}
\end{figure}

 
The electrostatic potential inside a uniformly charged spheroid can be written in a more compact form using cylindrical coordinates and setting $a=b=a_\perp$, $c=a_\parallel$, and $u_*=u/a_\parallel^2$ in Eq. (\ref{separable3ax}). 
With a little algebra it reads
\begin{equation}\label{phicompact}
\Phi^{\rm int}(r,z)=2\pi\rho_c\left[a_\perp^2\zeta_0(\alpha)-z^2\zeta_\parallel(\alpha)-r^2\zeta_\perp(\alpha)\right],
\end{equation}
where the three ``shape functions'' $\zeta_0, \zeta_\perp$ and $\zeta_\parallel$, are
\begin{eqnarray}\label{ausiliari}
\zeta_0(\alpha)&=&\frac{1}{2}\int_0^{+\infty}\frac{{\rm d}u_*}{(\alpha^2+u_*)\sqrt{1+u_*}}\quad\quad=\quad\frac{\sec^{-1}\alpha}{\sqrt{\alpha^2-1}},\\
\label{parallelo}\zeta_\parallel(\alpha)&=&\frac{\alpha^2}{2}\int_0^{+\infty}\frac{{\rm d}u_*}{(\alpha^2+u_*)(1+u_*)^{3/2}}=\frac{\alpha^2\left(\sqrt{\alpha^2-1}-\sec^{-1}\alpha\right)}{(\alpha^2-1)^{3/2}},\\
\zeta_\perp(\alpha)&=&\frac{\alpha^2}{2}\int_0^{+\infty}\frac{{\rm d}u_*}{(\alpha^2+u_*)^2\sqrt{1+u_*}}\quad=\frac{\sqrt{\alpha^2-1}+\alpha^2\sec^{-1}\alpha}{2(\alpha^2-1)^{3/2}}.
\end{eqnarray}
Note that, once made explicit they depend only on $\alpha$. Alternative expressions are given in term of  hyperbolic functions and logarithms in Ref. \cite{kiwamoto}, while their asymptotic forms for large and vanishing values of $\alpha$ are listed in Tab. (\ref{table:nonlin})\\
\begin{table}[ht] 
\centering 
\begin{tabular}{c c c c} 
\hline\hline 
Case & $\alpha\rightarrow0$ & $\alpha=1$ & $\alpha\rightarrow\infty$ \\ [0.5ex]	
\hline 
$\zeta_0(\alpha)$&$\ln(2/\alpha)$&1&$\pi/(2\alpha)$ \\ 
$\zeta_\parallel(\alpha)$&$\alpha^2\left[\ln(2/\alpha)-1\right]$&1/3&1 \\ 
$\zeta_\perp(\alpha)$&1/2&1/3 &$\pi/(4\alpha)$ \\ [1ex]
\hline 
\end{tabular} 
\label{table:nonlin} 
\caption{Asymptotic trends of $\zeta_0, \zeta_{\parallel}$ and $\zeta_{\perp}$ in the limit of large and vanishing $\alpha$, as well as $\alpha=1$ (sphere).} 
\end{table}
Once written in form (\ref{phicompact}), it is evident that the potential inside the homogeneous spheroid is a harmonic function of the coordinates $(r,z)$ only, of which $\zeta_{\parallel}$ and $\zeta_{\perp}$ are its two eigenfrequencies.\\
\indent Note that, any regular ellipsoidal distribution of charge or mass stratified on similar {\it concentric} homeoids, generates at its exterior a family of {\it confocal} ellipsoidal equipotential surfaces, see Refs. \cite{chandra87} and \cite{kellogg}, 
therefore equipotential and equidense surfaces do not coincide. For a discussion of the analogous problem for the gravitational collapse of spheroidal initially cold distributions of matter 
see e.g. \cite{lin65}, \cite{falle75}, \cite{binney77}, \cite{boily99} and references therein.\\
\indent The internal\footnote{Potential and electric field on the exterior of the ellipsoid are obtained simply by putting $\lambda$ as inferior extreme of integration in the shape functions.} electrostatic field, given by $\mathbf{E}^{\rm int}(r,z)=-\nabla\Phi^{\rm int}(r,z)$ is then, written by components
\begin{align}
\begin{cases}\label{field}
E_z^{\rm int}(z)=-\partial_z \Phi^{\rm int}(r,z)=4\pi\rho_c\zeta_\parallel(\alpha)z\\
E_r^{\rm int}(r)=-\partial_r \Phi^{\rm int}(r,z)=4\pi\rho_c\zeta_\perp(\alpha)r.
\end{cases}
\end{align}
In other words, this means that the perpendicular and parallel components of $\mathbf{E}^{\rm int}$ are linear functions of their coordinates $r$ and $z$ respectively.\\
\indent In Fig. \ref{potspheroid}, as an example, we show a section of the equipotential surfaces for the potential generated by a prolate ($\alpha=10$) and an oblate ($\alpha=0.1$) spheroid.\\
\indent Having established an expression for the electric inside the spheroid, the separated nonrelativistic equations of motion, for an infinitesimal element of volume of unit charge $\delta q$ and mass $\delta m$ read
\begin{align}\label{goveq}
\frac{{\rm d}^2}{{\rm d}t^2}\tilde z=4\pi\frac{\delta q}{\delta m}\rho_c\zeta_\parallel(\alpha)\tilde z,\nonumber\\
\frac{{\rm d}^2}{{\rm d}t^2}\tilde r=4\pi\frac{\delta q}{\delta m}\rho_c\zeta_\perp(\alpha)\tilde r
\end{align}
where $\tilde z=z/z_0$ and $\tilde r= r/r_0$ and $r_0$ and $z_0$ are the element's initial coordinates.\\
\indent Due to the linearity in $r$ and $z$ of the components of $\mathbf{E}^{\rm int}$, and to the symmetry of the problem, using the same arguments of Sect. \ref{contmod} we find that different elements of volume coming from two nested ellipsoidal shell can not cross each other. 
Thus a homogeneously charged spheroid starting with initial conditions characterized by vanishing kinetic energy expands retaining a homogeneous charge density, see also Ref. \cite{bat01}. 
However, since Eqs. (\ref{goveq}) are essentially those for a particle in a 2d harmonic repulsor with two different eigenfrequencies, we expect that the aspect ratio $\alpha$ changes as a function of time since the ellipsoid expands at different rates
 along the $z$  and $r$ directions.\\
\indent Since no overtaking can happen among different ellipsoidal shells, the Coulomb explosion of a uniform charged spheroid of total charge $Q$ and mass $M$ and initial aspect ratio $\alpha_0=a_{\perp,0}/a_{\parallel,0}$, is fully determined by the time evolution of its two semi-axes $a_\perp$ and $a_\parallel$. 
Essentially this is done by integrating the equations of motion (cfr. also Eqs. \ref{goveq}) for two infinitesimal elements of volume initially placed at one of the two poles $(z=a_{\parallel,0}),r=0$ and on the equatorial plane $(z=0,r=a_{\perp,0})$, that read
\begin{eqnarray}\label{cesphero}
\frac{{\rm d}^2}{{\rm d}t^2}a_\parallel=\omega_0^2\frac{a_{\perp,0}^2a_{\parallel,0}}{a_\perp^2}\zeta_{\parallel}\left(\frac{a_\perp}{a_{\parallel}}\right),\nonumber\\
\frac{{\rm d}^2}{{\rm d}t^2}a_\perp=\omega_0^2\frac{a_{\perp,0}^2a_{\parallel,0}}{a_\perp a_\parallel}\zeta_{\perp}\left(\frac{a_\perp}{a_{\parallel}}\right).
\end{eqnarray}
In the equations above $\omega_0=Q\sqrt{3/(Ma_{\parallel,0}a_{\perp,0}^2)}$ is the initial plasma frequency. In order to simplify the notation we use the normalized quantities 
$\tau=\omega_0 t$, $\tilde a_\perp =a_\perp/a_{\perp, 0}$ and $\tilde a_\parallel=a_\parallel/a_{\parallel, 0}$, so that Eqs. (\ref{cesphero}) become
\begin{eqnarray}\label{cespheronorm}
\frac{{\rm d}^2}{{\rm d}\tau^2}\tilde a_\parallel&=&\frac{\zeta_{\parallel}\left(\alpha_0\frac{\tilde a_\perp}{\tilde a_{\parallel}}\right)}{\tilde a_\perp^2},\nonumber\\
\frac{{\rm d}^2}{{\rm d}\tau^2}\tilde a_\perp&=&\frac{\zeta_{\perp}\left(\alpha_0\frac{\tilde a_\perp}{\tilde a_{\parallel}}\right)}{\tilde a_\perp \tilde a_\parallel}.
\end{eqnarray}
Assuming initial kinetic energy $K_0=0$, the set of (normalized) initial conditions reads
\begin{align}
\begin{cases}\label{incondspher}
\tilde a_{\parallel}(\tau=0)=1,\\
\dot{\tilde a}_{\parallel}(\tau=0)=0,\\
\tilde a_{\perp}(\tau=0)=1,\\
\dot{\tilde a}_{\perp}(\tau=0)=0.
\end{cases}
\end{align}
The first integral of (\ref{cespheronorm}), accounting for the conversion of potential energy into kinetic energy is formally
\begin{eqnarray}\label{formal1}
\frac{1}{2}\left(\frac{\d\tilde a_\parallel}{\d\tau}\right)^2=&\int_{1}^{\tilde a_\parallel(\tau)}{\frac{\zeta_{\parallel}\left(\alpha_0\frac{\tilde a_\perp}{\tilde a_{\parallel}}\right)}{\tilde a_\perp^2}\d\tilde a_\parallel},\nonumber\\
\frac{1}{2}\left(\frac{\d\tilde a_\perp}{\d\tau}\right)^2=&\int_{1}^{\tilde a_\perp(\tau)}\frac{\zeta_{\perp}\left(\alpha_0\frac{\tilde a_\perp}{\tilde a_{\parallel}}\right)}{\tilde a_\perp\tilde a_\parallel}\d\tilde a_\perp.
\end{eqnarray}
Noting that at every time $\tilde a_\parallel$ and $\tilde a_\perp$ can be written as function of the normalized time dependent aspect ratio $\tilde\alpha$, the latter become
\begin{eqnarray}\label{formal2}
\frac{1}{2}\left(\frac{\d\tilde a_\parallel}{\d\tau}\right)^2=&\int_{1}^{\tilde a_\parallel(\tau)}{\frac{\zeta_{\parallel}\left(\alpha_0\tilde\alpha\right)}{\tilde\alpha^2\tilde a_\parallel^2}\d\tilde a_\parallel},\nonumber\\
\frac{1}{2}\left(\frac{\d\tilde a_\perp}{\d\tau}\right)^2=&\int_{1}^{\tilde a_\perp(\tau)}\frac{\zeta_{\perp}\left(\alpha_0\tilde\alpha\right)\tilde\alpha}{\tilde a_\perp^2}\d\tilde a_\perp.
\end{eqnarray}
\begin{figure} [h!t]
        \centering 
         \includegraphics[width=\textwidth]{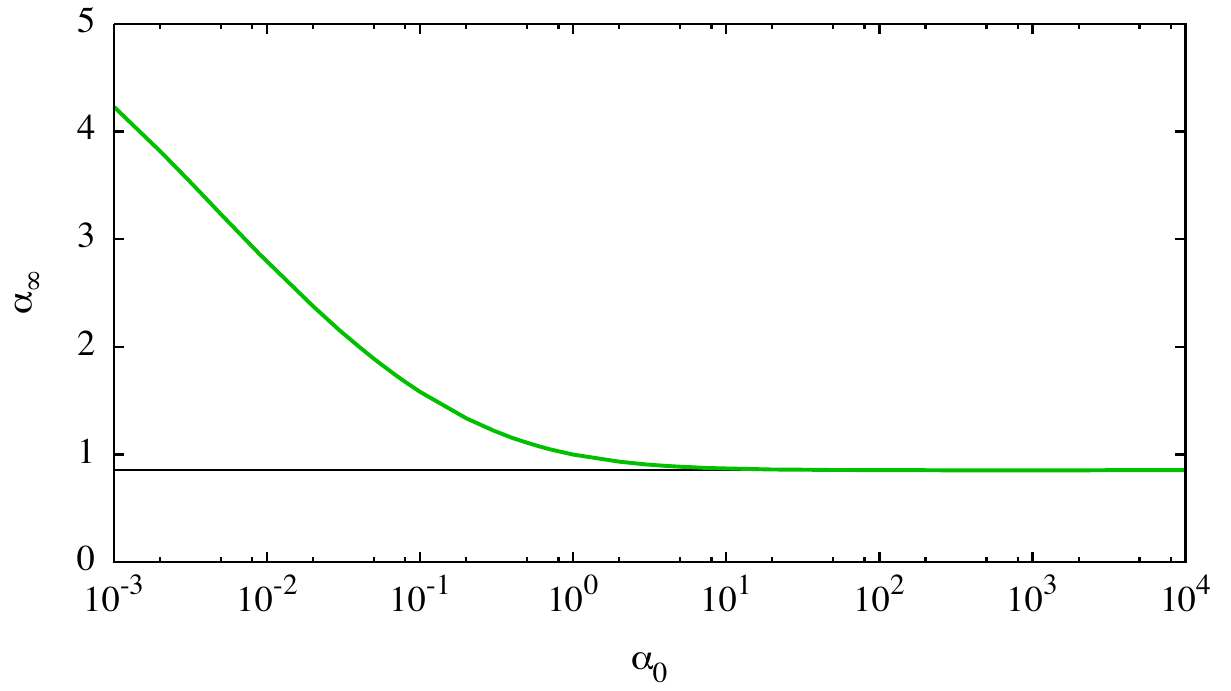}
         \caption{Asymptotic aspect ratio $\alpha_{\infty}$ as a function of $\alpha_{0}$. The thin black line shows the limit $\sim0.86$ for $\alpha_0\rightarrow\infty$.}
\label{aspect}
\end{figure} 
In the limit of $\tau\rightarrow\infty$, at the right hand sides of the equations above one has the asymptotic kinetic energies $\mathcal{E}_{\parallel,\infty}$ and $\mathcal{E}_{\perp,\infty}$ for the two elements of volume 
starting from one of the poles, and from the equator. Such energies are also the maximal kinetic energies that elements of volume can attain along the $r$ and $z$ coordinates, 
we can therefore assume that 
\begin{equation}\label{asino}
\alpha_\infty\propto\sqrt{\mathcal{E}_{\perp,\infty}/\mathcal{E}_{\parallel,\infty}}.
\end{equation}  
\indent Unfortunately, due to the dependence of $\zeta_{\parallel}$ and $\zeta_{\perp}$ on the time dependent aspect ratio $\alpha$, no analytical solution is available for the system of Equations (\ref{cespheronorm}) in terms of simple functions for a general value of $\alpha_0$, and one is forced to solve it numerically for instance with explicit finite difference schemes.\\
\indent However, in the two special cases $\alpha_0\rightarrow\infty$ and $\alpha_0\rightarrow 0$ it is possible to extract $\alpha_\infty$ making use of (\ref{formal2}) and (\ref{asino}) and the asymptotic forms of $\zeta_\perp$ and $\zeta_\parallel$, obtaining 
$\alpha_\infty\sim0.86$ and $\infty$ respectively. In Fig. \ref{aspect}, we show the asymptotic final aspect ratio for a large range of $\alpha_0$. Note that the convergence to the limit value for large $\alpha_0$ is particularly fast.\\
\indent Figure (\ref{limita})  shows the time evolution of $\alpha$ for some values of $\alpha_0$ between 0.1 and 10 obtained integrating numerically Eqs. (\ref{cespheronorm}). It is evident how initially prolate spheroids become oblate as they Coulomb-explode and initially oblate
spheroids become instead prolate. This means that in the first case the expansion occurs mainly in the transverse direction, while in the latter along the symmetry axis. Remarkably, while for extremely prolate configuration the $\alpha_\infty$ is unbound, for extremely initially 
oblate systems, the limit aspect ratios are bounded by a 
finite value (i.e. one can not obtain final arbitrarily prolate configuration).\\   
\indent In Fig. \ref{limlim}, for the same systems introduced above, the time dependent maximal energies along the transverse and parallel directions are plotted. In the case of a prolate spheroid ($\alpha_0<1$), where the transverse semi-axis increases much faster than the 
parallel one, the saturation to the maximal transverse energy to its asymptotic value is also faster than along the other direction. The converse is true for an initially oblate ($\alpha_0>1$) spheroid.\\ 
\indent It remains to derive an expression for the asymptotic differential energy distribution $n(\mathcal{E})$. Due to the linearity at every time of the components of the electric field in the coordinates $z$ and $r$ (cfr. Eq. (\ref{field})), 
 the two components $v_\perp(r)$ and $v_\parallel(z)$ of the velocity of each infinitesimal element of volume are at any time the linear functions of its coordinates $z,r$
\begin{eqnarray}\label{velotime}
v_\parallel(z)&=&(z/a_{\parallel})v_{\parallel,{\rm max}},\\
v_\perp(r)&=&(r/a_{\perp})v_{\perp,{\rm max}},
\end{eqnarray}
where $v_{\parallel,{\rm max}}$ and $v_{\perp,{\rm max}}$ are the maximum values of the velocity along the symmetry axis of the spheroid and in the equatorial plane respectively. 
As a consequence of that, the initial potential energy of the system maps into its asymptotic kinetic energy, and thus we can use the same arguments of Sect. \ref{contmod} to extract its final distribution.\\
\indent Since in a homogeneous spheroid the charge contained in a thin cone directed along $a_\perp$, of infinitesimal opening angle $\delta\vartheta$ increases with $z^3$, 
and analogously the charge in an infinitesimal cylinder at the equatorial plane increases with $r^3$, the two partial energy distributions $n(\mathcal{E}_\parallel)$ and $n(\mathcal{E}_\perp)$, along the symmetry axis and in the equatorial read
\begin{eqnarray}
n(\mathcal{E}_\parallel)&=&=\frac{3}{2}\sqrt{\frac{\mathcal{E}_{\parallel}}{\mathcal{E}_{\parallel,{\rm max}}^3}}\theta(\mathcal{E}_{\parallel,{\rm max}}-\mathcal{E}_\parallel),\\
n(\mathcal{E}_\perp)&=&=\frac{3}{2}\sqrt{\frac{\mathcal{E}_{\perp}}{\mathcal{E}_{\perp,{\rm max}}^3}}\theta(\mathcal{E}_{\perp,{\rm max}}-\mathcal{E}_\perp),
\end{eqnarray}
where $\mathcal{E}_{\parallel,{\rm max}}=\delta m v_{\parallel,{\rm max}}^2/2$ and $\mathcal{E}_{\perp,{\rm max}}=\delta m v_{\perp,{\rm max}}^2/2$.\\
\indent At every time $t$, it is evident from Eqs. (\ref{velotime}), that all the equivelocity surfaces (i.e. surfaces where the modulus of the expansion velocity is constant) have the same aspect ratio 
\begin{equation}
 \alpha_v(t)=\frac{a_{\perp}(t)}{a_{\parallel}(t)}\frac{v_{\parallel,max}(t)}{v_{\perp,max}(t)}.
\end{equation}
We stress the fact that such aspect ratio is different from that of the charge distribution (and it is the reason why the latter changes), see the sketch in Fig. \ref{aspetto}.
For the reason that no overtaking happens, the asymptotic velocity on each ellipsoidal shell of aspect ratio $\alpha_v(t)$ (and therefore the corresponding kinetic energy) depends on its initial potential energy. From the charge enclosed inside every equipotential surface one obtains   
\begin{align}
n(\mathcal{E})_{\alpha_0<1}=\frac{3}{2\mathcal{E}_{\perp,{\rm max}}}
\begin{cases}\label{incondsphera}
\sqrt{\frac{\mathcal{E}}{\mathcal{E}_{\parallel,{\rm max}}}},\quad{\rm for}\quad\mathcal{E}<\mathcal{E}_{\parallel,{\rm max}}\\
\sqrt{\frac{\mathcal{E}_{\perp,{\rm max}}-\mathcal{E}}{\mathcal{E}_{\perp,{\rm max}}-\mathcal{E}_{\parallel,{\rm max}}}},\quad{\rm for}\quad\mathcal{E}_{\parallel,{\rm max}}<\mathcal{E}<\mathcal{E}_{\perp,{\rm max}}
\end{cases}
\end{align}
for the prolate case, and 
\begin{align}
n(\mathcal{E})_{\alpha_0>1}=\frac{3}{2\mathcal{E}_{\perp,{\rm max}}}
\begin{cases}\label{incondspherb}
\sqrt{\frac{\mathcal{E}}{\mathcal{E}_{\parallel,{\rm max}}}},\quad{\rm for}\quad\mathcal{E}<\mathcal{E}_{\perp,{\rm max}}\\
\sqrt{\frac{\mathcal{E}}{\mathcal{E}_{\parallel,{\rm max}}}}-\sqrt{\frac{\mathcal{E}-\mathcal{E}_{\perp,{\rm max}}}{\mathcal{E}_{\parallel,{\rm max}}-\mathcal{E}_{\perp,{\rm max}}}},\quad{\rm for}\quad\mathcal{E}_{\perp,{\rm max}}<\mathcal{E}<\mathcal{E}_{\parallel,{\rm max}}
\end{cases}
\end{align}
for the oblate case.
\begin{figure} [h!t]
        \centering 
         \includegraphics[width=0.95\textwidth]{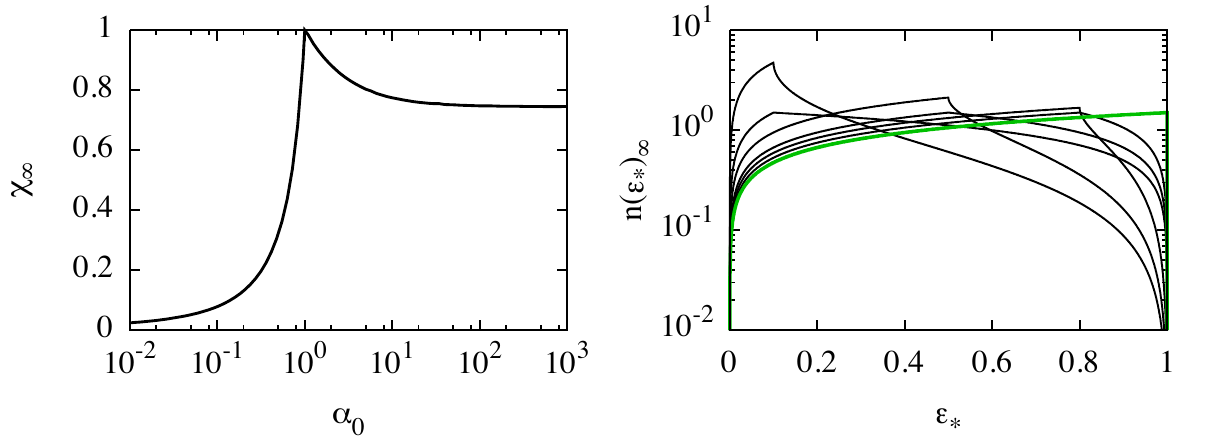}
         \caption{Left panel: ratio $\chi_{\infty}$ of the energy corresponding to the max of $n(\mathcal{E})$ to the maximal energy. Right panel: differential energy distribution $n(\mathcal{E})$ for different values of $\chi_{\infty}$, the thick green curve marks the $\alpha=1$ case.}
\label{neteo}
\end{figure} 
 In Fig.\ref{neteo}, left panel, the ratio $\chi_{\infty}=(\mathcal{E}_{\perp,{\rm max}}/\mathcal{E}_{\parallel,{\rm max}})_{\alpha_0<1}$; $=(\mathcal{E}_{\parallel,{\rm max}}/\mathcal{E}_{\perp,{\rm max}})_{\alpha_0>1}$ is shown. 
In the right panel the asymptotic differential energy distributions are shown for different values of $\chi_{\infty}$.\\ 
 \indent Note that if $\mathcal{E}$ is normalized to its maximum value $\mathcal{E}_{\rm max}=\mathcal{E}_{\parallel,{\rm max}}$ for the oblate case and $\mathcal{E}_{\perp,{\rm max}}$ for the prolate case, 
the latter equations consent one parameter fits with the end products of numerical simulations of the Coulomb explosion of homogeneous spheroids, as we will see in the next Section. Note also, that for $\alpha_0\neq1$ the scaled energy distribution vanishes for $\mathcal{E}_*=1$. 
This is not surprising as the latter energy is in all cases that of a subset of measure 0 of the whole charge distribution (i.e. the poles for the initially prolate systems and the equatorial ring, for the initially oblate ones).
\subsection*{Numerical simulations using particles}
\begin{figure} [h!t]
        \centering 
         \includegraphics[width=\textwidth]{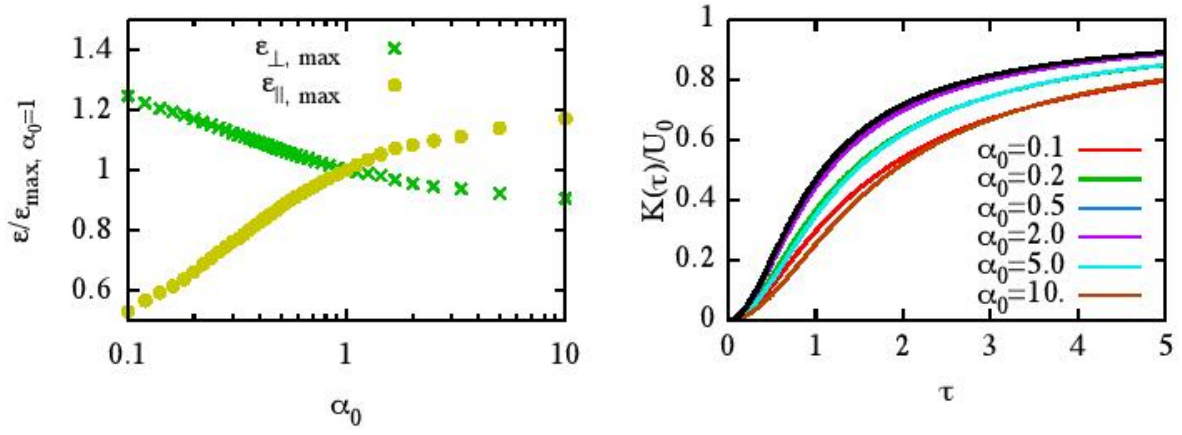}
         \caption{Left panel: maximal energies along the symmetry axis (circles) and in the equatorial plane (crosses) expressed in units of the maximal asymptotic energy for the sphere for $N-$body simulations with $N=2.5\times10^4$. Right panel: Time dependence of the kinetic energy $K$ in units of the total initial electrostatic potential energy $U_0$ for $\alpha_0=0.1, 0.2, 0.5, 2, 5$ and 10. The thick black curve marks the $\alpha_0=1$ case for which the energy conversion happens in the shortest time in units of $t_{\rm dyn}$.}
\label{limitspheroid}
\end{figure}

\begin{figure} [h!t]
        \centering 
         \includegraphics[width=\textwidth]{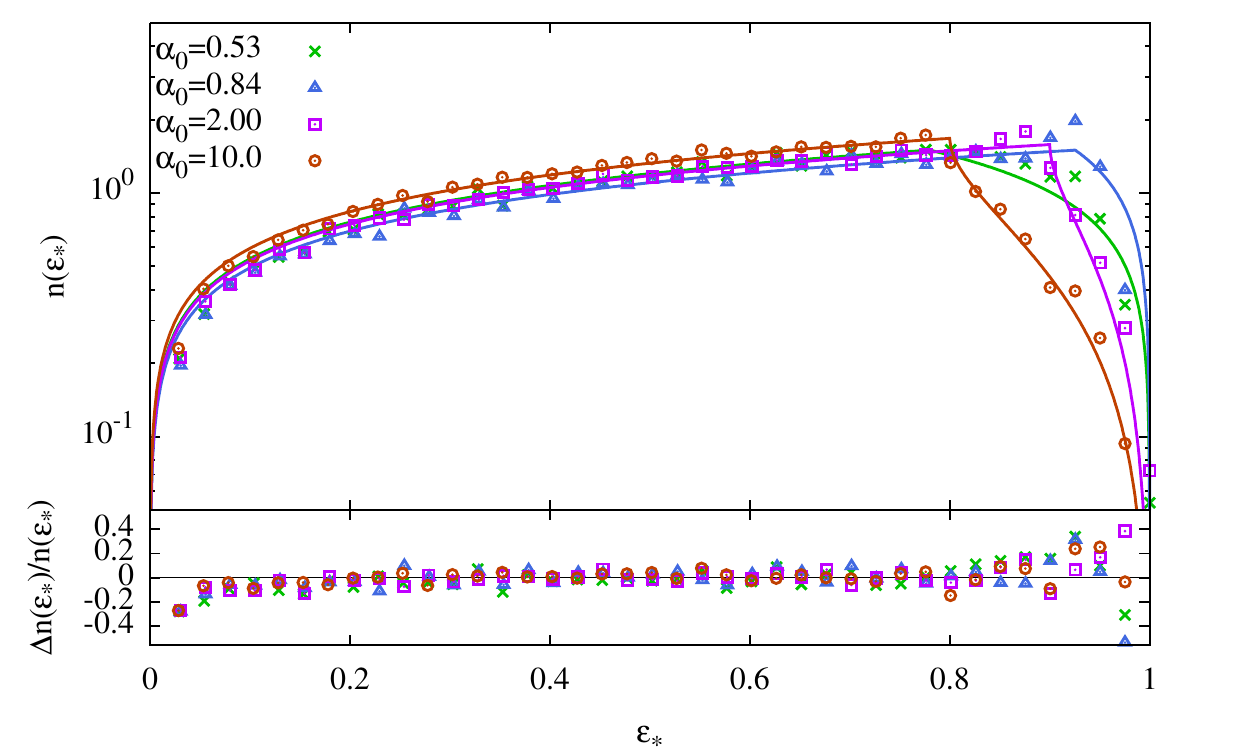}
         \caption{Upper panel: differential energy distribution $n(\mathcal{E}_*)$ of the end products at $\tau_{99\%}$ (points) for different direct $N-$body simulations ($N=2.5\times10^4$), with the indicated values of the initial aspect ratio $\alpha_0$. Fits with their analytic expressions (lines) are also shown. Each curve is normalized to its maximum energy, so that $\mathcal{E}_*=\mathcal{E}/\mathcal{E}_{\rm max}$. Lower panel: residuals of the fits $\Delta n(\mathcal{E}_*)/n(\mathcal{E}_*)$.}
\label{fitspheroid}
\end{figure}

\begin{figure} [h!t]
        \centering 
         \includegraphics[width=\textwidth]{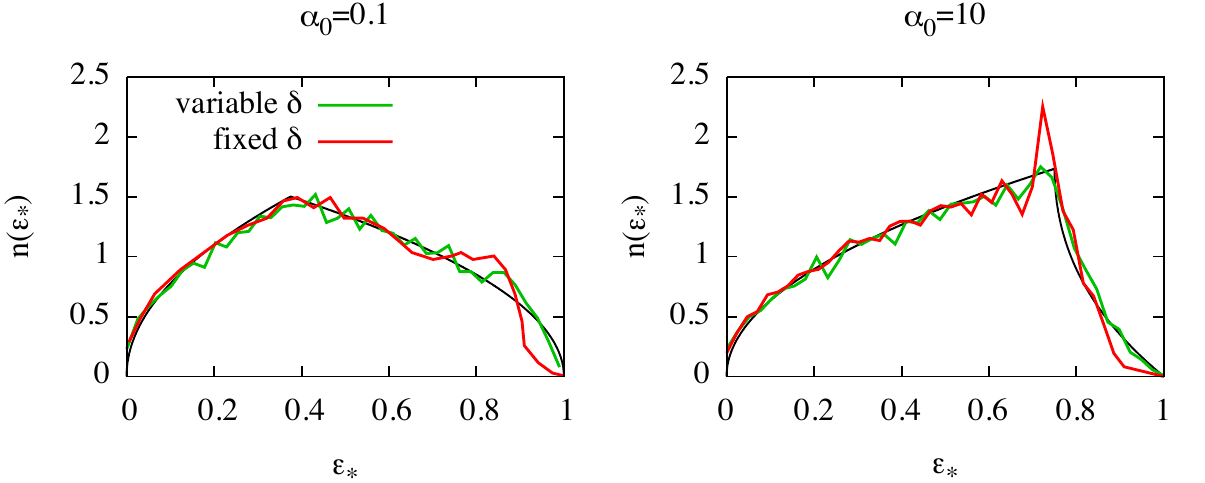}
         \caption{Differential energy distributions for the case with imposed initial minimal inter-particle distribution (red curves) and random distribution (green curves) for initially prolate ($\alpha_0=0.1$) and initially oblate ($\alpha_0=10$) spheroids. The thin black lines marks the analytic expressions. All cases have the same average particle number density and number of particles $N=10^4$.}
\label{spherhole}
\end{figure}

\begin{figure} [h!t]
        \centering 
         \includegraphics[width=\textwidth]{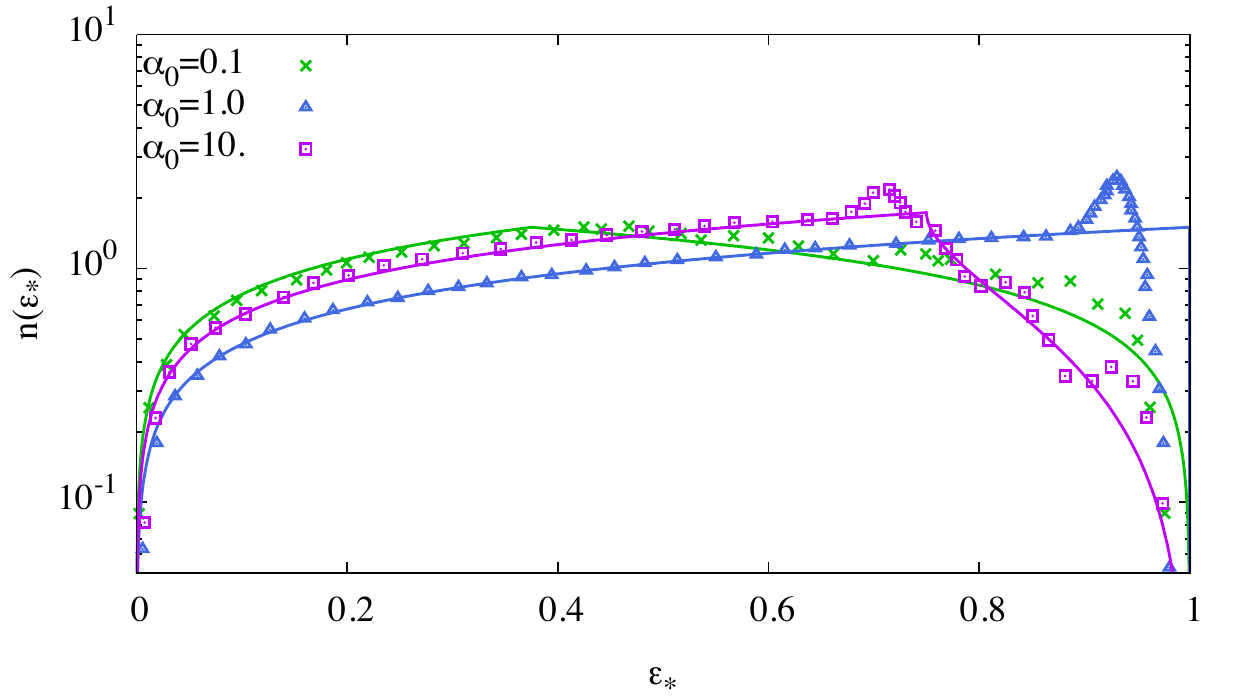}
         \caption{Normalized differential energy distributions $n(\mathcal{E}_*)$ for the cases with $\alpha_0=0.1$, 1 and 10 for the end products of PIC simulations. Note that the energies are rescaled to each case's maximal energy so that $\mathcal{E}_*=\mathcal{E}/\mathcal{E}_{\rm max}$. Note also, that in the lower energy region the numerical points do not fall considerably under the analytic curve as for the direct simulations. All systems have the same initial number density and number of particles $N=1.5\times10^6$.}
\label{nepic}
\end{figure}

We have run $N-$body simulations of single component spheroidal Coulomb explosion with both direct molecular dynamics and with {\it particle-in-cell} (PIC) schemes. 
For an extended treatment of the numerical methods here involved, again we direct the reader to Chap.(\ref{numerica}) and the references cited therein.\\ 
\begin{figure} [h!t]
        \centering 
         \includegraphics[width=0.9\textwidth]{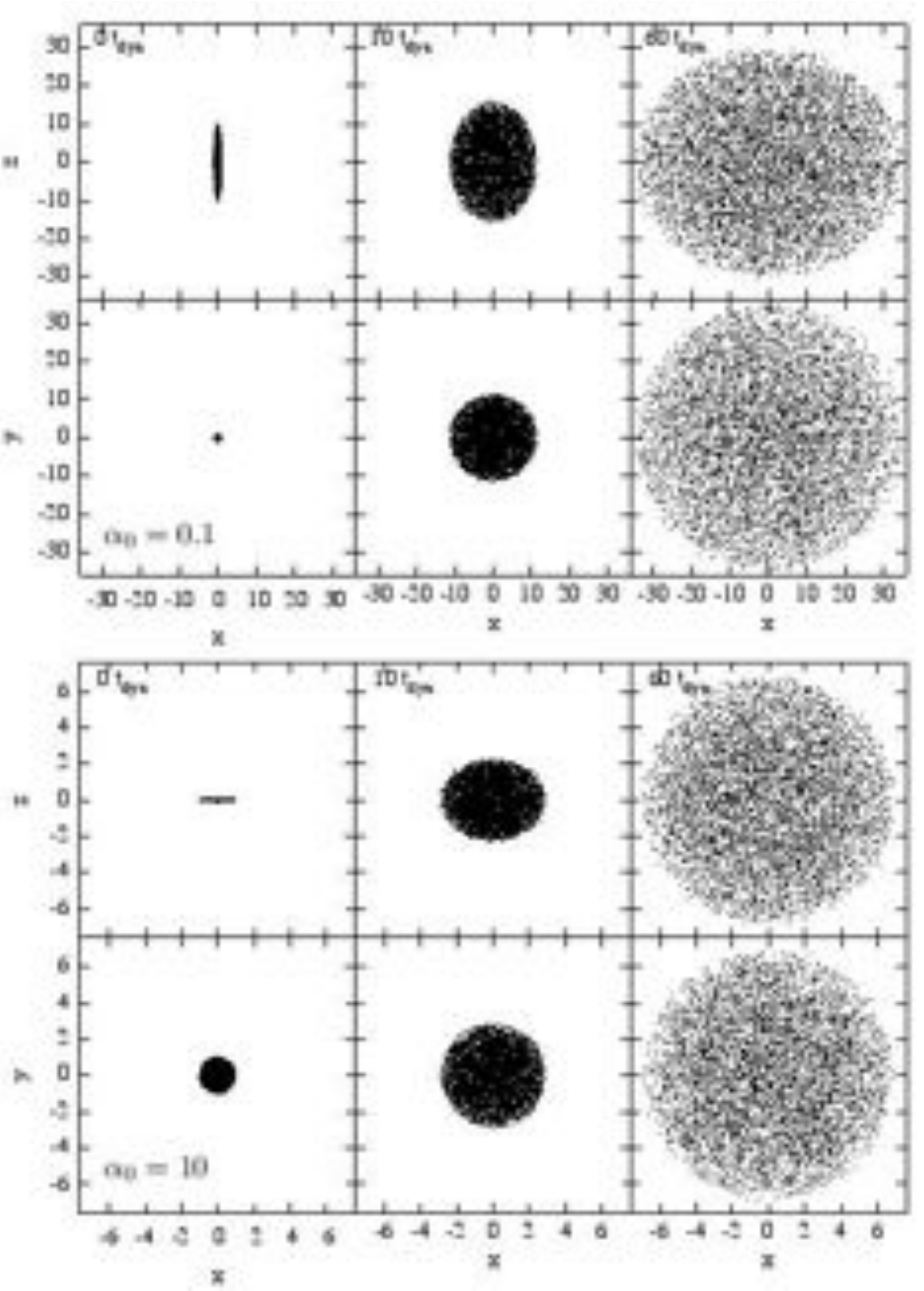}
         \caption{Projection at 3 different times of the positions in the $x,z$ (upper row) and $x,y$ (bottom row) planes, for of an initially cold prolate spheroid ($\alpha_0=0.1)$, top panel), and for an initially cold oblate spheroid ($\alpha_0=10$, bottom panel) with the same initial density. Particle's coordinates are given in units of $a_{\perp,0}$, in both cases $N=10^4$.}
\label{esplosionesph}
\end{figure} 
\indent In both sets of simulations we have spanned the range of initial aspect ratio $\alpha_0$ going from 0.1 to 10 and considered only systems starting with no kinetic energy. 
The dynamical time $t_{\rm dyn}$ used to normalize times is again defined as in Eq. (\ref{dyntime}) and, as usual, we run the calculations up to $\tau_{99\%}$.\\  
\indent As discussed in Ref.\cite{grech2011}, the products of our numerical simulations are in good agreement with the analytic predictions discussed above. 
An example, Figure \ref{esplosionesph} shows at different times the positions of the particles from two MD simulations of the Coulomb explosion of initially cold spheroids with $\alpha_0=0.1$ and $\alpha_0=10$, and the same number density and particle number $N=10^4$. Note how the expansion is faster in the transverse direction for the first case, while is faster along the $z$ axis in the latter.\\
\indent As one would expect from the theoretical time evolution of $\mathcal{E}_{\perp,{\rm max}}$ and $\mathcal{E}_{\parallel, {\rm max}}$ (see Fig. \ref{limlim}), for fixed initial particle density and dynamical time, 
for the $\alpha_0=1$ case the conversion of the initial potential energy $U_0$ in kinetic energy happens faster than for other values of $\alpha_0$. This is shown in the right panel of Fig. \ref{limitspheroid}.\\
\indent However, (see right panel of the same figure), for larger and smaller values of $\alpha_0$, higher energies (in units of $\mathcal{E}_{\rm max}$ for the sphere with same particle density) can be obtained along the symmetry axis (for initially oblate spheroids), and in the meridional plane (for initially prolate spheroids).\\
\indent Knowing its analytical expression in the continuum picture, it comes natural to compare the numerical $n(\mathcal{E})$ of the simulations' end products with its analytical counterpart. For direct MD simulations, in Fig. \ref{fitspheroid} we show the differential energy distribution 
at $\tau_{99\%}$ for the indicated values of $\alpha_0$ as well as the fitted curves given by Eqs. (\ref{incondsphera}) and (\ref{incondspherb}). 
The analytic expression derived for the continuum model fits the numerical $n(\mathcal{E})$ for a broad range of energies (on average roughly the $75\%$ of the whole spectrum). 
However, non negligible discrepancies are found at low and high values of  $\mathcal{E}_*$, where the residuals (see bottom panel) exceed 0.3. 
The reasons why one finds such deviations from the theoretical curve near the normalized cutoff energies $\mathcal{E}_*=0$ and $\mathcal{E}_*=1$ are essentially different.\\
\indent Particles occupying the low energy region of the spectrum are those sitting close to the centre of the system in the initial condition. In homogeneous spheroidal systems the number of particles increases linearly with the ellipsoidal radial coordinate $s$. 
This means that in a homogeneous distribution sampled with point particles at low $s$ (i.e. close to the centroid) there is a relatively small number of particles, implying that the numerical $n(\mathcal{E})$ is depopulated close to the minimum energy.\\
\indent If the deviation in the lower energy part of the spectrum has essentially a trivial origin, what happens near the cutoff energy is different and has the same origin as in the case of the homogeneous sphere discussed in Chap. \ref{chapmultice} and Ref.\cite{mika2013}.\\ 
\indent Note that for a generic homogeneous spheroidal system with $\alpha_0\neq1$, the theoretical $n(\mathcal{E})$ vanishes for $\mathcal{E}=\mathcal{E}_{\rm max}$, see Equations (\ref{incondsphera}) and (\ref{incondspherb}), 
the $\alpha_0=1$ case stands out as the the only one having the maximum 
value of $n(\mathcal{E})$ right at $\mathcal{E}_{\rm max}$. 
Discreteness effects leading to energy bunching, affecting essentially the contribution to the differential energy distribution at $\mathcal{E}_{\rm max}$, 
are for spheroidal systems mitigated by the fact than particles reaching $\mathcal{E}_{\rm max}$ do no come from a surface, as in the case of a sphere, but from the poles for an initially oblate spheroid, and from the equatorial ring for an initially prolate spheroid, hence, the less the fraction of the system contributing to $\mathcal{E}_{\rm max}$, the less the numerical $n(\mathcal{E})$ is prone to discreteness effects.\\
\indent Spanning a broad range of values of $\alpha_0$, we found that the peak in the numerical $n(\mathcal{E})$ disappears for $\alpha_0<0.25$ and $\alpha_0>2$, for initial conditions where no minimal inter-particle separation is fixed, 
while it is always present (in particular for initially oblate systems) if a fixed inter-particle distance $\delta$ between near neighbours is enforced, 
albeit not as in the spherical case. Fig. \ref{spherhole} shows the final energy distributions for systems with and without fixed minimum distance $\delta$, for initially prolate and oblate spheroids; the systems with fixed minimum $\delta$ (red curves) are clearly departing from the analytic function.\\  
\indent The final number energy distributions from the particle-in-cell simulations made with the code \textsc{calder} (see Refs. \cite{2003LPB....21..573P} and \cite{2003NucFu..43..629L} for its details, 
see also Chapter \ref{numerica} for a more general description of Particle-mesh codes), starting from analogous initial conditions 
also show the high energy peak as shown in Fig. \ref{nepic}. Since in this type of numerical approach, the potential and force are based on the solution of PDEs rather than on the direct sums of individual particles contributions, 
one would expect it to better reproduce the quantities obtained analytically from {\it continuum models}, having ``erased'' discreteness effects.\\ 
\indent Nevertheless, the fact that charge densities and electromagnetic fields are discretized on a mesh implies, contrary to what happens in direct simulations, that near the surface of the system the density is spuriously lower, even if the particles are distributed homogeneously in the continuum space. 
This resulting in an effectively decreasing density profile, and thus prone to generate shocks of which the high energy peaks in $n(\mathcal{E})$ are a clear signature.\\
\indent For the PIC simulations presented here and in Ref. \cite{grech2011}, we found that $n(\mathcal{E})$ is reproduced fairly well for the $\alpha_0=1$ case (for more than the 80\% of the normalized energy interval), 
except for the usual prominent peak near $\mathcal{E}_{\rm max}$, while the final differential energy distribution for the spheroids with $\alpha_0<1$ is found to be well fitted by its analytic expression on roughly the 70\% of the energy interval, that is on average less than what happens for the end products of direct simulations. 
In general the results for the cases with $\alpha_0>1$ deviate much from the analytical predictions, see Fig. \ref{nepic}.\\
\indent We also stress the fact that, contrary to what happens in the simulations with the direct code, the low energy tails of $n(\mathcal{E})$ is well reproduced for all the systems simulated with the PIC code, since the combined effect of a larger number of particles and the way the force is computed do not allow for its depopulation.
\section{Uniform triaxial systems}
The imaging of complex biomolecules with x-ray lasers (see Refs. \cite{2006NatPh...2..839C} and \cite{imaging}, see also \cite{neuze}) presents the problem of recovering the tridimensional structure 
from 2-dimensional images (projections) of different samples whose orientation with respect to the laser focus is unknown.\\
\indent If one assumes as a first crude approximation that such molecules are triaxial ellipsoids, the energies of their fragments may in principle give us information on their orientation. This makes us some interest to investigate the Coulomb explosion of triaxial systems.
\subsection*{Continuum model}
The potential at a given point $\mathbf{r}=(x,y,z)$ generated by a triaxial ellipsoid filled with homogeneous charge density is a harmonic function of the $(x,y,z)$ coordinates of the form
\begin{equation}
\Phi_{x,y,z}=\Phi_0-\frac{1}{2}\left(\omega_a^2x^2+\omega_b^2y^2+\omega_c^2z^2\right),
\end{equation}
where $\Phi_0$ is a constant and the three (time dependent) eigenfrequencies $\omega_a,\omega_b,\omega_c$ depend on the three semi-axes $a,b$ and $c$ through incomplete elliptic integrals (see Ref.\cite{chandra87}).\\
\indent In analogy to what has been done in cylindrical coordinates for a rotational ellipsoid, in Cartesian coordinates the Coulomb explosion of a triaxial ellipsoid of initial charge density $\rho_{c,0}$ and initial semi-axes $a_0, b_0$ and $c_0$ oriented along $x,y$ and $z$, is fully described by three coupled ODE (envelope equations) for the three infinitesimal volume elements of mass $\delta m$ and charge $\delta q$ initially placed at $(a_0,0,0)$; $(0,b_0,0)$ and $(0,0,c_0)$.
\begin{figure} [h!t]
        \centering 
         \includegraphics[width=\textwidth]{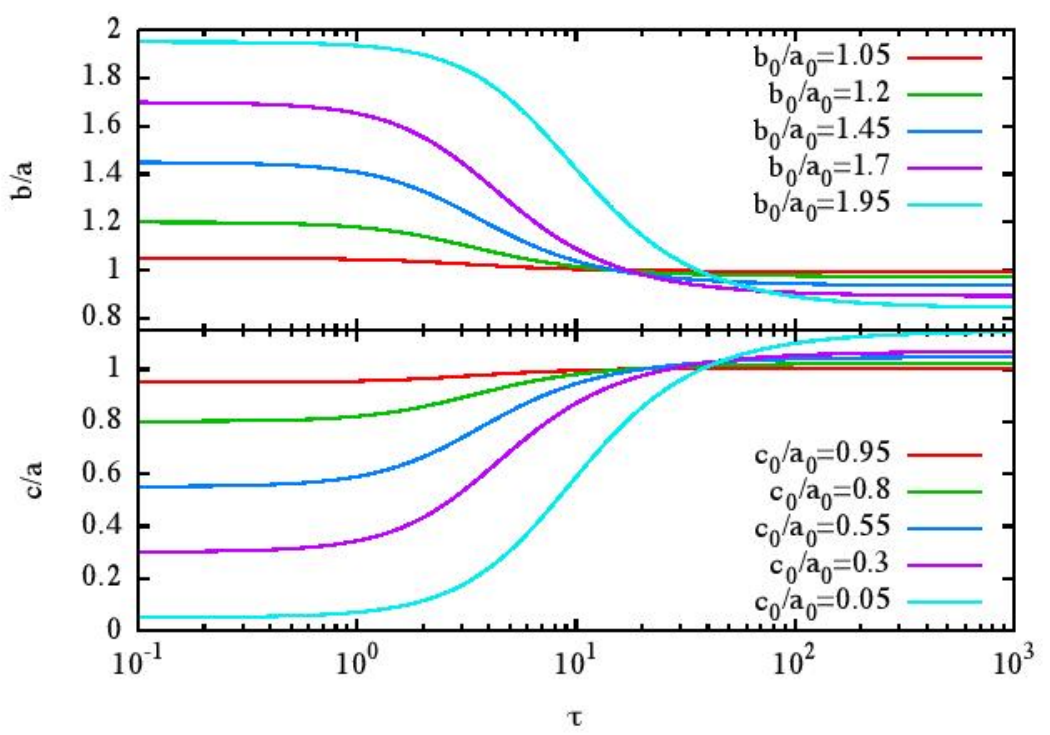}
         \caption{Time evolution of the ratios $b/a$ (top panel) and $c/a$ (bottom panel) for different triaxial uniform ellipsoids with intermediate semi-axis $a$, obtained by integrating numerically the three envelope equations. Times are normalized to the inverse of the initial plasma frequency $\omega_0$.}
\label{exptrix2}
\end{figure} 
Setting $C=2\pi\rho_{c,0}a_0b_0c_0\delta q/\delta m$ the three equations read
\begin{eqnarray}
\label{1trix}\frac{\de^2a}{\de t^2}&=&Ca\int_0^{+\infty}\psi(a,b,c,u)\frac{\de u}{(a^2+u)},\\
\label{2trix}\frac{\de^2b}{\de t^2}&=&Cb\int_0^{+\infty}\psi(a,b,c,u)\frac{\de u}{(b^2+u)},\\
\label{3trix}\frac{\de^2c}{\de t^2}&=&Cc\int_0^{+\infty}\psi(a,b,c,u)\frac{\de u}{(c^2+u)},
\end{eqnarray}
where $\psi(a,b,c,u)$ is defined by
\begin{equation}
\psi(a,b,c,u)\equiv\frac{1}{\sqrt{(a^2+u)(b^2+u)(c^2+u)}}.
\end{equation}
Unfortunately, in Cartesian coordinates it is not possible to derive simple limit expressions for the asymptotic axial ratios $b_{\infty}/a_{\infty}$ and $c_{\infty}/a_{\infty}$ due to their implicit dependence on elliptic functions in the auxiliary variable $u$. 
\begin{figure} [h!t]
        \centering 
         \includegraphics[width=0.9\textwidth]{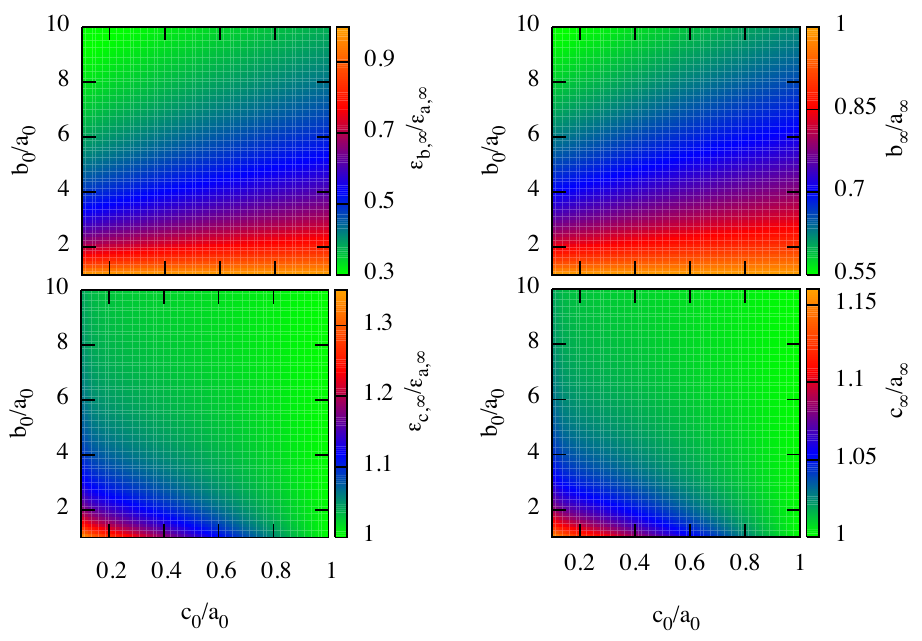}
         \caption{Left panels: final axial ratios $c_\infty/a_\infty$ and $b_\infty/a_\infty$ for combinations of initial ratios $c_0/a_0$ and $b_0/a_0$ in the interval $(0.1, 1)$ and (1, 10) respectively. Right panels:  maximal kinetic energies ratios $\mathcal{E}_{b,\infty}/\mathcal{E}_{a,\infty}$ and $\mathcal{E}_{c,\infty}/\mathcal{E}_{a,\infty}$ along the same directions.}
\label{etrix}
\end{figure} 
The analogous problem arises also in polar ellipsoidal coordinates, and there is in general less analytic work to do.\\
\indent Intuitively, it must be pointed out that triaxial ellipsoids that depart only slightly from prolate or oblate spheroids, are expected to behave not very differently 
from their regular counterparts. By integrating numerically Eqs. (\ref{1trix}), (\ref{2trix}) and (\ref{3trix}) for the time evolution of $a,b$ and $c$, one obtains the asymptotic final axial ratios $b_{\infty}/a_{\infty}$ and $c_{\infty}/a_{\infty}$ 
as well as the maximal kinetic energies energies reached along the three axes during the the explosion.\\
\indent Figure \ref{exptrix2} shows for some initial values of the three semi-axes the time evolution of the ratios $b/a$ and $c/a$. It is clearly evident how the intermediate semi-axis (in these cases $a_0$) remains the intermediate as the system expands, 
while the initially longer is asymptotically the shorter and vice versa. As a general trend, for fixed initial charge density, charge and mass, the more the initial configuration is markedly triaxial, the slower (in units of the dynamical time scale $\omega_0^{-1}=\sqrt{Ma_0b_0c_0/3}/Q$) is the convergence to the limit axial ratios.\\ 
\indent Obviously, the largest among the asymptotic maximal energies along the three semi-axes, $\mathcal{E}_{a,\infty},\mathcal{E}_{b,\infty}$ and $\mathcal{E}_{c,\infty}$, is that along the direction of the initially shortest semi-axis, 
as it is evident from the left panels of Fig. \ref{etrix} where the ratios $\mathcal{E}_{b,\infty}/\mathcal{E}_{a,\infty}$ and $\mathcal{E}_{c,\infty}/\mathcal{E}_{a,\infty}$ are shown.
\begin{figure} [h!t]
        \centering 
         \includegraphics[width=\textwidth]{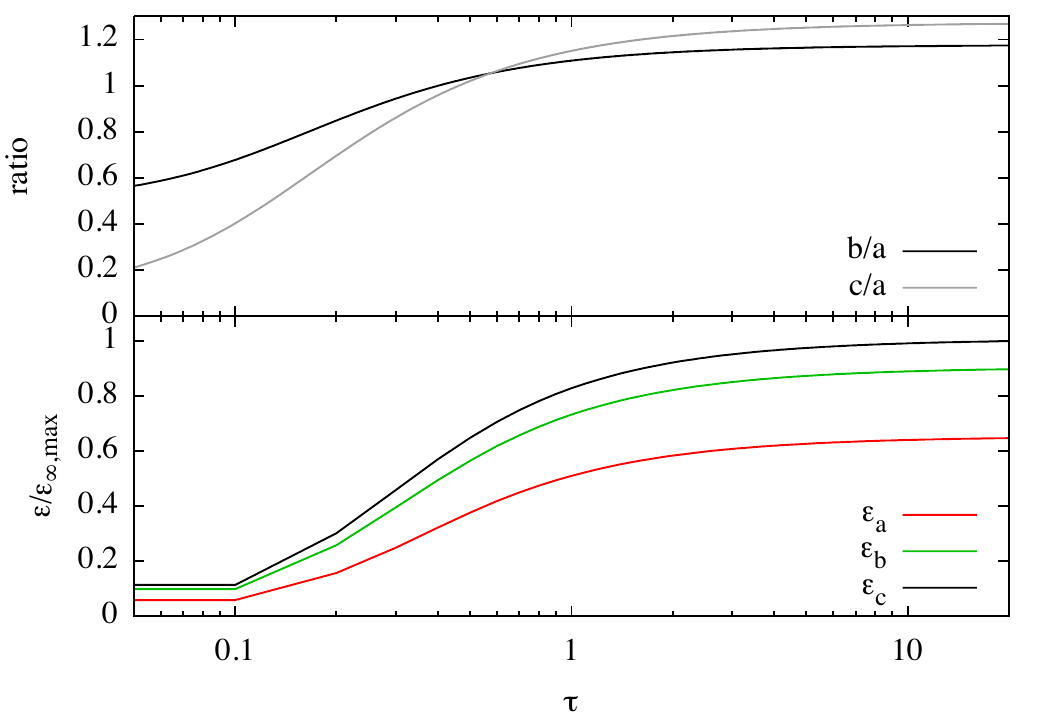}
         \caption{Top panel: evolution of the axial ratios $b/a$ and $c/a$ (top panel) for an $N-$body simulation starting from a markedly triaxial initial configuration ($b_0/a_0=0.5$ and $c_0/a_0=0.1$). Bottom panel: maximal energies along $a,b$ and $c$ in units of the asymptotic maximal value of the three $\mathcal{E}_{\infty,{\rm max}}=\mathcal{E}_{c,\infty}$.}
\label{exptrix}
\end{figure} 
The ratios of the corresponding semi-axis are also shown (right panels), and present the same trend with the initial combinations of $a_0,b_0$ and $c_0$.\\ 
\indent In contrast to the spheroidal systems, to derive the asymptotic expression for $n(\mathcal{E})$ is not possible in terms of simple functions since an integration over a triaxial ellipsoidal volume is involved, 
therefore we can in principle derive them only ``empirically" from the end products of our $N-$body simulations. 
\subsection*{Numerical simulations using particles}
Following the same criterion for axisymmetric systems, we performed direct $N-$body simulations of Coulomb explosion of initially cold triaxial ellipsoids.\\
\indent We spanned the range of ratios (1, 10) in $b_0/a_0$ and (0.1, 10) in $c_0/a_0$ 
keeping fixed the system's number density $n_0=3N/4\pi a_0b_0c_0$ and total charge and mass. In none of the runs we impose a minimal inter-particle distance. As for the spheroidal systems, the dynamical time scale $t_{\rm dyn}$ of the simulation is defined by Eq. (\ref{dyntime}).\\ 
\begin{figure} [h!t]
        \centering 
         \includegraphics[width=\textwidth]{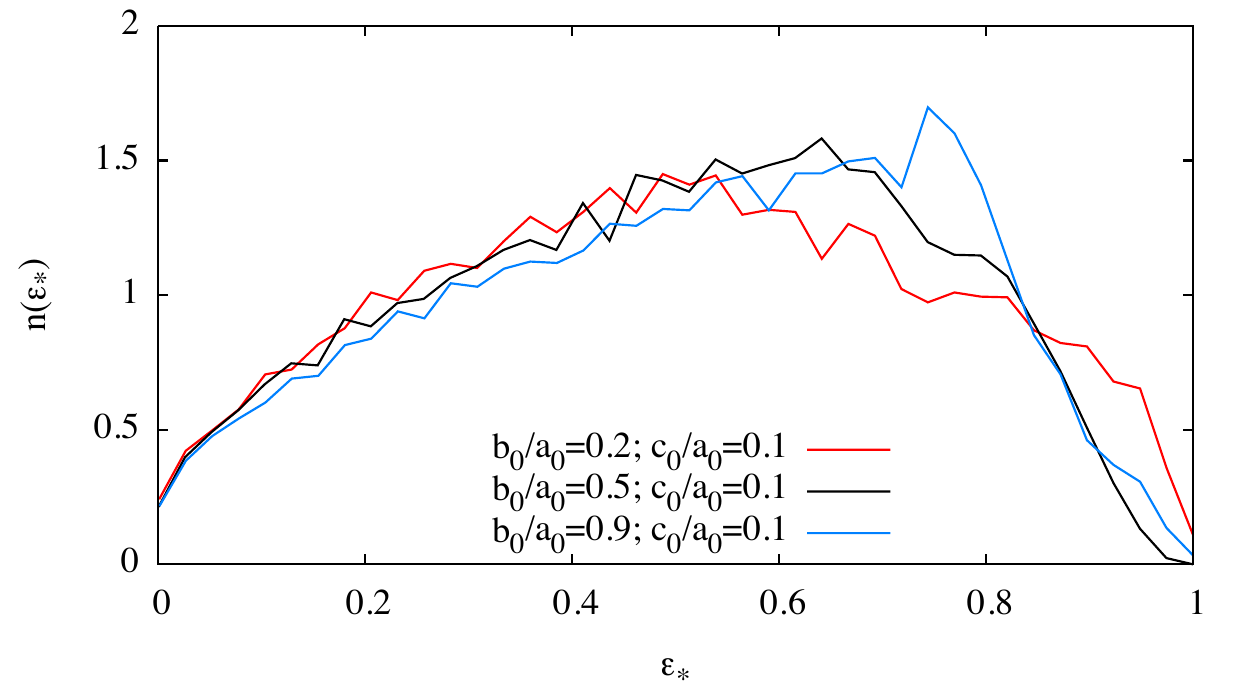}
         \caption{Normalized differential energy distribution $n(\mathcal{E}_*)$ at $\tau_{99\%}$ with respect to $\mathcal{E}_{\rm max}$ for three different combinations of $b_0/a_0$ and $c_0/a_0$ that are slightly prolate, markedly triaxial and slightly oblate. In all cases $N=2.5\times10^4$.}
\label{netrix}
\end{figure} 
\indent The evolution of the maximal energies along the three semi-axis as well as that of the ratios of the latter reproduces nicely what found in the continuum model. In Fig. \ref{exptrix} we show the time evolution of such quantities for an initially markedly triaxial ellipsoid.\\
\indent In Fig. \ref{netrix} the $n(\mathcal{E}_*)$ is plotted for the end products  of three $N-$body simulations starting from three particular initial conditions quasi-prolate ($b_0/a_0=0.2$, $c_0/a_0=0.1$), strongly triaxial ($b_0/a_0=0.5$, $c_0/a_0=0.1$) 
and quasi-oblate ($b_0/a_0=0.9$, $c_0/a_0=0.1$). It is evident, that the final differential energy distributions resemble as expected to those of the parent axisymmetric systems when $a_0\simeq\ b_0>c_0$ or $a_0\simeq\ b_0<c_0$ 
while for particularly triaxial initial configurations, the $n(\mathcal{E})$ appears to be a combination of those of the two (regular) types, oblate and prolate.\\ 
\indent It must be noted, that none of the analyzed systems in the ranges of initial axial ratios presents a spike at high energies in $n(\mathcal{E})$ as observed for spherical and spheroidal systems. 
\section{Non uniform axisymmetric systems}
\begin{figure} [h!t]
        \centering 		
         \includegraphics[width=\textwidth]{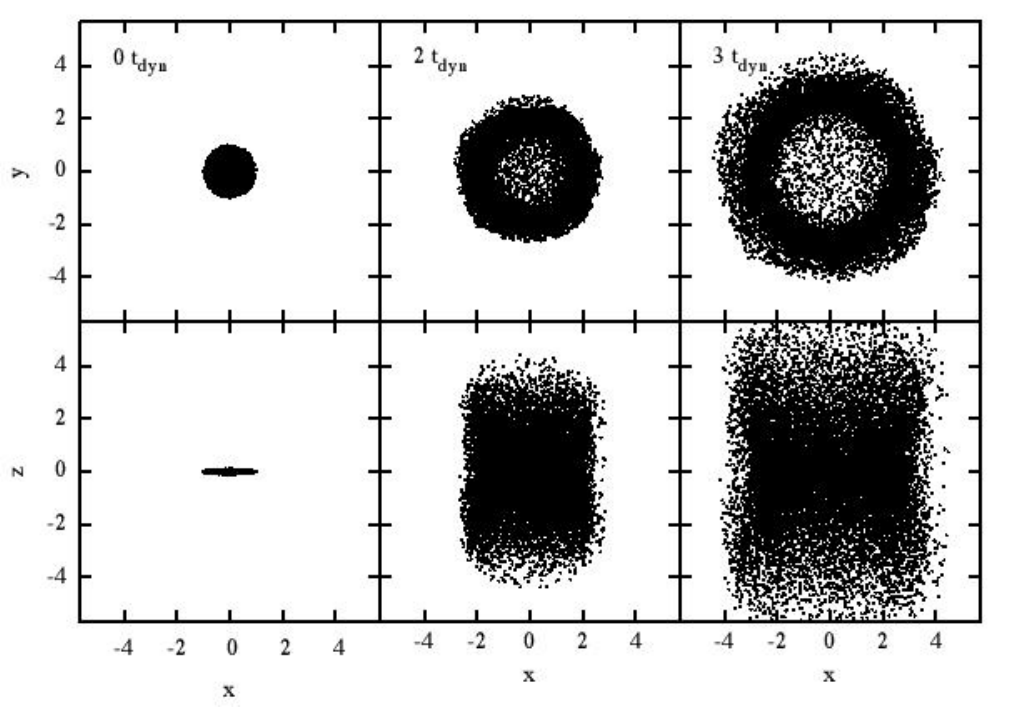}
         \caption{Projections of the particles positions in the $x,y$ (upper row) and $x,z$ (bottom row) planes for a system of $N=15000$ particles starting with $\alpha_0=0.1$. Positions are given in units of the initial major semi-axis $a_{\perp}$. Note how an initial dense core results in particles bunching at a dense equatorial ring.}
\label{tori}
\end{figure}

To conclude, in the same line of Chapter \ref{chapmultice}, we now briefly treat the Coulomb explosion of spheroids with nonuniform density profile. Such problem is of some relevance for instance 
in the field of particle acceleration from the interaction of strong lasers with nanostructured targets with non uniform densities (see e.g. \cite{2010NIMPA.620...63G} and references therein), and as also pointed out in Ref. \cite{2003PhRvL..91n3401K}.\\
\indent As it happens to spherical systems with non uniform initial density profiles, 
\begin{figure} [h!t]
        \centering 
         \includegraphics[width=\textwidth]{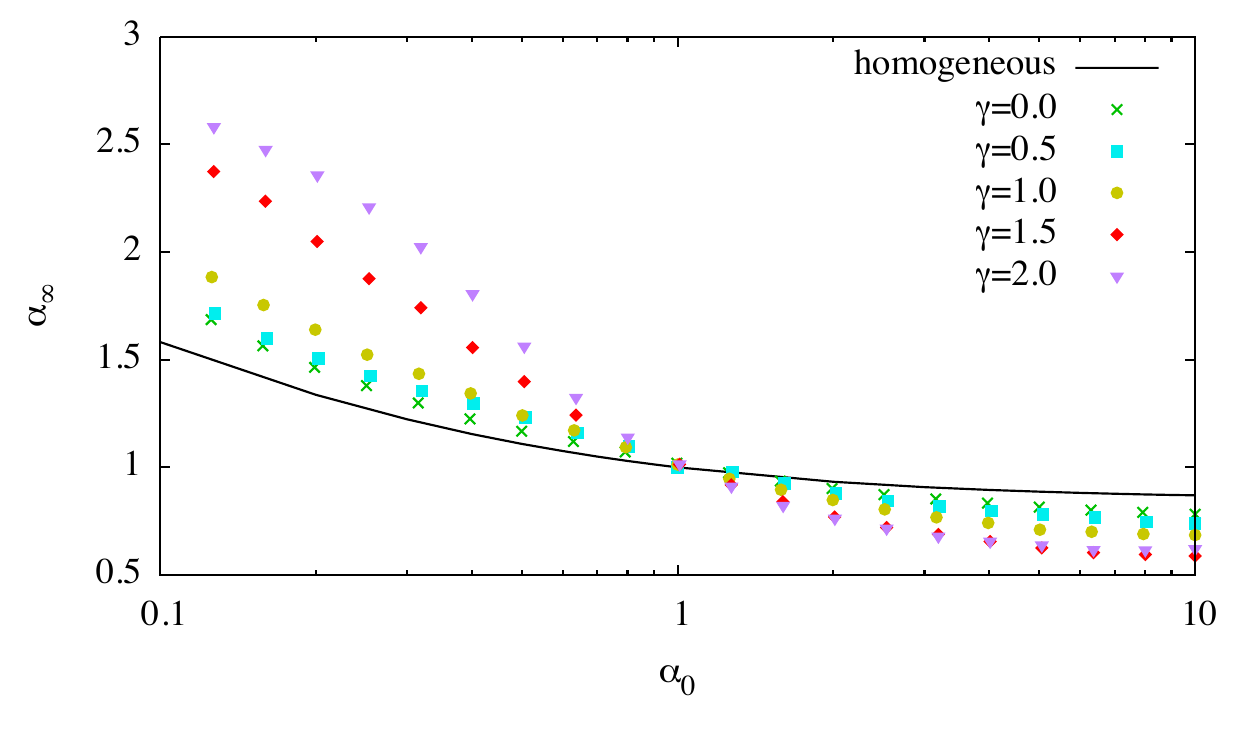}
         \caption{Aspect ratio at $\tau_{99\%}$ as function of $\alpha_0$ in the range (0.1, 10). Different symbols refer to different values of the logarithmic density slope $\gamma$. The solid black line marks as in Fig. \ref{aspect} the theoretical relation for the case of a homogeneous density profile.}
\label{alfagamma}
\end{figure} 
also spheroidal systems with non uniform density are expected to undergo shell crossing as they Coulomb explode. 
The additional complication due to their, non preserved axial ratios makes the problem even more prohibitive and, in principle, complete analytic treatment is impossible and one is forced to rely on numerics\\ 
\indent In our MD simulations, we extracted the initial positions of the particles, from the family of triaxial models with density given by
\begin{equation}\label{denhen3}
\rho(s)=\frac{Q(3-\gamma)}{4\pi abc}s^{-\gamma}(1+s)^{\gamma-4};\quad 0\leq\gamma<3,
\end{equation} 
where the ``ellipsoidal'' radius $s$ is defined by
\begin{equation}
s\equiv\sqrt{(x/a)^2+(y/b)^2+(z/c)^2}
\end{equation}
and $\gamma={\rm d}{\log}(\rho)/{\rm d}s$ is the (angularly averaged) logarithmic density slope. As usual, $Q$ is the total charge and $a,b$ and $c$ are as usual the three semi-axes.\\
\indent The density given by Eq. (\ref{denhen3}) is nothing but the triaxial generalization of of Denhen's one parameter $\gamma-$models (see e.g. Ref. \cite{1996ApJ...460..136M}), also used to in the field of beam physics to model charged particles bunches in accelerators, see Refs. \cite{2003LNP...626..154K} and \cite{2005NYASA1045...12K}.
We made this choice since it allows one to control (even in this case) the importance of the density cusp at the system's centroid, from flat cored systems ($\gamma=0$) to highly concentrated systems ($\gamma=3$).\\
\begin{figure} [h!t]
        \centering 		
         \includegraphics[width=\textwidth]{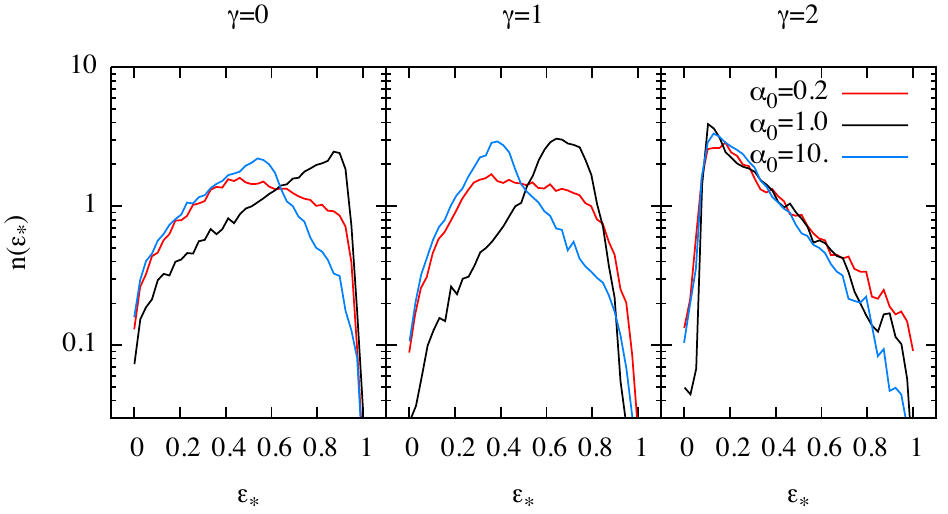}
         \caption{Differential energy distribution $n(\mathcal{E}_*)$ for $\gamma=0,$ 1 and 2 and from left to right and different value of the initial aspect ratio $\alpha_0$.}
\label{negamma}
\end{figure} 
\indent For reasons of simplicity, and having established that homogeneous triaxial ellipsoids do no show a considerably different behavior from that of the spheroids, here we limit ourselves to the axially symmetric $a=b=a_\perp$, $c=a_\parallel$ case. Note that in principle the density given by Eq. (\ref{denhen3}) falls to 0 only at $s\rightarrow+\infty$, 
here we have defined {\it truncated} models with a surface defined by 
\begin{equation}
s_t=\sqrt{(x/a)^2+(y/b)^2+(z/c)^2}=1.
\end{equation}
Figure (\ref{alfagamma}) shows for different values of  $\gamma$ the aspect ratio of the end products of the simulations as a function of the aspect ratio of their initial conditions. 
In the interval of initial aspect ratio $0.1\leq\alpha_0\leq10$ shown here, the aspect ratio of the end products of the expanding initially prolate $\gamma-$models takes for every $\gamma$ larger values with respect to those reached by 
spheroids of uniform density starting from the same $\alpha_0$, by contrast if the model is initially oblate, $\alpha_\infty$ falls always under the value attained by the correspondent homogeneous spheroid independently of $\gamma$.\\
\indent As a general trend, at fixed $\alpha_0$, $\alpha_\infty$ is smaller for larger values of $\gamma$ if $\alpha_0>1$ and vice versa, larger for $\alpha_0<1$. For a very prominent initial density cusp at the centroid $2\leq\gamma<3$, 
the relation $\alpha_\infty$ $vs$ $\alpha_0$ seems to flatten to limit values both at high and low values of $\alpha_0$ (not shown here). On the other hand, the differential energy distributions of the end products, shown in Fig. \ref{negamma} 
are qualitatively similar to those of the correspondent homogeneous spheroids when $\gamma\sim0$ (no central density cusp). Systems with mild cusps $0<\gamma<2$, have quite complex final differential energy distributions with several changes of slope, 
both for initially prolate and oblate initial conditions due to the interplay between shell crossing and time dependent aspect ratio. In the spherical cases, $n(\mathcal{E})$ are indeed very similar to those shown in Chapter \ref{chapmultice} 
for other non homogeneous radial density profiles. For large central density cusps ($2\leq\gamma<3$), independently on $\alpha_0$, $n(\mathcal{E})$ peaks at low energies and falls to high energies with similar power law decays also for $\alpha=1$.\\ 
\indent Remarkably, see Fig. \ref{tori}, for initially very flattened ($\alpha<0.5$) cuspy systems, the spheroidal symmetry is not retained during the expansion, instead, after few $t_{\rm dyn}$, the systems assume a toroidal shape, 
with most of the particles on a thick equatorial ring.
\section{Summary}
The main results presented in this chapter, in part published in Ref. \cite{grech2011}, can be summarized as follows: by means of a simple semi-analytical model for the Coulomb explosion of a uniformly charged spheroid in the limit of nonrelativistic regime we have studied the expansion of non-neutral aspherical nanoplasmas. 
The model allows one to express quantities such as the maximum energy a particle can reach at a given time,
the time-dependent particle energy distributions, as a function of the initial spheroid aspect ratio $\alpha_0$, charge density $\rho_{c,0}$, and
total conserved charge $Q$.\\
\indent Our theoretical predictions for aspect ratio evolution and final number energy distribution are found to be in good agreement with numerical simulations with both direct molecular dynamics and particle-mesh approaches. 
The ''discreteness`` peak in the number energy distribution observed in MD simulations of spherical and homogeneous system is also present, albeit of a lesser magnitude, also for spheroidal systems and in a certain measure for triaxial ellipsoid.\\
\indent In addition we performed simulations of spheroidal systems with non uniform charge densities finding a similar behavior to that of their spherical counterparts treated in the previous chapter. However, in all cases the high energy peaks in due to the particle overtaking are less prominent the more the symmetry departs from the spherical one.\\
\indent We stress the fact that, albeit highly idealized, a pure Coulomb explosion model is indeed useful and to some extent ``realistic" in the vertical ionization regime, where electrons are expelled
from the target on a time much shorter than the characteristic
time of ion motion, attainable using ultra-intense
lasers or x-ray pulses.\\
\indent Moreover, with the recent progress in nanotechnology, Coulomb explosion of specifically designed nanostructured targets can be considered. Our results
may thus give us simple design guidelines how to optimize target properties; for example, for inertial fusion applications or to maximize ion collision events for neutron production.\\
\indent Finally, our results can also be helpful to model laser-solid-target interaction for ion acceleration, which is characterized by the emission of short,
compact, and highly charged ion bunches. Propagation of these bunches (e.g., through a vacuum) is strongly affected by space charge effects. By approximating the accelerated ion bunches as uniformly charged spheroids, the results presented here may allow us to derive the conditions required for limited energy and angular dispersions.
\chapter{Dynamics of molecular hydride clusters irradiated by intense XFEL pulses at LCLS}\label{chapLCLS}
Atomic clusters have received much attention in recent
years as they are tunable targets for intense
laser-matter interaction. This applies to ``conventional'' laser pulses as well as to VUV and X-ray pulses available from new and upcoming free-electron laser machines.\\
\indent Molecular clusters add another degree of freedom and may thus be a tool to approach the radiation damage processes of large organic molecules exposed to X-ray radiation and they are as well interesting with respect to the possibility to drive fusion of deuterium via Coulomb explosion of deuterated samples, see \cite{2006PhRvE..74a6403B} \cite{2010JPhB...43m5603Z}, \cite{2013PhRvL.111h2502B} and \cite{2013PhRvE..87b3106B}.\\
\indent In this chapter we apply the results on the heterogeneous clusters presented in the previous chapters to study the real case of hydrides clusters (i.e. clusters of molecules containing an element of the first row and one or more hydrogens) irradiated by short and intense X-ray pulses causing multiple $K-$shell photoionizations. In particular, we discuss, in the light of our model, the experimental findings at SLAC's Linac Coherent Light Source (LCLS) on methane clusters exposed to soft X-ray ($\hbar\omega\sim1{\rm keV}$).
\section{The experiments at LCLS}
\begin{figure} [h!t]
        \centering 		
         \includegraphics[width=0.8\textwidth]{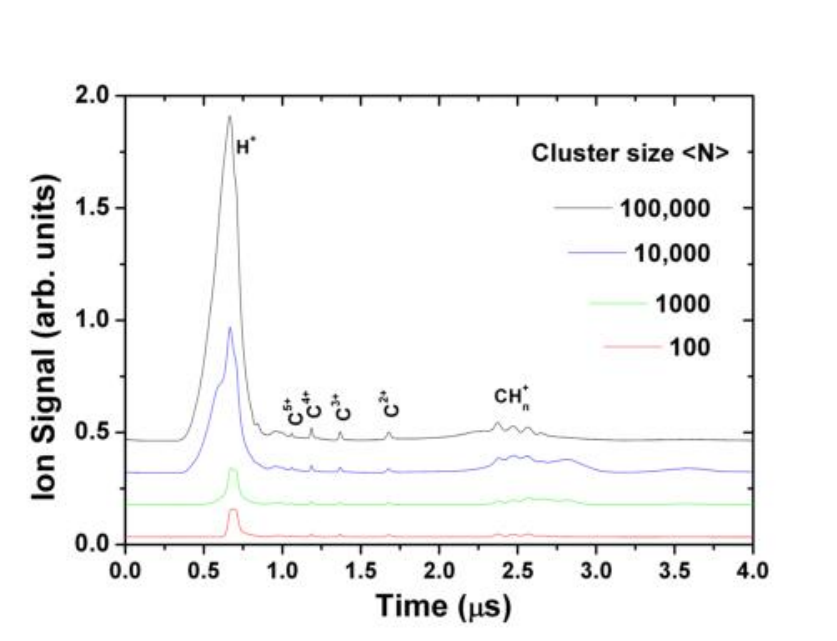}
         \caption{Time of flight spectrum in microseconds for the fragments of $(\ce{CH4})_N$ clusters with $N=10^2,10^3,10^4$ and $10^5$. irradiated by short X-ray pulses (roughly 20 femtoseconds) with $\hbar\omega=1$ keV. Figure taken from Ref. \cite{nirmala1}.}
\label{tof}
\end{figure}

\begin{figure} [h!t]
        \centering 		
         \includegraphics[width=0.8\textwidth]{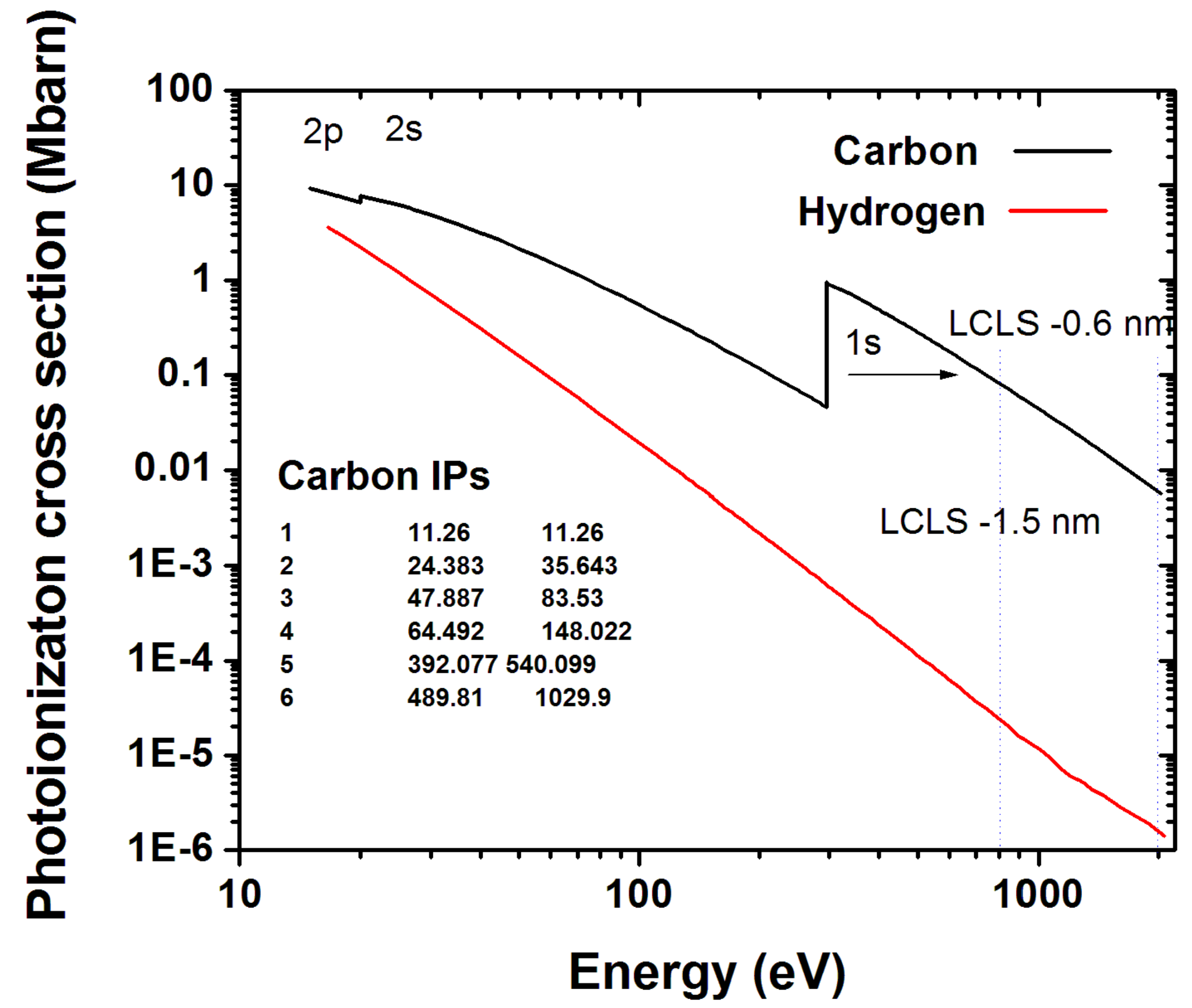}
         \caption{Total photoionization cross section for a carbon atom (black curve) and for hydrogen (red curve) as function of the photon energy in eV. The Carbon curve has always larger values than that for hydrogen and it is dominated by the $K-$shell contribution for $\hbar\omega$ larger than roughly 300 eV. In a \ce{CH4} molecule, the cross section of the $K-$shell of carbon does not significantly differ from that of the atomic species, while it is reasonable to assume that the cross section of the valence shell (molecular orbitals) is of the same order as that of atomic hydrogen.}
\label{sezionidurto}
\end{figure}

Recent experiments \cite{nirmala1} performed with the x-ray Free Electron Laser (xFEL) of the LINAC Coherent Light Source in Stanford, where jets of methane clusters have been irradiated with short pulses ($\sim 10 - 100$ fs) of soft x-rays, showed a large proton signal in contrast to an almost vanishing yield of carbon ions or molecular fragments containing carbon in the time of flight spectrum, as it is evident in Fig \ref{tof} where the different curves show the fragments yield for different cluster sizes in the jet.\\
Such effect appears to be independent on the cluster size, in the range observed here, and hints towards a charge migration from the two species in the molecules. Photons are prevalently absorbed via $K-$shell photoionization in carbon, that at such 
photon energies has a cross section of the order of 250 kilobarn at photon energy of 1 keV, almost two orders of magnitude larger than the cross section for photoionization of the valence electrons of the molecule, (see Fig. \ref{sezionidurto}, 
see also \cite{Doe:2009:Online}).\\
\indent Although the ionization mechanism is different, what is observed here is a species segregation like that seen for the same cluster irradiated with VUV sources \cite{2006AIPC..827..109D} and \cite{2013SPIE.8777E..0JT}, where the light protons overtake the heavy ions once reaching high velocities. 
What is puzzling is the low signal of the carbon ions implying that a large fraction of the carbon component emerges as neutrals. 
\section{The model}
Prompted by these findings, we performed a parametric numerical study of the interaction with short X-ray pulses of molecular clusters, in order to investigate the effect in a more general frame. 
We do not restrict ourselves to the case of methane clusters, but we consider the series of first row hydride molecules \ce{H2O}, \ce{NH3} and \ce{CH4}. As an ``atomic limit" of our system we use rare gas clusters of neon (\ce{Ne}) 
since such species has the same electronic configuration of the aforementioned hydrides and comparable $K-$shell ionization cross section. We report here the key assumptions and summarize the numerical model detailed in Chapter \ref{numerica}. 
\subsection*{Setting the stage}
In the same spirit of Refs. \cite{2007PhRvA..76d3203G}, \cite{2013CP....414...65G} and \cite{GNO09}, we use a hybrid {\it quantum-classical} model where ions and electrons are considered as classical particles and propagated with simple $N-$body schemes, while quantum processes, such as photoionization or Auger decays, are treated with the Monte Carlo method based on rates computed from photoionization cross sections, laser intensity and energy amplitudes of the transition.\\ 
\indent Such approach, contrary for instance to that used in Ref. \cite{BER04} where electrons were approximated by a continuum fluid, allows us to obtain a better description of the electron component for processes such as Auger decay or recombination as well as to book-keep the electronic shell occupation of the ions, and in addition, to stick to easier to implement propagation algorithms.\\ 
\indent In this study, we assume theoretical X-ray pulses with Gaussian time envelope and uniform space envelope. The latter choice is motivated by the fact that in the experiments discussed above, the laser focus diameter is $\sim2$ $\mu$m and the typical cluster radius 
is of the order of a few nm. We fix the photon energy $\hbar\omega=1{\rm keV}$ while we span a range of full-width-at-half-maximum for the time envelope, from 1 to 50 fs. Peak intensities $I_0$ span the interval $10^{16}-10^{20}{\rm W/cm^2}$. 
The $K-$shell photoionization cross sections $\sigma_{\rm 1s}$ and other atomic parameters of the elements here considered are summarized in Tab. \ref{parameters}.\\ 
\indent For such laser parameters, the interaction is always in the so called {\it weak field regime}, corresponding to values larger than unit attained by the Keldysh parameter \cite{drake}
\begin{equation}
\gamma_{\rm K}\equiv\sqrt{\frac{\Delta E}{2E_{\rm pond}}},
\end{equation} 
where $\Delta E$ is the ionization energy for the $K-$shell electron and the ponderomotive energy $E_{\rm pond}$ is given as
\begin{equation}
E_{\rm pond}=\frac{e^2I_0}{2m_{e}c\omega^2},
\end{equation}
where $e$ and $m_e$ are the electron charge and mass, $c$ is the speed of light and $\omega$ is the laser carrier frequency.
For the range of intensities explored here $\gamma_{\rm K}\simeq1.5$, that is strictly larger than 1, implying that we are in the regime of non-perturbative single photon tunneling ionization, (see, e.g. Ref. \cite{2006JPhB...39R..39S} 
and references therein).\\
\indent From the outcome of the experiments with methane clusters, it is evident that the occurrence of a strong hydrogen peak in the spectra is not influenced by the cluster size. We have considered therefore only clusters containing a number of molecules (or atoms) in the range $N_*=50 - 1000$, 
so that the computational effort due to the number of particles in the simulation is kept reasonably low.\\
\indent The probability used in the Monte Carlo sampling, that a photoionization event happens between times $t$ and $t+\Delta t$ for the $j$-th shell of the $i$-th atom or molecule in the cluster, 
as well as the probability of an Auger transition, are computed with the schemes discussed in Sect. \ref{quantumeffects}, while, due to the low size interval considered here the dynamics is resolved with a direct scheme where the 
Coulomb interaction potential is smoothed at distances $r\leq a_0$ with a linear spline, see Sect. \ref{sectsoft}.
\subsection*{The choice of the initial conditions}
The photon energies considered here are much larger than the typical binding energy of the electrons in the molecular orbitals, which are of the order of 4.5 eV, as well as of that the intermolecular van der Walls bonds in the cluster 
(of the order of 0.5 - 2 eV). In addition, the time scales of the exposure to the laser pulse ($\sim 10$ fs) and therefore of the charging process, are shorter than any chemistry related process in the cluster, we thus neglect completely the latter. Therefore, all simulation particles are initialized with $v=0$.\\
\indent The ``molecules" are simply implemented by placing their constituting elements at equilibrium distance. To include the effect of field ionization (i.e. atoms or molecules are further ionized by the global cluster electrostatic field), each atom has associated an electron represented by a real simulation particle that is initially 
placed at the same position. Note that smoothing the Coulomb potential implies that the latter electron is confined by an harmonic-like potential when placed at $r\leq a_0$ from the mother atom. With respect to the Auger processes, 
the effective electron (if still present) has always the priority on the ``virtual electrons" (i.e. those not yet initialized but just accounted in the electronic configuration of the ion) for being emitted as Auger electron, whenever such event takes places.\\
\indent Atoms or molecules are displaced in the clusters according to the typical truncated icosahedral shell structure, see also \cite{GNO09} and  \cite{2007PhRvA..76d3203G} and references therein, 
so that on each shell the charge density is constant.
\begin{table}[ht] 
\centering 
\begin{tabular}{c c c c c} 
\hline\hline 
Specie & $E_{\rm bind}\quad\left[{\rm eV}\right]$ & $\sigma_{\rm 1s}\quad\left[{\rm kbarn}\right]$ & $\tau_{\rm auger}^{\rm molecule}\quad\left[{\rm fs}\right]$ & $\tau_{\rm auger}^{\rm atom}\quad\left[{\rm fs}\right]$ \\ [0.5ex]	
\hline 
\ce{CH4}&284.2&44.07&7.76& 11.06\\ 
\ce{NH3}&409.9&76.99&5.36& 7.52\\ 
\ce{H2O}&543.1&121.9&4.46& 5.07\\ 
\ce{Ne}&870.2&248.2 &\quad& 2.90 \\ [1ex]
\hline 
\end{tabular} 
\label{parameters} 
\caption{$K$ shell parameters of the heavy atom in the molecular and atomic species used in the simulations. Cross sections $\sigma_{\rm 1s}$ and binding energies $E_{\rm bind}$ are assumed to be identical in the two cases while the average $K$ shell hole lifetimes sensibly differ, therefore in the last column we put as a reference the value of $\tau_{\rm auger}$ for the pure atomic species. All values listed here refer to neutral systems.} 
\end{table}
\section{Numerical simulations and results}
Previous numerical studies have addressed the problem of atomic or composite clusters exposed to strong laser pulses, aiming either at the description of ions dynamics (such as for instance in \cite{BER04}, 
see also the simplified models introduced in Chapter \ref{chapmultice}) or at a detailed description of the electron spectra, at expenses of a more simplified treatment of the ion motion, that in some studies are even approximated by a smooth charge 
background or {\it jellium model}, see e.g. \cite{2013CP....414...65G}. The reason of this being that ions move significantly on time scales considerably larger than that of the electrons. The typical kinetic energy of a $K-$shell photoelectron extracted from a carbon atom by a 1 keV photon is of the order of 700 eV, implying that it crosses a distance of $50$ atomic units of distance (radius of a typical nanocluster) in roughly 0.2 femtoseconds, while the typical ion motion scale is of the order of ten femtoseconds.\\
\indent In this thesis we tackle the problem with a rather simplified approach, in a way that the interplay of the ionic and electronic components can be treated as function of few tunable parameters. Moreover, using an approach based on rate equations and Monte Carlo sampling,
 allows one to model the coupling to the laser field in a computationally ``cheaper'' way than a full quantum treatment.\\
\indent We have studied both the time evolution and the dependence on the lasers intensity of global quantities such as the total charge of the system and the average and maximal energies of the different species of particles, as well as spectra such as the charge states of the ions or their energy distribution $n(\mathcal{E})$.
\subsection*{The interplay between photoionization and Auger decay, qualitative picture}
\begin{figure} [h!t]
        \centering 		
         \includegraphics[width=\textwidth]{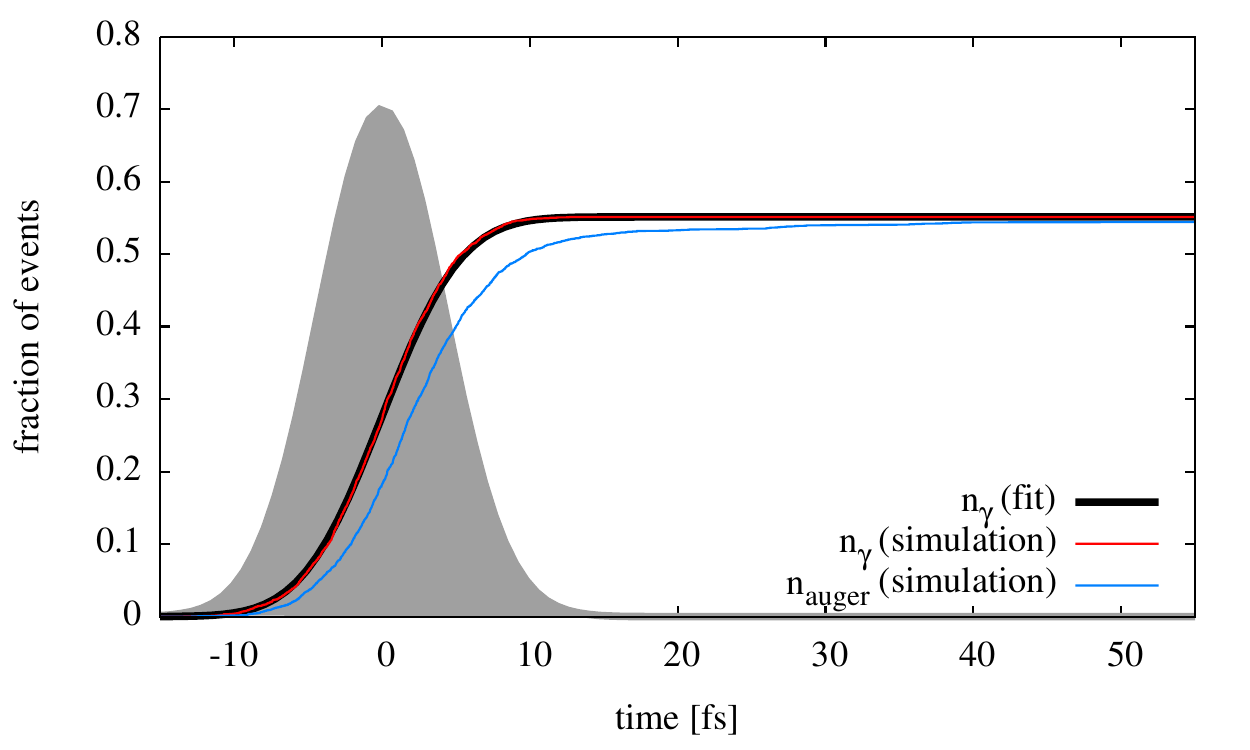}
         \caption{Number of absorbed photons (and released photoelectrons) as function of time (thin red line) fitted with Eq. (\ref{errorf}) (heavy black line) and number of Auger electron produced by the decay of the inner shell vacancies (blue curve) in a $(\ce{CH4})_{297}$ cluster irradiated by a pulse with $I_0=10^{18} {\rm W/cm^2}$, $T=10$ fs and photon energy of 1 keV so that $c_*{I_0\sigma}/{\hbar\omega}\sim0.22$. The gray shaded area represents the time envelope of the pulse.}
\label{photele}
\end{figure}

\begin{figure} [h!t]
        \centering 		
         \includegraphics[width=0.6\textwidth]{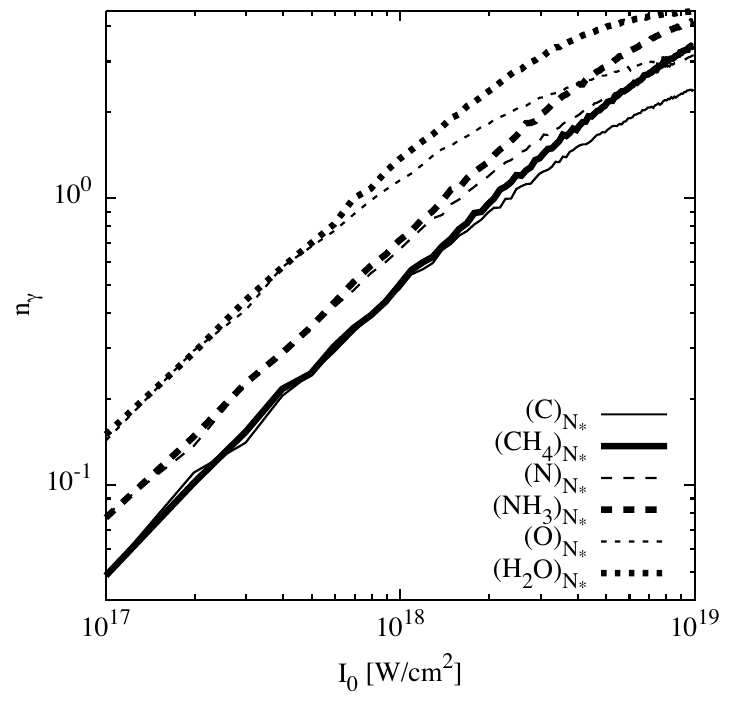}
         \caption{Number of absorbed photons per unit $n_{\gamma}$ for molecular systems (heavy lines) and parent atomic systems (thin lines) for different laser peak intensities. For values of $I_0$ smaller than roughly $8\times 10^{17} {\rm W/cm^2}$, no difference between the two cases can be noticed. Each point of the curves is averaged over 30 statistical realizations with identical initial conditions.}
\label{figabs}
\end{figure}

Photoionization and consequent Auger decay produce two families of electrons. Their initial energies may in principle have a complicated spectrum since different shells can be ionized and many channels of Auger decay are possible with different energies. However, photons of a few keV energy ionize mainly the $K$ shell (1s orbitals) for the elements of the first row
under consideration. In our test simulations for \ce{CH4} clusters, less
than 3\% of the photoelectrons come from the valence shell
at peak intensities $I_0>10^{19} {\rm W/cm^2}$, making reasonable to neglect its contribution. Hence, for the sake of
simplicity, all results presented have been obtained with
photoionizing exclusively $K-$shell electrons, this means that photoelectrons can be produced with two possible energies only, one relative to the case of the full $K$ shell and the other for the case of the one fold ionized $K$ shell.\\
\indent On the other hand, the typical Auger life-times of $K$ shell vacancies, see Tab. \ref{parameters}, are always shorter than the pulse length assumed in this study, so that in principle autoionization may refill the 
inner shell within the time the laser pulse is effective, in a way that the two processes are strictly entwined.\\
\indent Assuming for the pulse a Gaussian time envelope and constant space profile, means that in principle the number of absorbed photons per atom $n_{\gamma}$ theoretically scales with time with an error function
\begin{equation}\label{errorf}
n_{\gamma}(t)=c_*\frac{I_0\sigma}{\hbar\omega}\sqrt{\frac{\pi}{2}}\left[{\rm erf}\left(\frac{t}{\sqrt{2}\sigma}\right)+1\right],
\end{equation}
where $\sigma=T/2\sqrt{2\ln 2}$ is the Gaussian's standard deviation and $c_*$ is to be interpreted as a constant containing the information relative to the number of atoms (or molecules) in the cluster $N_*$, the photoabsorption cross section and the 
Auger lifetime $\tau$. Note that, this is valid only for combinations of cluster sizes and intensity such that no saturation takes place (i.e. the total number of absorbed photons is smaller than $N*$). Equation (\ref{errorf}) can be used in principle for a one parameter ($c_*$) fit of the correspondent numerical curve as it is shown in Fig. \ref{photele}
where the quantity $n_{\gamma}$ extracted from the realization average of 20 runs with identical initial conditions is perfectly fitted by its theoretical counterpart.\\
\indent It is evident in this specific case, how Auger decay sets in early within the pulse, so that at $t=0$, the time when the peak in intensity is attained, half of the molecules 
that have been photoionized also produced an Auger electron. For the species of interest, once released such electrons have typical kinetic energies $K_{\rm auger}$ corresponding to the $\Delta E$ between unstable and decayed configurations, of the order of 250  eV, allowing them to escape the cluster if
\begin{equation}\label{trapping}
K_{\rm auger}>\frac{eQ_t}{R_t},
\end{equation}
where $Q_t$ is the total charge of the ionized cluster when its radius increased to the value $R_t$.\\
\indent The $K-$shell photoabsorption cross sections $\sigma_{\rm 1s}$ differ little between the atomic and molecular species (here we have assumed them to be identical). Therefore, the different Auger lifetimes of the molecular and atomic species marks the most important difference between molecular and atomic clusters exposed to the same X-ray pulses, since this implies that the inner shells are ``refilled" on different time scales in the two cases. 
However, for given photon energy and pulse length, the effect becomes appreciable only at large intensities $I_0$, as it is clearly evident from Fig. \ref{figabs}.\\
\indent At fixed pulse length, the higher the laser intensity is, massive photoionization in the cluster and subsequent decay of the $K-$shell holes, cooperate in building in both atomic and molecular systems, a deeper cluster potential. Having assumed a constant particle density on every shell, and since the ionization occurs isotropically (photoionization cross sections $\sigma_{\rm 1s}$ of the elements here considered (cfr. Tab. \ref{parameters})), we can idealize it being harmonic for $r<R_t$ and Coulombian for $r>R_t$.
The cluster electrostatic field induces the so called field ionization, whenever it happens to be strong enough to strip the ions of their external electrons. The effect is in general more effective on the ions sitting at the cluster's surface, where the electric field is stronger.\\
\indent Electrons produced this way are typically confined by the space charge of the ions and have kinetic energies lower than $eQ/R$.
\begin{figure} [h!t]
        \centering 		
         \includegraphics[width=0.75\textwidth]{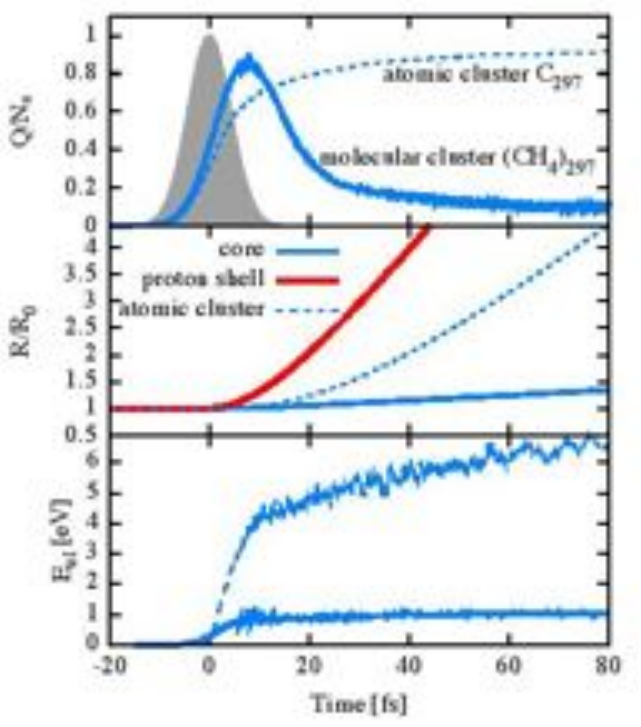}
         \caption{Top: time dependent total charge $Q$ per unit enclosed by the surface of the ion core of a $(\ce{CH4})_N$ cluster (solid line) and total charge per unit enclosed by the surface of an atomic carbon cluster (dashed line) both irradiated by a gaussian pulse with $I_0=10^{18} {\rm W/cm^2}$, $T=10 {\rm fs}$ and photon energy of 1 keV. The gray shaded area shows the time envelope of the pulse. Middle: radii of the ion core (blue solid line) and proton shell (red solid line) and radius of the pure carbon cluster in units of the initial cluster radius $R_0$. Bottom: average kinetic energy of the trapped electrons in the cluster in the atomic and molecular case.}
\label{radialexpl}
\end{figure} 
\subsection*{Molecular cluster versus pristine atomic clusters}
Figure \ref{radialexpl} illustrates the time evolution of characteristic parameters total charge per unit $Q/N_*$, radius $R$ and ``pseudo temperature" of the confined electron nanoplasma (i.e. the average electron kinetic energy $E_{\rm el}$), 
for an atomic cluster $(\ce{C})_{297}$ and a molecular cluster $(\ce{CH4})_{297}$ under the influence of the laser pulse (gray shaded area). 
It is immediately clear that the dynamics of the pristine carbon cluster (dashed line) and the methane cluster (solid line) is completely different.\\ 
\indent We observe that, as one would expect, the carbon atoms get successively charged through photoionization leading to more than 90\% carbon ions (top panel). 
The cluster ions create a deep binding potential from which most Auger electrons can not escape, forming a nanoplasma together with the field ionized electrons.\\
\indent The maximum kinetic energy of the trapped electrons is limited by the depth of the cluster potential and the average kinetic energy (bottom panel, dashed line) 
\begin{figure} [h!t]
        \centering 		
         \includegraphics[width=\textwidth]{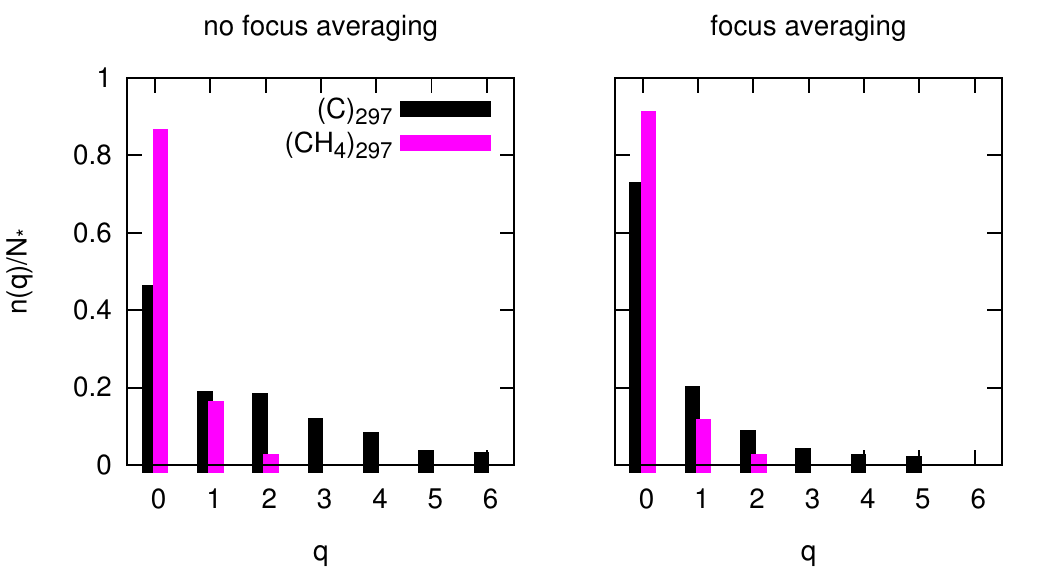}
         \caption{Left panel: charge states at $t=500$ fs for the same systems and laser parameters of Fig. \ref{radialexpl} normalized respect to the total number of heavy ions in the cluster $N_*$. Right panel: same quantity as left but spatially averaged over focus.}
\label{cstate}
\end{figure} 
is indicative of the nanoplasma temperature\footnote{It shall be pointed out that technically speaking, one can not 
in principle speak of electron ``temperature'', as the time scales at which we are looking at the systems might be considerably shorter respect to the typical time scales at which the electron plasma attains a 
Maxwellian energy distribution due to Coulomb collisions \cite{georgescu2008}, for our combinations of density, total cluster charge and heavy ions mass. In addition, it is important to mention that confined Coulomb systems and in general systems of particles interacting with long range forces undergo 
other processes of energy relaxation due to collective oscillations and other non collisional phenomena, see e.g. \cite{2003LNP...626..154K}, \cite{teles} and \cite{pfdc1}}. 
This is the normal behavior as is well known from rare-gas clusters exposed to X-ray pulses, see e.g. Refs. \cite{2006JPhB...39R..39S} and \cite{GNO11}.\\
\indent The molecular cluster, however, does not follow this scheme: While initially similarly charged as in the pristine
cluster, the total charge $Q(R)$ enclosed by the radius of the core experiences a dramatic drop at around 9 fs, so that this region is in the end almost neutral on
average (Fig. \ref{radialexpl} middle panel, solid line). At the same time, the kinetic
energy of the trapped electrons and, hence, the temperature
of the nanoplasma remains comparatively low (Fig. \ref{radialexpl}, bottom panel,
solid line).\\
\indent Both phenomena originate in the ejection of fast protons from the molecular cluster (see upper (red) line in middle panel of 
Fig. \ref{radialexpl}).
Although the carbon $K-$shells are initially photoionized, the charge distribution of the doubly charged methane after
Auger decay is such that the carbon ion is screened and the positive charge is dominantly localized on two hydrogen atoms which are likely to be ejected from the entire cluster as protons. 
These protons take away the excess positive charge created by photoionization which is, of course, not possible in the 
\begin{figure} [h!t]
        \centering 		
         \includegraphics[width=0.7\textwidth]{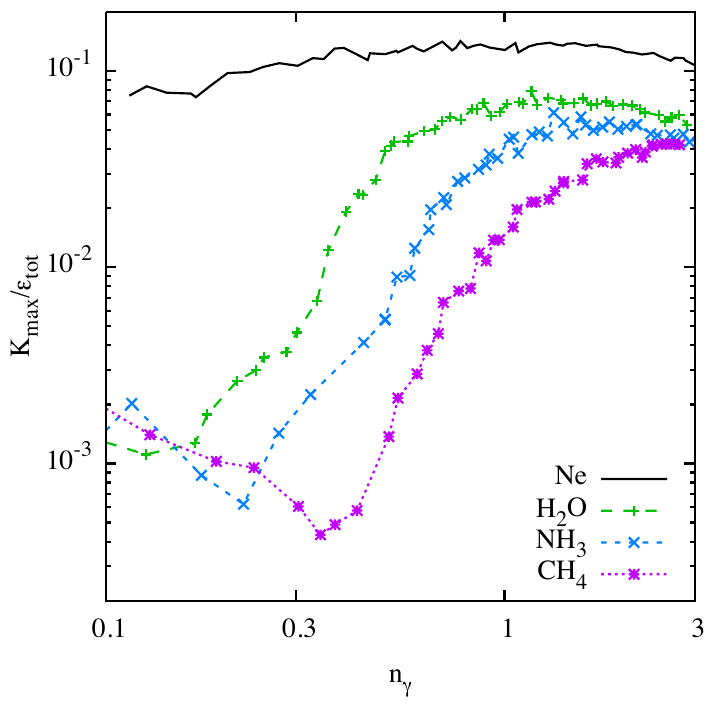}
         \caption{Kinetic energy of the fastest ion $K_{\rm max}$, 0.5 ps after the peak of the pulse ($T=10$ fs), versus the average
number of photons absorbed per atom/molecule $n_\gamma$. Cluster size is $N_*=689$. The solid line marks the ``atomic limit" of the neon cluster. Note that the kinetic energies are normalized in units of the total energy absorbed by the cluster $\mathcal{E}_{\rm tot}$.}
\label{maxe}
\end{figure} 
pristine carbon cluster. The remaining positive charge in the cluster is small giving rise to a weak potential which can only trap low-energy electrons. Therefore, the temperature of the nanoplasma is by a factor of 6 smaller than for the pure carbon cluster after $t\sim65$ fs; 
this holds true also for the Coulomb explosion of the carbon ions with the charging of the carbon ions even more dramatically reduced, almost by an
order of magnitude. In Figure \ref{cstate} the charge states spectrum defined as
\begin{equation}
n(q)=\sum_{i=1}^{N_*}\delta_{qq_i},
\end{equation}
are presented for the methane and pure carbon clusters at 0.5 ps after the peak of the pulse. In the left panel the spectra are averaged over 50 realizations assuming constant space envelope, while in the right panel a Lorentzian space envelope is assumed at the pulse's focal spot (see Eq. \ref{lorenzo}), and the spectra are averaged over 200 realizations distributed in space with normal distribution. 
In both cases, for the molecular cluster almost 85\% of the carbon emerges as neutrals and the rest is singly charged. In the spectrum of the pure atomic system, up to six fold 
\begin{figure} [h!t]
        \centering 		
         \includegraphics[width=0.7\textwidth]{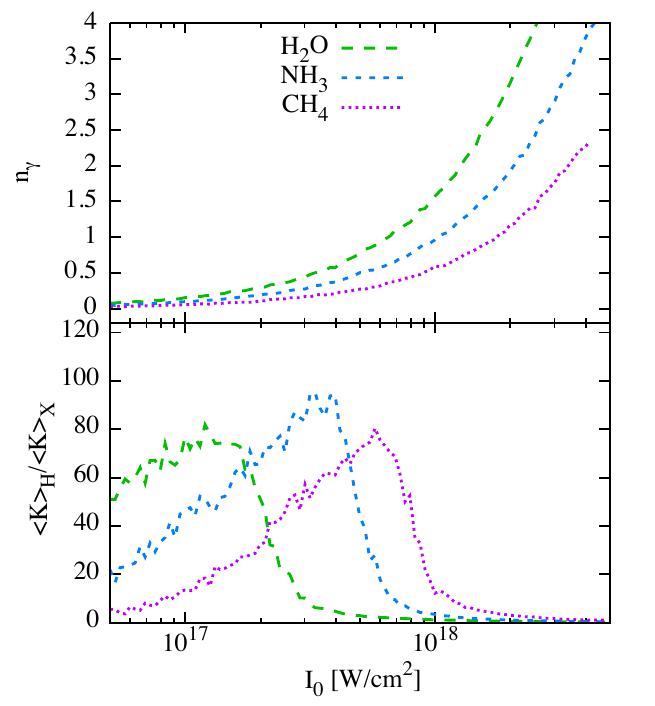}
         \caption{Top panel: fraction of absorbed photons per molecule $n_\gamma$ a function of the x-ray intensity $I_0$ for the same systems shown in Fig. \ref{maxe}. Bottom panel: Ratio of the average kinetic energy of
protons and heavy atoms.}
\label{peake}
\end{figure} 
charged ions are present when the space envelope is assumed constant, while only up to five fold charged ions 
are observed in the other case, but still it appears to be qualitatively different than the spectrum of the molecular cluster.\\
\indent The radical difference between the final charge states of carbon ions in molecular and atomic clusters is not imputable to a different photoabsorption (see Fig. \ref{figabs}), but instead to a more efficient 
recombination\footnote{A time scale of a few hundreds of femtosecond is not sufficient to equilibrate electrons and ions, the latter are characterized by a non-thermal energy distribution. 
The recombination has to be intended here as electrons being {\it classically} bound to the neighbouring ion.} of the nanoplasma electrons for the molecular system due to the presence in the molecular cluster of a larger 
number of plasma electrons per ion with lower average kinetic energy.\\
\indent As a general remark, for the other cases of \ce{NH3} and \ce{H2O} the effect is qualitatively similar to what we have discussed for the methane clusters; as we will see in the following, 
the difference in behavior of the atomic and molecular clusters becomes more marked for intermediate intensities. 
\subsection*{Intensity and pulse length dependence of the ion dynamics}
In the limit of short pulses and extreme intensities the dynamics of the ions and protons is obviously expected to be qualitatively similar to the idealized case of multi-component Coulomb explosion treated in Sect. \ref{multisystem}. For a given pulse length $T$, one may expect that the absolute difference in velocity of heavy and light ions gets larger with increasing intensity and, therefore, higher charging of the cluster. Figure \ref{maxe}, however, reveals that the ratio of the kinetic energy for the fastest heavy ion $K_{max}$, in units of the total energy absorbed by the cluster
\begin{equation}
\mathcal{E}_{\rm tot}=\hbar\omega N_*n_\gamma,
\end{equation}
exhibits a non monotonic behavior as a function of the fraction of photons absorbed per atom/molecule $n_\gamma$ with a dip at a certain critical number.\\
\begin{figure} [h!t]
        \centering 		
         \includegraphics[width=0.95\textwidth]{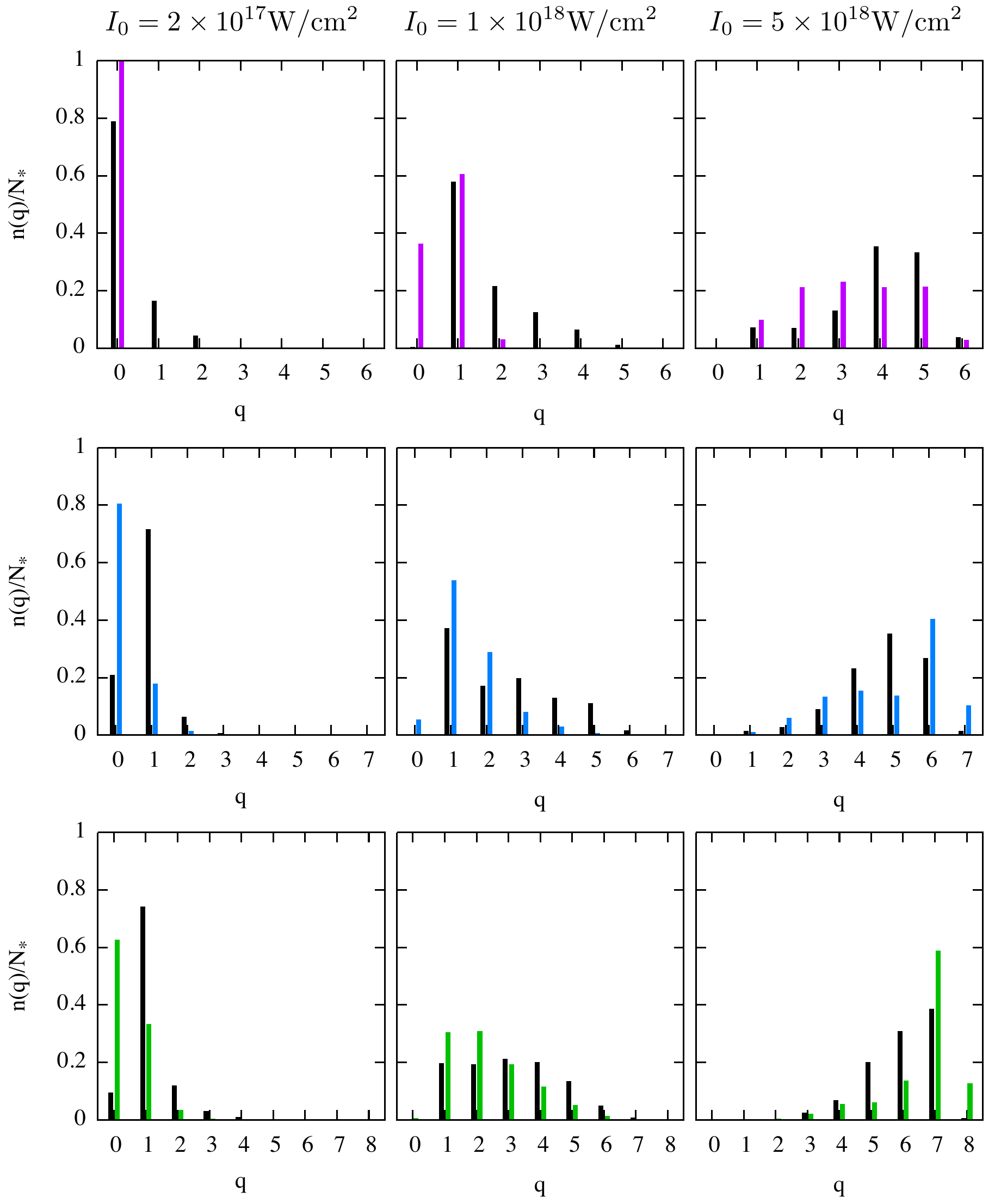}
         \caption{Normalized charge-state distribution $n(q)/N_*$ for carbon ions $\ce{C}^{q+}$ from $(\ce{CH4})_{689}$ (top row), for nitrogen ions from $(\ce{NH3})_{689}$ (middle row) and oxygen ions from $(\ce{H2O})_{689}$ (bottom row). The black bars mark the case of the corresponding pristine atomic systems, and the x-ray intensities $I_0$ are indicated on top.}
\label{cstate2}
\end{figure} 
\indent The latter depends moderately on the species considered, but is otherwise a universal feature
of hydride clusters in obvious contrast to the isoelectronic atomic neon cluster (black solid line in Fig \ref{maxe}). Such feature in the curves for the hydride clusters is a signature of the dynamical segregation of protons and heavy ions as 
can be seen in bottom panel of Fig. \ref{peake}, where a more global quantity,
namely, the ratio of the average energy of all protons versus that of all heavy ions
\begin{equation}
\mathcal{R}=\frac{\langle K\rangle_\ce{H}}{\langle K\rangle_\ce{X}},\quad \ce{X}=\ce{O};\ce{N};\ce{C}
\end{equation}
is shown as a function of peak laser intensity $I_0$. Again, one sees a qualitatively similar non monotonic behavior of all three hydrides, even though the photo-absorption goes monotonically in all three cases (top panel same  figure). There is a maximal segregation at a well-defined intensity for each of the three hydrides since they differ in their respective ionization energy for the 1s electrons, Auger rates, and photoionization cross sections, as  listed in Tab. \ref{parameters}. Hereby, the shift of the peak positions is in principle due to the different
photoionization cross sections.\\
\indent The final ion charge states $n(q)$ taken at $t=500$ ps are shown in Fig. \ref{cstate2} for the three hydrides cluster as well as their atomic counterparts, for $N_*=689$ and $I_0=2\times10^{17},1\times10^{18}$ and $5\times10^{18} {\rm W/cm^2}$. 
As expected, a much lower charging of heavy ions as compared to the pristine cluster of the heavy-atom species at the lowest and intermediate intensities.\\
\indent At $I_0=2\times10^{17}$, most heavy atoms remain neutral or singly charged in the pristine as well as in the hydride cluster. This changes drastically for intermediate intensities of about $10^{18}{\rm W/cm^2}$, 
where the fraction of neutral atoms surviving the light pulse illumination is small to vanishing in the pristine cluster. In the
hydride cluster (\ce{CH4} and \ce{NH3} cases), on the other hand, about 80\% neutral heavy atoms result from recombination with the cold electrons after proton segregation in the surface layer, which has been fully 
charged due to efficient field ionization. The water cluster has an almost negligible fraction of neutrals ($\sim 2\%$), due to its higher photoionization cross section and shorter auger lifetime. Nevertheless the spectrum is shifted towards lower charge states than that of pure oxygen cluster..\\
\indent For higher intensities, we expect the proton segregation to cease (as seen in Fig. \ref{maxe}) and, as a consequence, similar charge spectra for the pristine and the hydride cluster. 
This is indeed true with respect to a vanishing yield of neutral atoms. However, the form of the charge distribution is still somewhat different. 
\begin{figure} [h!t]
        \centering 		
         \includegraphics[width=0.95\textwidth]{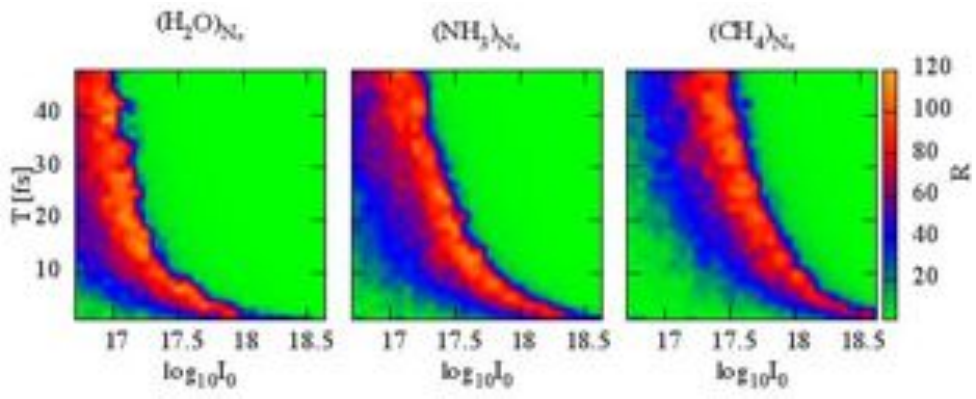}
         \caption{From left to right for $(\ce{H2O})_{N_*}$, $(\ce{NH3})_{N_*}$ and $(\ce{CH4})_{N_*}$, the ratio $\mathcal{R}$ of the average kinetic energy of
protons and heavy atoms is shown as function of the laser's peak intensity $I_0$ and length $T$ is shown. In this case $N_*=689$.}
\label{timedep}
\end{figure} 
In particular, \ce{NH3} and \ce{H2O} cluster present even larger percentages of higher charge states than their parent atomic clusters, 
in contrast to what observed for \ce{CH4} and \ce{C} clusters. The latter is due to the fact that at $I_0=5\times10^{18}{\rm W/cm^2}$ for these species, the number of absorbed photons in the atomic and molecular case is considerably different (cfr Fig. \ref{figabs}), and in addition less protons are ``available'' to carry away the excess positive charge.\\    
\indent Figure \ref{timedep} shows that the dynamical segregation of protons and heavy ions does not happen only for the pulse length considered here, but takes place for a broad interval of pulse lengths. 
The peak in the ratio $\mathcal{R}$ broadens and shifts towards lower $I_0$ for increasing $T$, this is a signature of the fact that the emergence of such segregation effect is principally due to the amount of charge created in the system. 
\section{A synthetic model}
From the analysis carried out to this point it is clear that the proton segregation is strongly influenced by the laser's intensity for fixed pulse length, as it is revealed by the total charge yields and particle-averaged kinetic energies. However, it is still needed to clarify whether this segregation is a local effect due to the heavy-light
character of the hydride molecules or whether is a consequence of the cluster nature of the entire system. In order to clarify this, we have set up a simple model where $N=10^4$ singly charged ions are distributed homogeneously in a sphere of initial radius $R_0$, neither supporting any molecular substructure nor allowing any intra-atomic or intramolecular
electronic processes. Three fourths of the ions have the mass of the proton, while the other one fourth has a 20 times
higher mass (roughly the mass of Neon). $N-Q$ electrons are placed at randomly selected ions. With this initial configuration, ions and electrons, interacting via Coulomb 
\begin{figure} [h!t]
        \centering 		
         \includegraphics[width=0.9\textwidth]{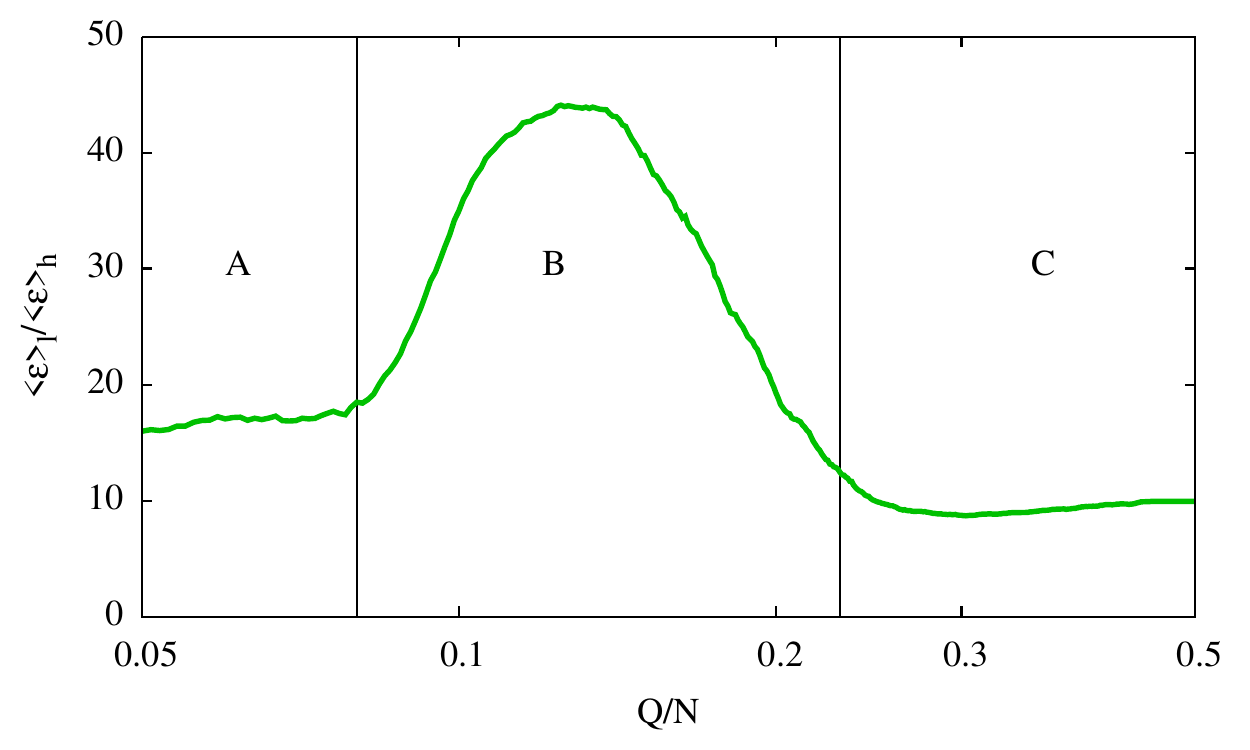}
         \caption{Ratio of the average energies of  light $\langle\mathcal{E}\rangle_l$ and heavy $\langle\mathcal{E}\rangle_h$ ions with scenarios A, B and C discussed in the text.}
\label{ratioionssimple}
\end{figure} 
forces regularized at short distance, are propagated up to 1 picosecond.\\
\indent Because of its net positive charge, the system undergoes expansion. The ratio of the average final energies of light and heavy ions, exhibits one
central maximum see Fig. \ref{ratioionssimple} (region B, $0.08<Q/N<0.25$) similarly, as for
the fully microscopic calculations for hydride clusters discussed in the previous Sections. In Figure \ref{eionssimple} we show the final energy of the light and heavy ions as a function of their initial radial position $r/R_0$. It is clear how the dependence of the final energy of an ion on its initial radial position in the cluster  influences its final energy and induces the heavy-light
\begin{figure} [h!t]
        \centering 		
         \includegraphics[width=0.77\textwidth]{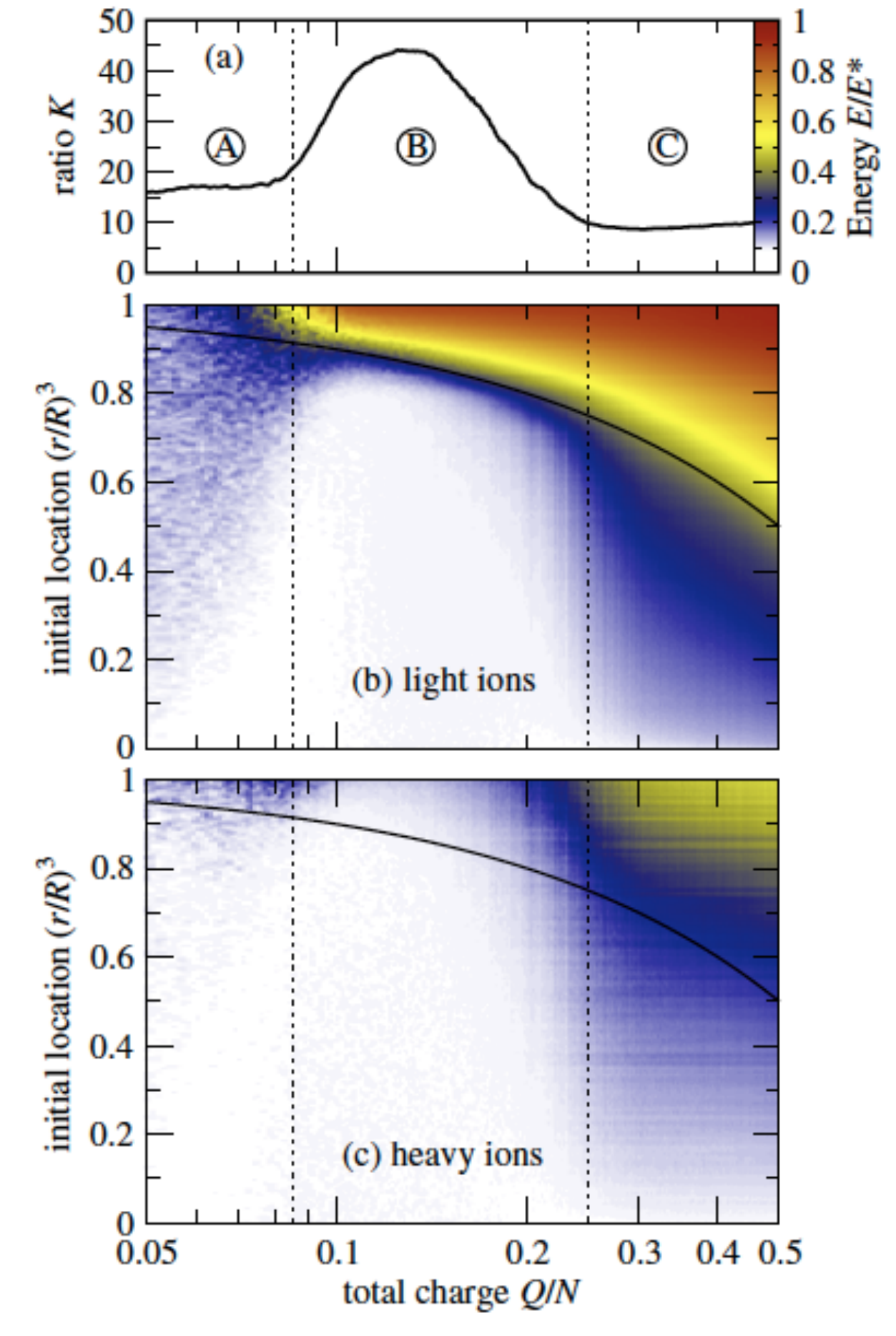}
         \caption{Normalized ion kinetic energies $\mathcal{E}_*$ of a heterogeneous
cluster model with $N=10^4$ particles in a sphere of initial radius $R_0$ as a
function of the total charge $Q=N$ of the system and their initial position in units of $R_0$. The energies of the heavy ions (bottom panel) are normalized to $Q/R_0$ while those of the light ions (middle panel) to $Q/4R_0$. The thin green line marks the part of the cluster which is supposed to explode if the charge would concentrate in a shell at the surface. The top panel reproduces the ratio $R$ also shown in Fig. \ref{ratioionssimple}. Figure adapted from Ref. \cite{dicintio2013}}
\label{eionssimple}
\end{figure} 
ion segregation.\\
\indent In region A ($Q/N\leq0.08$), as well as in region C $(Q/N\geq0.25)$, protons and heavy ions originate from all initial positions
in the cluster with an increasing energy towards the surface. The mean kinetic energy is larger in C than in A due to the stronger charging, but heavy-light segregation does not take place in either of the two regions. In region B, however, the charging $Q$ is sufficiently strong to trigger field ionization of surface ions, as it has been discovered for homogeneous clusters \cite{GNO09}. As a consequence, one expects the cluster core to be screened by the field ionized
electrons up to the radius indicated by the green line in left and right panels of Fig. \ref{eionssimple}.\\
\indent However, the protons in the hydride cluster core are light enough (or more precise: have a sufficiently large charge-to-mass ratio) to have started moving before the repelling forces are compensated by the screening electrons. Hence, protons leave the cluster core and, as a result, the surplus of screening electrons prevents heavy ions in the surface layer even beyond the screening radius of a homogeneous cluster from exploding. In contrast, protons escape with high final energy from the surface layer.\\
\indent In region C, the initial charge $Q$ further increases which weakens the screening 
effect through field ionized electrons for two reasons. First, the fraction of screening electrons available versus initial charge $Q$ decreases. Second, the temperature of the screening electrons is higher than in B due to the deepening of the trapping potential with increasing $Q$. Hence, the surface layer no longer forms efficiently and a Coulomb explosion, as in region A, although more violently, results.\\
\indent We may conclude that heavy-light segregation is not a local effect of heavy-light molecules. Rather, it happens in a surface layer of the heterogeneous cluster triggered by field ionization.\\
In contrast, ``dynamical acceleration" in such clusters see e.g.
\cite{2005PhRvL..95s5003H} and \cite{2001PhRvL..87c3401L}, does not require electron screening and relies exclusively on ion-ion repulsion throughout the cluster.
\section{Summary}
We now summarize this chapter, whose results are also published in Ref. \cite{dicintio2013}. First of all, it must be pointed out that the behavior of molecular hydride clusters is qualitatively very different from pristine atomic clusters as the first have an additional channel of energy loss, namely the light protons, to release the energy absorbed from the laser pulse.\\
\indent As a general qualitative fact, there is a higher percentage of neutrals and low charge states in the final $n(q)$ of the heavy ions in the molecular case than in the pure atomic case.\\
The dynamics of two species of ions is strongly dependent on the laser intensity $I_0$. Three regimes can be distinguished where, the weakly charged cluster expands as a whole as a quasi-neutral plasma (hydrodynamic expansion), the cluster is charged enough do induce significant field ionization at the surface, electrons collapse to the core neutralizing it and a charged shell of protons explodes faster carrying away most of the energy absorbed by the system, and finally the cluster almost is entirely charged and behaves as in the limit of multi-species pure Coulomb explosion.\\
\indent The intensity dependent heavy ion-proton segregation in the intermediate regime is found to be a universal feature of hydride clusters and can be classified as a cluster effect.\\
\indent In conclusion, we speculate that, due to the experimentally relevant intensity window where such segregation effect happens, the results discussed here could turn useful for the study of radiation damage of large bio-molecules as it may occur during coherent diffractive imaging with intense X-ray pulses, see Refs. \cite{neuze}, \cite{2002Natur.420..482W}, and \cite{HAU04,2006NatPh...2..839C}.
\chapter{Numerical methods}\label{numerica}
In this chapter we present the different numerical methods used in this thesis to study the dynamics of finite size plasmas generated by ionization of cluster targets. In addition we discuss the techniques based on Monte Carlo sampling and rate equations, used to treat photoionization, and  the Auger decay and recombination processes.\\
\indent It must be also pointed out, that many of the concepts concerning numerical simulations using particles discussed here, also apply in the field of gravitational $N-$body simulations, as the Coulomb electrostatic force and Newtonian gravity share the same $1/r^2$ nature.\\
\indent In particular, the codes used for the simulations discussed in this work, and developed by the Author during its Ph.D., are presented in detail with the results of some of their performance tests.
\section{$N-$body methods for systems of charged particles}\label{metodinbody}
As previously introduced in Chapter \ref{chapmultice}, in the simple electrostatic picture where the contribution of self-induced magnetic fields and radiative losses are neglected and the velocities are always non relativistic, a plasma can be thought as an ensemble on $N$ particles of charge $q_i$, mass $m_i$ with positions and velocities $\mathbf{x}_i$ and $\mathbf{v}_i$, interacting via Coulomb forces and it fully described by the Hamiltonian
\begin{equation}\label{nbodyham}
H=\sum_{i=1}^N\left( \frac{\mathbf{p}_i^2}{2m_i}+\frac{1}{2}\sum_{j\neq i=1}^{N}\frac{q_iq_j}{||\mathbf{r}_j-\mathbf{r}_i||}\right)
\end{equation}
where $\mathbf{p}_i=m_i\mathbf{v}_i$. Given the initial conditions $\mathbf{R}_0=(\mathbf{r}_1,\mathbf{r}_2,...,\mathbf{r}_N)_{t=0}$ and $\mathbf{V}_0=(\mathbf{v}_1,\mathbf{v}_2,...,\mathbf{v}_N)_{t=0}$, the configuration of the system at time $t$ is formally obtained by integrating the system of $6N$ coupled ODEs (Hamilton equations)
\begin{eqnarray}\label{hamjaceq}
 \frac{{\rm d}\mathbf{r}_i}{{\rm d} t}=\frac{\partial H}{\partial\mathbf{p}_i};\quad \frac{{\rm d}\mathbf{p}_i}{{\rm d} t}=-\frac{\partial H}{\partial\mathbf{r}_i};\quad i=1,N.
\end{eqnarray}
In the limit of large $N$ and vanishing inter-particle correlations, the evolution of the system is that of its one-particle phase space distribution defined on the 6-dimensional one particle phase space $f(\mathbf{r},\mathbf{v},t)$, through the Collisionless Boltzmann Equation (CBE), or Vlasov Equation in the jargon of Plasma Physics, see Appendix \ref{kt},
\begin{equation}\label{vlasov2}
\frac{{\rm D}f}{{\rm D}t}\equiv\frac{\partial f}{\partial t} + \mathbf{v} \cdot \frac{\partial f}{\partial\mathbf{r}}+\nabla\Phi\cdot\frac{\partial f}{\partial\mathbf{v}}=0.
\end{equation}
The electrostatic potential $\Phi(\mathbf{r})$ is related to $f$ by the Poisson equation
\begin{equation}
\Delta\Phi(\mathbf{r})=4\pi\rho(\mathbf{r})=4\pi q\int_{\Omega}f(\mathbf{r},\mathbf{v}){\rm d}^3\mathbf{v},
\end{equation}
where $\Omega$ is the region of the phase space occupied by the system and $q$ is the unit charge\footnote{Note that this approach can be obviously extended to multi-species plasmas where each species of particle of mass $m_i$ and charge $q_i$ has its own $f_i$ and $\Phi$ is computed by solving the Poisson equation due to all the species.}.\\
\indent  In this picture what one has in principle to solve is a system of two PDEs (Vlasov and Poisson equations) and an integral equation for the density called Vlasov-Poisson system. When the effect of (external and self-induced) electromagnetic fields and the coupling of matter with radiation are taken into account, the term $\nabla\Phi\cdot{\partial f}/{\partial\mathbf{v}}$ in Eq. (\ref{vlasov2}) is substituted by the full Lorenz force per unit of mass $\mathbf{F}=q/m(\mathbf{E}+\mathbf{v}/c\times\mathbf{B})$ and 3 of the 4 Maxwell's equations have to be added to the Vlasov equation. In this case one speaks of a Vlasov-Maxwell system.\\
\indent Numerical models based on Equation (\ref{vlasov2}) deal essentially with the discretization of the DF
\begin{equation}\label{vlasovdiscr}
f_*(\mathbf{r},\mathbf{v},t)=\sum_{i=1}^n \delta\left[\mathbf{r}-\mathbf{r}_i(t)\right]\delta\left[\mathbf{v}-\mathbf{v}_i(t)\right]
\end{equation}
where $n$ is the effective number of {\it macroparticles} used in the simulation that is some orders of magnitude smaller than $N$, and the electrostatic potential at $\mathbf{r}_i$ is given by
\begin{equation}\label{potdiscr}
\Phi(\mathbf{r}_i,t)=\int g(\mathbf{r}_i-{\mathbf r}){\rm d}^3{\mathbf r} \int f_*({\mathbf r},\mathbf{v},t) {\rm d}^3{\mathbf v}=\sum_{j\neq i}^n q_j g(\mathbf{r}_i-\mathbf{r}_j),
\end{equation}
where now $q_j$ is the charge of the $j-$the macroparticle and $g(\mathbf{r}_i-\mathbf{r}_j)=1/|\mathbf{r}_i-\mathbf{r}_j|$ is the Green's function of the Laplacian operator $\Delta$, and $|...|$ is the standard Euclidean norm. Given $\Phi(\mathbf{r})$ and the acceleration as $\mathbf{a}(\mathbf{r})=\nabla\Phi(\mathbf{r})$, what a numerical code does, is essentially to propagate in time with finite difference integration methods (see e.g. Ref. \cite{nr3} for a complete description) the $n$ particles, by solving their equations of motion, that is substantially returning to the picture of Eqs. (\ref{hamjaceq}).\\
\indent What characterizes the different plasma (as well as gravitational) $N-$body codes is the scheme that allows one to compute the electrostatic potential $\Phi$ (as well as the electric fields). The main families of codes based on particles are:
\subsubsection*{Direct $N-$body codes}
This category of codes (see e.g. Ref. \cite{aarseth94} for a complete review), also known as {\it particle-particle} codes (PP) uses the conceptually simplest approach, the potential $\Phi_i=\Phi(\mathbf{r}_i)$ and the acceleration $\mathbf{a}_i$ at the position of the $i-$th simulation particle are computed directly, summing over the contributions of all the other particles of the system as
\begin{eqnarray}
\Phi_i=q_i\sum_{j\neq i=1}^N\frac{q_j}{||\mathbf{r}_i-\mathbf{r}_j||},\nonumber\\
\mathbf{a}_i=\frac{q_i}{m_i}\sum_{j\neq i=1}^N\frac{q_j(\mathbf{r}_i-\mathbf{r}_j)}{||\mathbf{r}_i-\mathbf{r}_j||^3}.
\end{eqnarray}
Such operation involves for each step two nested cycles over the number of particles implying that the number of operations (numerical complexity) for the force calculation, and hence the computational time, scales with $N$ as $O(N^2)$. Because of that, codes based on direct force computation are not suitable to treat more than $\sim5\times10^4$ particles, even on modern processors,  since for larger systems the time taken by the force computation, that is the bottleneck of the scheme, becomes prohibitive even within a single time step $\Delta t$.\\
\indent  However, if one is interested in simulating systems of small size where actually the dynamical collision among particles play, an important role, the direct codes have in this case the advantage that the forces on the particles can be better estimated (depending only on the machine precision) than with the other methods discussed next.\\
\indent There is unfortunately an additional drawback, due to the singular nature of the Coulomb interaction for vanishing separation, the potential and the force computed for a couple of particles placed at a very small distance is large and therefore, the velocity change attained within a single time step can be spuriously high. In addition, due to the finiteness of the machine representable reals, the pair force may diverge even for no zero separation. As we will see in Sect. (\ref{sectsoft}), these complications are generally circumvented by modifying the Coulomb potential at small separation by the introduction of the softening length $\epsilon$, (see e.g. \cite{2001MNRAS.324..273D} and \cite{2011EPJP..126...55D}, see also \cite{1996AJ....111.2462M}), so that in Eq. (\ref{vlasovdiscr}) $q_i\delta\left[\mathbf{r}-\mathbf{r}_i(t)\right]$ is replaced by $\mathcal{D}\left[\mathbf{r}-\mathbf{r}_i(t)\right]$ where $\mathcal{D}$ is a distribution with scale length $\epsilon$ that smoothes $q$ on a finite or infinite support. Consequently, the softened potential that it generates satisfies $\Delta\Phi^{\rm soft}=4\pi\mathcal{D}$, and the softened acceleration is given as usual by $\mathbf{a}^{\rm soft}=-\nabla\Phi^{\rm soft}$.
\subsubsection*{Particle-Mesh codes (PM)} 
\begin{figure} [h!t]
        \centering 		
         \includegraphics[width=0.95\textwidth]{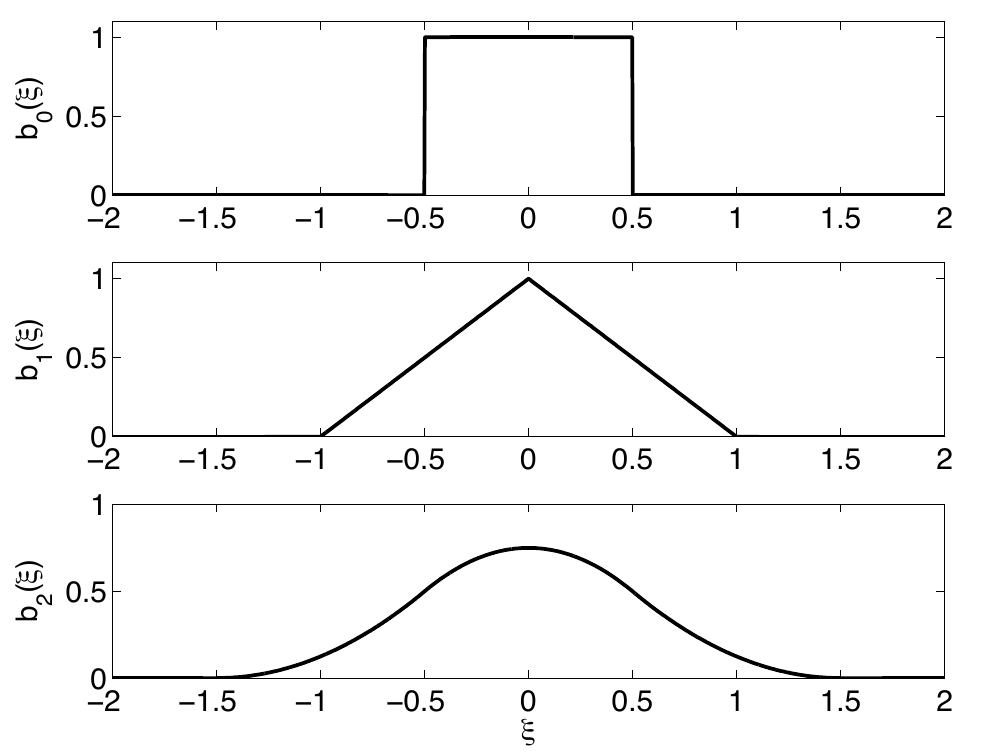}
         \caption{From top to bottom, first three spline functions $b_k(\xi)$.}
\label{schemespline}
\end{figure}

A second category of $N-$body schemes, sometimes also dubbed Particle in Cell (PIC), used when the number of particles in the simulation larger than $10^4$ and when collisional effect are negligible, that is extensively discussed in Ref. \cite{hockney}, is so-called particle-mesh. In this case, $\Phi$ is first computed using $\Delta\Phi=4\pi\rho$ on a 3-dimensional cartesian grid superimposed to the system where the density $\rho$ is displaced, and then the potential and its gradient are interpolated at the particles positions. To assign the density $\rho_{i,j,k}$ to a given mesh points $P_{i,j,k}$ first a shape function $S{(x,y,z)}$ is introduced so that the contribution $W^P{(x_l-x_i,y-y_j,z-z_k)}$ of the $l-$th particle to the mesh cell $P_{i,j,k}$ is
\begin{eqnarray}
 W^P{(x_l-x_i,y-y_j,z-z_k)}=\nonumber\\
\int_{x_i-\Delta x/2}^{x_i+\Delta x/2}\int_{y_j-\Delta y/2}^{y_j+\Delta y/2}\int_{z_k-\Delta z/2}^{z_k+\Delta z/2}S{(x_l-x_i,y-y_j,z-z_k)}{\rm d}x_l{\rm d}y_l{\rm d}z_l,
\end{eqnarray}
where $\Delta x, \Delta y$ and $\Delta z$ are the mesh spacings along the three axis. The density  $\rho_{i,j,k}$ then reads
\begin{equation}\label{cdpl}
\rho_{i,j,k}=\frac{1}{\Delta x\Delta y\Delta z}\sum_{l=1}^N q_l W^P{(x_l-x_i,y_l-y_j,z_l-z_k)}.
\end{equation}
The two main choices of $S$ that are generally used are the {\it nearest-grid-point} (NGP) method that assigns each particle's charge $q_l$ and coordinates $(x_l,y_l,z_l)$ to the nearest point of the grid with the shape function
\begin{eqnarray}\label{NGP}
S^{\rm NGP}{(x_l-x_i,y_l-y_j,z_l-z_k)}=\nonumber\\
=\frac{1}{\Delta x\Delta y\Delta z}\theta\left(\frac{1}{2}-\frac{|x_l-x_i|}{\Delta x}\right)\theta\left(\frac{1}{2}-\frac{|y_l-y_j|}{\Delta y}\right)\theta\left(\frac{1}{2}-\frac{|z_l-z_k|}{\Delta z}\right), 
\end{eqnarray}
and the {\it cloud-in-cell} (CIC) method that instead associates to the particle a finite size and a cubic shape with side of length $\Delta p$, and assigns to $P_{i,j,k}$ a fraction of charge corresponding to the portion of its volume overlapping with it. In this case
\begin{equation}\label{cic}
S{(x_l-x_i,y_l-y_j,z_l-z_k)}^{\rm CIC}=\frac{1}{(\Delta p)^3}b_0\left(\frac{x_l-x_i}{\Delta p}\right)b_0\left(\frac{y_l-y_j}{\Delta p}\right)b_0\left(\frac{z_l-z_k}{\Delta p}\right),
\end{equation}
where $b_0$ is the zeroth order $b$-spline function defined as
\begin{align}
b_0(\xi)=
\begin{cases}
1\quad {\rm for}\quad |\xi|<1/2\\
0\quad {\rm otherwise.}
\end{cases}
\end{align}
Higher orders $b_l$ are obtained recursively via
\begin{equation}
b_l(\xi)=\int_{-\infty}^{+\infty}b_0(\xi-\xi^\prime)b_{l-1}(\xi^\prime)\d\xi^\prime. 
\end{equation}
As an example, $b_0$, $b_1$ and $b_2$ are shown in Fig. \ref{schemespline}.  Using a higher order $b_k$ in (\ref{cic}), instead of $b_0$ refines the charge deposition algorithm. For $l=1$ one has the so called {\it triangular-shape-cloud} (TSC).\\
\indent Once the charge density is placed on the grid with one of the described methods, the main step (that constitutes actually the core of a PM-code) is represented by the solution of the Poisson equation to obtain $\Phi_{i,j,k}$ as a function of $\rho_{i,j,k}$, that 
has then to be derived and interpolated at the particle's positions. Two different methods exist to solve $\Delta\Phi=4\pi\rho$, the finite difference method and the solution in Fourier space. In the first, the Poisson equation is discretized and reorganized in the form
\begin{equation}\label{discretepoisson}
\Phi_{i+1,j,k}+\Phi_{i-1,j,k}+\Phi_{i,j+1,k}+\Phi_{i,j-1,k}+\Phi_{i,j,k+1}+\Phi_{i,j,k-1}-6\Phi_{i,j,k}=4\pi\rho_{i,j,k}(\Delta s)^2
\end{equation}
where $\Delta x=\Delta y =\Delta z =\Delta s$ (i.e. a cubic grid with equal mesh spacing in all directions is assumed); then by rearranging the elements of the resulting 3d arrays for $\Phi$ and $\rho$ on two 1d arrays, with the index relabeling $l=iN_g+jN_g+k$, where $N_g$ is the number of grid points in each direction, one is left with 
\begin{equation}\label{discretepoisson2}
\Phi_{l+2N_g+1}+\Phi_{l+N_g+1}+\Phi_{l-(2N_g+1)}+\Phi_{l-(N_g+1)}+\Phi_{l+1}+\Phi_{l-1}-6\Phi_{l}=4\pi\rho_{l}(\Delta s)^2
\end{equation}
that is solved, for instance with the relaxation technique (see e.g. \cite{nr3}), as a linear system of the form
\begin{equation}\label{sistem}
\mathcal{M}\cdot\mathbf{a}=\mathbf{b}
\end{equation}
where $\mathcal{M}$ is the tridiagonal matrix of the coefficients, vector $\mathbf{b}$ contains the ``known information'' (i.e. the density) and $\mathbf{a}$ the unknown potential.\\
\begin{figure} [h!t]
        \centering 		
         \includegraphics[width=\textwidth]{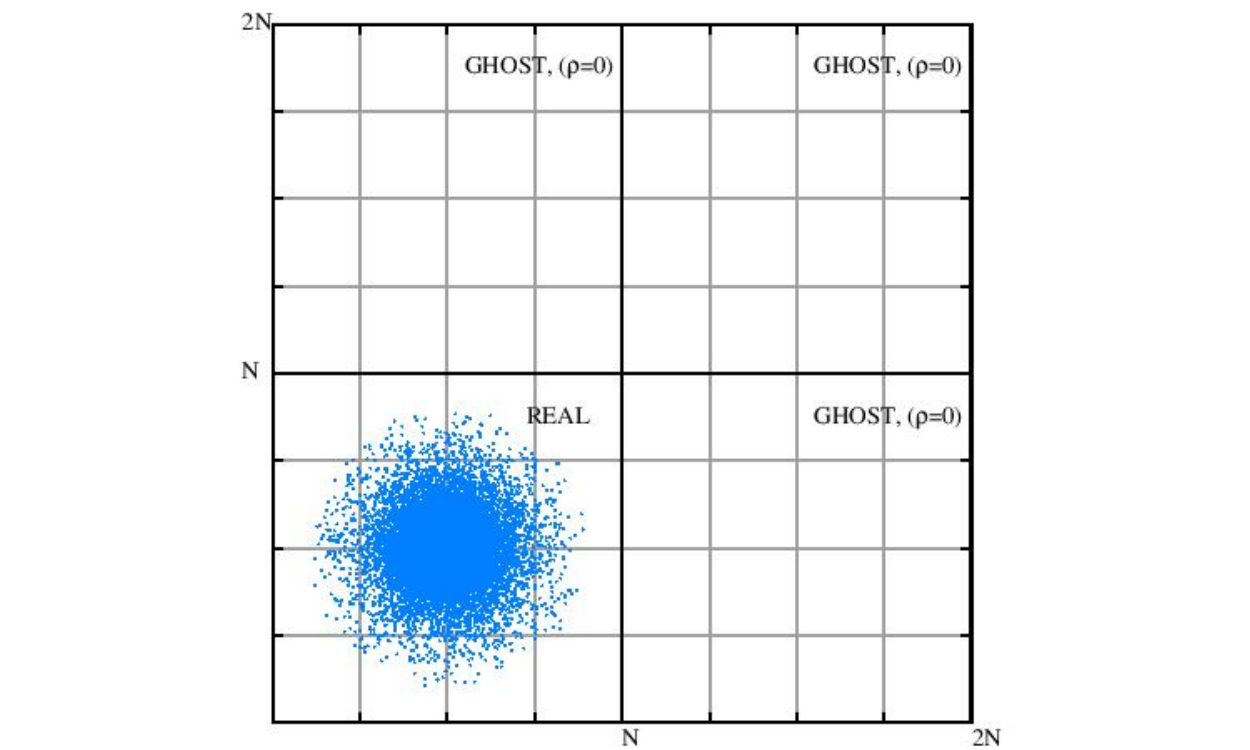}
         \caption{Sketch of the extended grid for the simulation of an isolated object. For simplicity only two dimensions have been represented.}
\label{schemegrid}
\end{figure} 
\indent The second method, is based on the fact that in Fourier space the Poisson equation reads 
\begin{equation}\label{poissonfourier}
\hat\Phi=\hat\rho\hat g
\end{equation}
where the hats denote the Fourier-transformed potential, density and Green function of the Laplace operator. The latter in real space is $1/r$, and it is defined on the $N_g\times N_g\times N_g$ mesh, as
\begin{equation}\label{grnfctn}
g_{i,j,k}=\frac{1}{\sqrt{\left[\Delta x(i-1)\right]^2+\left[\Delta y(j-1)\right]^2+\left[\Delta z(k-1)\right]^2}}.
\end{equation}
Note that for $i=j=k=1$, $g$ diverges. To overcome this inconvenient usually one takes 
\begin{equation}
g_{1,1,1}=\frac{1}{\sqrt{(\Delta x^2+\Delta y^2+\Delta z^2)}}
\end{equation}
or, alternatively 
\begin{equation}
g_{1,1,1}=\frac{2}{3\sqrt{(\Delta x^2+\Delta y^2+\Delta z^2)}}.
\end{equation} 
Such regularization introduce an effective {\it softening} of the Coulomb interaction on a length scale given by the size of the grid step.\\
\indent The simplest approach is again to take cubic grid with $\Delta s=\Delta x=\Delta y=\Delta z$, with this choice the discrete Fourier transforms of $\rho_{i,j,k}$ and $g_{i,j,k}$ are 
\begin{eqnarray}
\hat\rho_{\alpha,\beta,\gamma}=\sum_{i,j,k=1}^{N_g}\rho_{i,j,k}\exp\left[-{\rm i}\frac{2\pi}{N_g}(\alpha i+\beta j+ \gamma k)\right],\nonumber\\
\hat g_{\alpha,\beta,\gamma}=\sum_{i,j,k=1}^{N_g}g_{i,j,k}\exp\left[-{\rm i}\frac{2\pi}{N_g}(\alpha i+\beta j+ \gamma k)\right].
\end{eqnarray}
It is important to state that $\hat g$ has a diagonal structure in Fourier space, making the solution of Eq. (\ref{poissonfourier}) computationally faster than in the real-space based approach.
The potential $\Phi_{i,j,k}$ is finally obtained by back-transforming $\hat\Phi_{\alpha,\beta,\gamma}=\hat\rho_{\alpha,\beta,\gamma}\hat g_{\alpha,\beta,\gamma}$ and reads
\begin{equation}
\Phi_{i,j,k}=\frac{8\pi^3}{(N_g\Delta s)^3}\sum_{i,j,k=1}^{N_g}\hat\rho_{\alpha,\beta,\gamma}\hat\Phi_{\alpha,\beta,\gamma}\exp\left[{\rm i}\frac{2\pi}{N_g}(\alpha i+\beta j+ \gamma k)\right].
\end{equation}
Several numerical algorithms scaling in the best cases with $N_g$, as $O(N_g\log_2N_g)$, are nowadays publicly available to solve this problem, and of course machine optimized routines do exist. However, a detailed description, even of the most commonly used, is far outside from the scope of this chapter, so we redirect the reader to the specialized literature.\\
\indent The above discussed method gives the solution of the Poisson equation for an {\it infinitely} extended system with periodically symmetric density on a $N_g\times N_g\times N_g$ grid were $N_g$ is a power of 2. This is not our case and to treat isolated systems the most common approach is that used for instance in the code \textsc{superbox}, (see Ref. \cite{fell08} and references therein), where by doubling grid size in each direction, the domain is extended by 7 ghost boxes, similarly as what depicted in Fig. \ref{schemegrid} for the two-dimensional case, where $\rho$ is imposed to be 0 and the Green function is written as
\begin{eqnarray}
g_{2N_g-i,j,k}&=&g_{2N_g-i,2N_g-j,k}=g_{2N_g-i,j,2N-k}=g_{2N_g-i,2N_g-j,2N_g-k}\nonumber\\
&=&g_{i,2N_g-j,k}=g_{i,2N_g-j,2N_g-k}=g_{2N_g-i,j,2N_g-k}=g_{i,j,k},
\end{eqnarray}
so that it is periodic on the extended grid. The potential $\Phi_{i,j,k}$ is computed essentially in the same way as for infinite systems. In both cases, potential and acceleration at the particles' positions are obtained by interpolating from grid based quantities using standard techniques such as cubic splines or Lagrange polynomials, see \cite{nr3}.\\ 
\indent When self induced and external magnetic fields are present, in some PM codes such as for example \textsc{calder} \cite{2003LPB....21..573P}, that we have used in some of our numerical simulations, the Poisson equation is not solved at every step but instead, using the {\it density decomposition} technique introduced by Esirkepov in \cite{2001CoPhC.135..144E}, is solved only once at the beginning of the computation, and the Maxwell Equations\footnote{Typically since the system of 4 Maxwell equations is {\it superdetermined}, see \cite{jackson}, only two of them need to be solved, usually Amp\`ere and Faraday equations.} 
\begin{figure} [h!t]
        \centering 		
         \includegraphics[width=\textwidth]{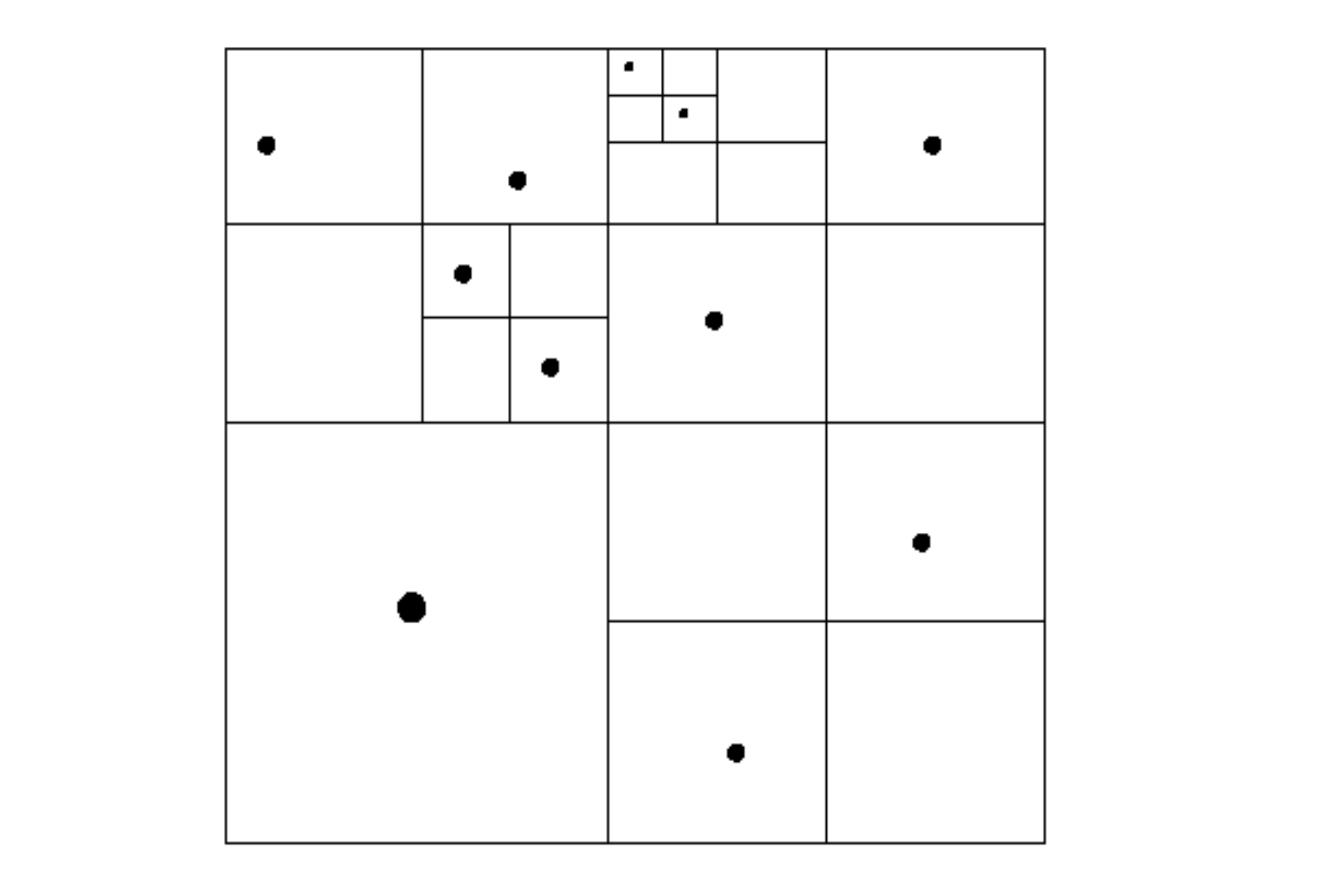}
         \caption{Sketch of the domain decomposition in a tree code. Again, the third dimension has been omitted.}
\label{schemetree}
\end{figure} 
are coupled instead to the continuity equation enforcing in this way the charge conservation.\\
\indent Finally, in some other PM codes, the co called {\it particle-particle-particle-mesh} (PPPM or P${^3}$M), the interaction between the particles is refined by computing also the direct force due to neighbours inside each cell.    
\subsubsection*{Tree-codes}
Although it has not been directly used in this study, we have to mention this latter family of codes particularly suited for highly clustered or dishomogeneous systems. Based on 1986 Barnes and Hut's scheme (see Refs \cite{1986Natur.324..446B}, \cite{hern87} and \cite{hockney}), the tree codes divide the force and the potential at a particle's position in two contributions, due to the neighbours, computed by direct sum and due to the distant particles, computed instead with multipole expansion. The domain of the simulation (i.e. the cubic volume space containing all the system, called {\it tree}) is divided hierarchically, first in 8 cubic cells with sides half the length of the initial cube side, the {\it branches}, and then recursively until each particle has its own cube or {\it leaf}, see Fig. \ref{schemetree}. Cubes containing no particle are discarded from the count. The introduction of the dimensionless parameter $\theta$, the so called {\it opening angle} (in general, $0.3<\theta<1.0$), discriminates between direct and multipole computation of potential and electrostatic field. If for a given particle placed at radius $\mathbf{r}$ distance $R$ from 
the centre of mass $\mathbf{r}_{\rm cm}$ of a cell of side $l$, 
\begin{equation}
\frac{l}{R}<\theta,
\end{equation}
 the contribution of the cell to $\Phi$ and $a$ in $\mathbf{r}$ is computed via multipole expansion, otherwise the cell is partitioned in sub-cells that are analyzed with the same method, in case that {\it leaves}-cells, containing only one particle are reached, the interaction is obviously computed by direct sum.\\
\indent The potential at position $\mathbf{r}$ due to a cell containing $n_c$ particles, of total charge $Q_{\rm tot}=\sum_{i=1}^{n_c}q_i$, is given in multipole expansion truncated at the quadrupole term (see e.g. Ref. \cite{jackson}), by
\begin{equation}
\Phi(\mathbf{r})=\frac{Q_{\rm tot}}{|\mathbf{r}-\mathbf{r}_{\rm cm}|}+\frac{1}{2|\mathbf{r}-\mathbf{r}_{\rm cm}|^5}(\mathbf{r}-\mathbf{r}_{\rm cm})\cdot\mathcal{Q}\cdot(\mathbf{r}-\mathbf{r}_{\rm cm}).
\end{equation}
The components of traceless quadrupole tensor $\mathcal{Q}$ are
\begin{equation}
Q_{i,j}=\sum_{k=1}^{n_c} q_k \left[3(r_{k,i}-r_{{\rm cm},i})(r_{k,j}-r_{{\rm cm},j})-|\mathbf{r}_k-\mathbf{r}_{\rm cm}|^2\delta_{i,j}\right],
\end{equation}
where $\mathbf{r}_k=(r_{k,x},r_{k,y},r_{k,z})$ are the particles positions inside the cell of centre of mass $\mathbf{r}_{\rm cm}$ and $\delta_{ij}$ is the Kr\"onecker delta with $i,j$ running over $x,y,z$. The correspondent electric field $\mathbf{E}(\mathbf{r})=-\nabla\Phi(\mathbf{r})$ is
\begin{equation}
\mathbf{E}(\mathbf{r})=\frac{Q_{\rm tot}}{|\mathbf{r}-\mathbf{r}_{\rm cm}|^2}\hat e-\frac{1}{|\mathbf{r}-\mathbf{r}_{\rm cm}|^4}\mathcal{Q}\cdot\hat e+\frac{5}{2}(\hat e\cdot\mathcal{Q}\cdot\hat e)\frac{\hat e}{|\mathbf{r}-\mathbf{r}_{\rm cm}|^4}
\end{equation}
where $\hat e=(\mathbf{r}-\mathbf{r}_{\rm cm})/|\mathbf{r}-\mathbf{r}_{\rm cm}|$.\\ 
\indent Tree codes, in their most straightforward implementation, have a numerical complexity in the force calculation that scales with the number of particles $N$ as $O(N\log N)$.
\section{The $N-$body codes used for the simulations}
We now discuss the details of the $N-$body codes used for the simulations presented in this thesis. Since we had to model both systems with relatively small number of particles (i.e. small atomic or molecular clusters with $\sim1000$ units) and with very large numbers of particles (i.e. large ionized clusters containing $10^6-10^7$ ions), two force and potential calculation schemes have been employed among those discussed above, the direct sum and the particle-mesh. The first being suited for small systems where inter-particle effects can not be neglected, while the latter is instead better to describe large systems where the force on most of the particles is dominated by the mean field. 
\begin{figure} [h!t]
        \centering 		
         \includegraphics[width=\textwidth]{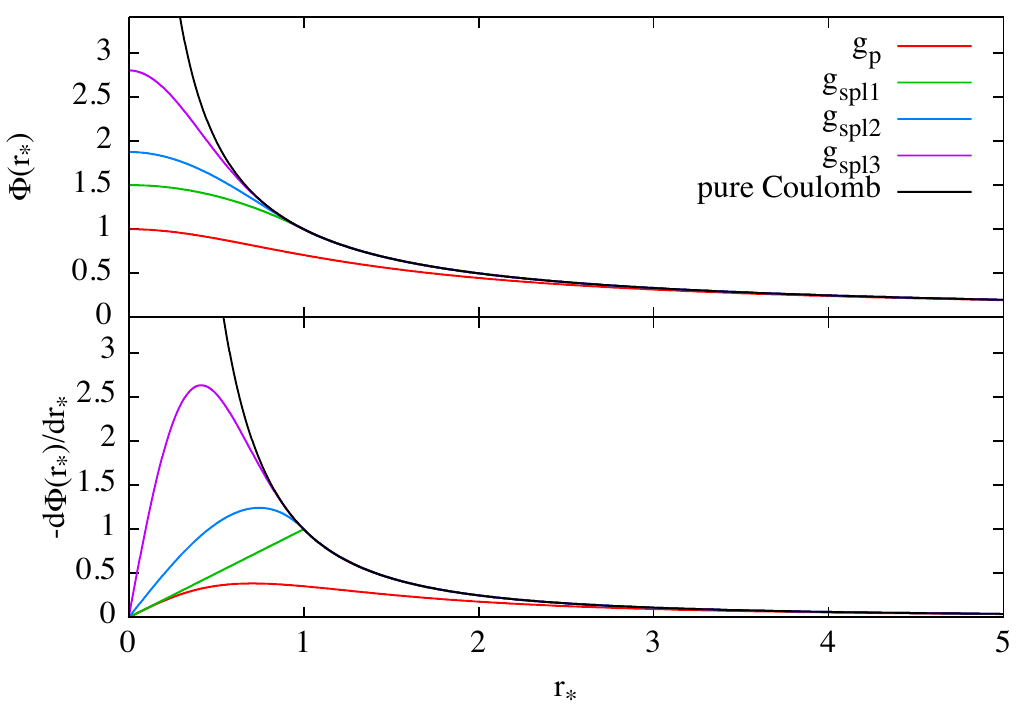}
         \caption{Potential (top) and acceleration (bottom) for different choices of the softening kernel (color lines) compared to the real Coulombic interaction (black line) as function of $r_*=r/\epsilon$. Quadratic and cubic splines $g_{\rm spl2}$ and $g_{\rm spl3}$, that have more complicate polynomial expressions are also shown.}
\label{figsoft}
\end{figure} 
\subsection*{The softening of potential and forces in direct codes}\label{sectsoft}
As anticipated before, due to the singular nature of the $1/r$ interaction fro $r=0$, the numerical computation of the potential and force due to a point-like particle present problems for small values of $r$.\\
\indent In the direct codes, where a large number of pair interaction computation is involved, as well as in tree codes for the near neighbours contributions, the potential at distance $r$ from a charge $q$ is substituted by
\begin{equation}
\Phi^{\rm soft}(r)=\frac{q}{\epsilon}g_*(r/\epsilon), 
\end{equation}
where $\epsilon$ is the softening length and $g_*$, the so called softening kernel, is a function that determines how the interaction is regularized, that has to approach $1/r$ for large separations. In our direct code we implement two of the most widely adopted choices
\begin{equation}\label{plummersoft}
g_p=\frac{1}{\sqrt{r^2/\epsilon^2+1}},
\end{equation}
that gives a softened form $\Phi$ corresponding to that generated by a spherical distribution of total charge $q$ with density $\rho=3q/(4\pi\epsilon^3)(1+r^2/\epsilon^3)^{-5/2}$, and
\begin{align}
g_{\rm spl1}=
\begin{cases}\label{1splinesoft}
-(r^2/\epsilon^2-3)/2;\quad r/\epsilon<1\\
\frac{\epsilon}{r};\quad r/\epsilon\geq1,
\end{cases}
\end{align}
that corresponds to the linear spline. In Fig. \ref{figsoft} we show the softened potential and acceleration regularized using different functions against the real interaction.\\
\indent Higher order spline functions as well as different fictitious densities with finite or infinite support can be used, (see Ref. \cite{2011EPJP..126...55D}). Alternatively, if the number of particles in the simulations is not prohibitively large, the problems introduced by the misrepresentation of $1/r$ for small $r$ can be partially overcome using an adaptive or particle dependent time step.\\
\indent The value of $\epsilon$ used in the simulations is chosen each time by bench-marking between physical consistency and numerical stability criterions. In the numerical calculations discussed in this thesis, where each particle of the run represent an ion or an electron and the softening kernel $g_{\rm spl1}$ of Eq. (\ref{1splinesoft}) is used, such as for example the simulations of molecular clusters irradiated by X-ray pulses, we adopt $\epsilon=a_0$, where $a_0$ is the Bohr radius. With this choice the interaction for $r>a_0$ is the true Coulomb one, while for $r\leq a_0$ is substituted by a harmonic oscillator.\\
\indent In the direct simulations where the numerical particles represent only a Monte Carlo sampling of a given macroscopic charge density (for instance our simulations of expanding spheroidal ion bunches, see Chap. \ref{chapspheroid}), where we adopt instead the softening kernel $g_p$ that implicitly smooths out the total charge density, which is constituted by a sum of deltas, and reduces unwanted discreteness effects, we take as optimal $\epsilon$ half of the minimum inter-particle distance of the initial condition, which allows also to chose sufficiently large time steps, see also Sect. \ref{sectts}.\\
\indent It must be pointed out that the substitution of the Coulomb potential with one of its softened versions based on an infinite kernel may introduce unphysical effects even at the level of the mean field felt by a test particle of the simulation. 
\subsection*{The normalization of the equations of motion and the choice of the optimal time step}\label{sectts}
The normalization of particles' equations of motion is an important point of the $N-$body simulation, since it affects also the choice of the optimal time step $\Delta t$. In the majority of the simulations performed for this thesis, where atomic physics processes are considered, the natural choice is to express all the physical quantities in atomic units (a.u.) defined so that $m_e=e=\hbar=\kappa_0=1$, with this choice, the Bohr radius for the hydrogen atom $a_0$ is also 1 and becomes the unit in which the lengths are expressed, while from the fine structure constant $\alpha=\kappa_0e^2/\hbar c$ we have the speed of light as $c=1/\alpha\simeq137$ which means that velocities are then expressed in units of $\alpha c$ and energies in terms of the so called Hartree energy $E_h=\alpha^2m_ec^2$.\\
\indent On the other hand, in simulations of systems dominated by the mean field or in those extended to large times a different approach is used. First a natural time scale $t_*$ in which the times will be expressed is set up from physical considerations or with dimensional arguments, for instance in the case of the Coulomb explosion of a homogeneous sphere of charge $Q$, mass $M$ and initial radius $R$ a typical choice\footnote{In principle for some particular problems involving electrons oscillations it might be worth taking instead the inverse Langmuir frequency of the electrons $t_*=\omega_0^{-1}=\sqrt{m_e\epsilon_0/4\pi n_ee^2}$ as scale time and the Debye length $\lambda_D$ as scale length.} is
\begin{equation}
t_*=\left[1+\frac{\ln(\sqrt{2}+1)}{\sqrt{2}}\right]\sqrt{\frac{MR^3}{Q^2}},
\end{equation}
that is the time it takes to the sphere to increase its radius to $2R$, second, by introducing a scaled length $r_*$ to express the lengths one finally has the velocity scale as $v_*=r_*/t_*$. In general the radius containing half of the particles of the simulation is a good choice of $r_*$.\\
\indent Choosing the time step $\Delta t$ for the integration of the equations of motion depends on the kind of normalization used by the code and also by other factors such as the intention to neglect or not phenomena happening on time scales considerably smaller than $t_*$. In the type of problems treated here, normalizations in a.u. are always associated with $\Delta t\sim\omega_B^-1/10$, where $\omega_B$ is the electron frequency of the hydrogen atom in the fundamental state of the Bohr hydrogen atom. Whenever a macroscopic timescale is chosen, we typically use $\Delta t\sim t_*/100$, that combined with the leapfrog integration scheme described in Sect. \ref{leap}, gives in most case an energy conservation within the $1.5\%$.
\subsection*{The propagation scheme}\label{leap}
Several methods do exist in order to integrate the system of ODEs (\ref{hamjaceq}), In both our direct and particle mesh codes we use the so called {\it leapfrog} scheme, also known as Verlet algorithm, (see e.g. Ref \cite{gm1991}). This particular second order method is based on the splitting of Hamiltonian (\ref{nbodyham}) in its two terms, kinetic and potential, so that a particle of mass $m$, momentum $\mathbf{p}=m\mathbf{v}$ and position $\mathbf{r}$ is moved in phase space at each time step $\Delta t$, in two half steps alternatively due to $K$ and to $U$:
\begin{align}
\mathbf{r}_{n+1/2}&=\frac{\Delta t}{2}\frac{\mathbf{p}_{n}}{m}+\mathbf{r}_n;\\
\mathbf{p}_{n+1}&=\Delta tm\mathbf{a}_{n+1/2}+\mathbf{p}_n;\\
\mathbf{r}_{n+1}&=\frac{\Delta t}{2}\frac{\mathbf{p}_{n+1}}{m}+\mathbf{r}_{n+1/2},
\end{align}
where $\mathbf{a}_{n+1/2}$ is the acceleration computed at half time step.\\
\indent Albeit being very simple to implement, the leapfrog scheme has the important properties of being time reversible (contrary to some higher order schemes such as Runge-Kutta or Hermite, see \cite{nr3}) and, at least theoretically within machine precision, to conserve in time average the invariants of the dynamical system identified by Eqs. (\ref{hamjaceq}) such as the Hamiltonian itself and the angular momentum $L$. Algorithms of this kind are called symplectic.\\
\indent It must be pointed out that if one uses relativistic mechanics or if more generally $a$ depends on $v$, the symplecticity properties are not fulfilled anymore and large errors on the particles' orbits can be introduced even with small and adaptive $\Delta t$. In the case of relativistic simulations, the most popular approach is the Boris scheme \cite{boris}, that needs only another intermediate step to update the relativistic factor $\gamma_{n+1/2}$ but unfortunately introduces spurious drifts in particles' orbits for particular configurations of electric and magnetic fields for which $\mathbf{E}+\mathbf{v}\times\mathbf{B}=0$. General velocity dependent accelerations can be treated with the recently introduced method by Rein and Tremaine \cite{2011MNRAS.415.3168R}, that is symplectic and involves just a coordinate change to a non inertial frame.\\
\indent In our implementation, whenever the problem involves relativistic velocities and or non zero magnetic field, we adopt instead the corrected Boris scheme introduced for the first time in the context of merging of relativistic beams, see \cite{vay08}. At the cost of a large number of substeps, this method manages to circumvent the problems plaguing the standard relativistic methods and is time reversible. A particle of mass $m$, charge $q$, velocity $\mathbf{v}_n$, position $\mathbf{r}_n$ and relativistic factor $\gamma_{n}=\sqrt{1-v_n^2/c^2}$ at step $n$ and undergoing the action of (in general) time dependent fields $\mathbf{E}$ and $\mathbf{B}$, is advanced at step $n+1$ in the following way:
\begin{align}
\mathbf{r}_{n+1/2}=\frac{\Delta t}{2}\mathbf{v}_{n}+\mathbf{r}_n;\\
\mathbf{u}_{n+1/2}=\gamma_n\mathbf{v}_n+\frac{q}{m}\frac{\Delta t}{2}\left(\mathbf{E}_{n+1/2}+\mathbf{v}_{n}\times\mathbf{B}_{n+1/2}\right);\\
\mathbf{u}^\prime=\mathbf{u}_{n+1/2}+\frac{q}{m}\frac{\Delta t}{2}\mathbf{E}_{n+1/2},\quad\gamma^\prime=\sqrt{1+u^{\prime2}/c^2};\\
\mathbf{\tau}=\frac{q}{m}\frac{\Delta t}{2}\mathbf{B}_{n+1/2},\quad u_*=\mathbf{u}^\prime\cdot\frac{\mathbf{\tau}}{c},\quad\sigma=\gamma^\prime-\tau^2;\\
\gamma_{n+1}=\sqrt{\frac{\sigma+\sqrt{\sigma^2+4(\tau^2+u_*^2)}}{2}},\quad\mathbf{t}=\frac{\mathbf{\tau}}{\gamma_{n+1}},\quad s=\frac{1}{1+t^2};\\
\mathbf{u}_{n+1}=s\left[\mathbf{u}^\prime+(\mathbf{u}^\prime\cdot\mathbf{t})\mathbf{t}+\mathbf{u}^\prime\times\mathbf{t}\right],\quad\mathbf{v}_{n+1}=\frac{u_{n+1}}{\gamma_{n+1}};\\
\mathbf{r}_{n+1}=\frac{\Delta t}{2}\mathbf{v}_{n+1}+\mathbf{r}_{n+1/2},
\end{align}
where as usual quantities indicated in boldface are vectors and in normal font scalars.
\subsection*{Building the grid in the PM code}
In our implementation of the particle mesh algorithm, essentially based on the \textsc{superbox} code, \cite{fell08}, the grid on which the charge density is displaced with a nearest grid point method (Equations \ref{cdpl} and \ref{NGP}) is a cubic cartesian one. Since we are principally describing systems undergoing expansion we allow its spacing $\Delta s$ to increase with time. This is done as follows, first at $t=0$ the centre of mass of the system (c.o.m.) is identified and the coordinates of the particles changed so that it coincides with $(0,0,0)$. Then the radius of the far most particle $r_{\rm max}$ is taken as the side of the part of the grid containing the real system and the first mesh constructed along the three axes with the same $\Delta S = 4 r_{\rm max}/N_g$ so that for $l=x$, $y$ or $z$
\begin{equation}
r^{l}_i=-r_{\rm max}+(i-1)\Delta s,
\end{equation}
the space (real plus ghost zones) is partitioned as sketched in Fig. \ref{schemegrid} in $N_g^3$ nodes $(r^{x}_i,r^{y}_j,r^{z}_k)$.\\
\indent The Green function is then computed as in Eq. (\ref{grnfctn}). The procedure is repeated at every time step without identifying the c.o.m.. Since a normalized discrete Fourier transform is used, and the grid spacing is uniform in all directions, one does not need to recompute the Green function, but just to multiply for the opportune function of $\Delta s$ the normalized value of the grid based potential $\Phi_{i,j,k}$ obtained by the Poisson solver routine before interpolating it at the particles positions.\\
\indent Using a time dependent mesh has implications on the choice of the simulations time step, we require that at every time 
\begin{equation}
\Delta t < \frac{\Delta s}{\Delta \sigma_v}
\end{equation} 
where $\Delta \sigma_v$ is the variation of the particles' velocity dispersion between two consecutive time steps. In general this is obtained by taking $\Delta t = 0.1\Delta s/\Delta \sigma_v$.
\section{Including quantum processes in classical dynamical simulations: a Monte Carlo approach}\label{quantumeffects}
We have discussed until this point, only the numerical techniques used to treat the dynamics of systems of charged particles. However, in this thesis, we are mainly interested on the response of such systems when exposed to strong laser radiation, in particular short and intense x-ray impulses with typical durations of a few femtoseconds, intensities in the range $10^{16} - 10^{20}$ ${\rm W/cm^2}$ and photon energies of the order of a few keV.\\
\indent When a cluster is irradiated with such strong laser fields (for monographic reviews see Refs. \cite{posthumus}, \cite{SAA02} and \cite{2006JPhB...39R..39S}, see also \cite{2002Natur.420..482W} and \cite{2012PhRvL.108m3401T}), its constituting atoms or molecules experience many photoionization events (i.e. electrons are stripped and released with kinetic energies $K_{\rm photo}=\hbar\omega-E_{\rm bind}$ where $\omega$ is the incoming photon frequency and $E_{\rm bind}$ is the electron binding energy). In addition, when the inner electronic shells are ionized first, and the ion has still other electrons, such unstable ionic configuration (say state $a$) decays to a stable configuration (state $b$) via (in general chains of) the so called Auger process \cite{ROH08}  that refills the inner shell and release a second electron with kinetic energy $K_{\rm auger}=E_a-E_b$ to the continuum.
\begin{figure} [h!t]
        \centering 		
         \includegraphics[width=0.8\textwidth]{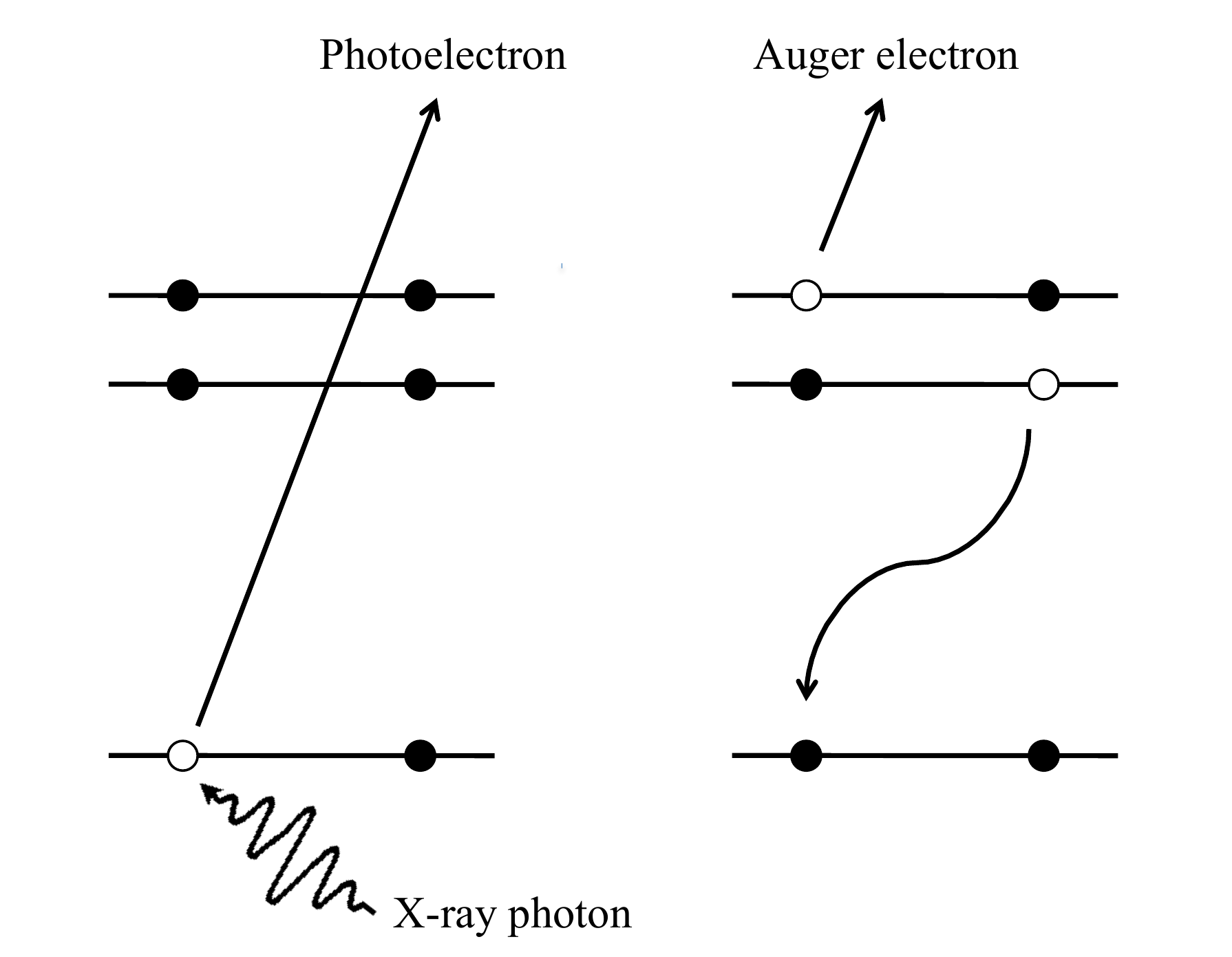}
         \caption{Sketch of $K-$shell photoionization followed by an Auger decay in a carbon atom.}
\label{figaug}
\end{figure} 
 To carry out a full quantum description is impossible even on modern machines since it involves the solution a many electron Schr\"odinger equations in three spatial dimensions. On the other hand, using semiclassical methods or techniques based on mean field approaches such as Density Functional Theory (DFT, see \cite{atomsmolecules}), implies a high degree of simplification that prevent to study in detail the individual dynamics of the electrons, unless one introduces additional artifacts \cite{2003PhRvA..68c3201B}.\\
\indent In order to find an acceptable compromise, we follow the same line of Refs. \cite{2007PhRvA..76d3203G}, \cite{GNO09}, \cite{georgescu2008} and \cite{2013CP....414...65G}, where the above mentioned  {\it quantum processes} are treated ``statistically"; roughly speaking, for the $i-$th atom or ion of the simulation first a transition probability $\mathcal{P}$ is computed (for instance from the laser intensity, the photoionization cross-section and the time step in the case of ionization), then a random number $p$ is extracted from a normal distribution and finally the ion's electronic configuration is changed accordingly,  whenever $p\leq\mathcal{P}$. Subsequently the electrons released by photoionization or Auger decay are initialized in the simulation with their kinetic energies $K_{\rm photo}$ or $K_{\rm auger}$ and then treated as classical particles.\\
\indent By doing this, each many-particle system treated in a single simulation is just a Monte Carlo realization of the problem considered, and therefore integrated as well as time dependent quantities (such as the final charge states or the time evolution of the total kinetic energy), have to be averaged over many realizations starting with the same initial conditions, in order to have statistical meaning or to be compared for different parameter sets.\\
\indent In the next subsections we show how the photoionization and Auger decay probabilities $\mathcal{P}^{{\rm photo}}$ and $\mathcal{P}^{{\rm auger}}$ are calculated.
 \subsection*{Photoionization}
The probability used in the Monte Carlo sampling, that a photoionization event happens between times $t$ and $t+\Delta t$ for the $j$-th shell of the $i$-the atom or molecule in the cluster is given by
\begin{equation}\label{probphotoion}
{\mathcal P}^{{\rm photo}}_{j,i}=I(t,\mathbf{r}_i)\frac{\sigma_{j,i}\Delta t}{\hbar\omega};\quad j=1s,2s,2p,...
\end{equation}
where $I(t,\mathbf{r}_i)$ is the laser intensity at time $t$ at the ion's position $\mathbf{r}_i$, and $\sigma_{j,i}$ the photoionization cross section for which we refer to the values given in Ref. \cite{Doe:2009:Online}. If the $j-$th shell has one or more vacancies, ${\mathcal P}^{{\rm photo}}_{j,i}$ is reduced by the factor 
\begin{equation}
\chi=\frac{N_e-N_v}{N_e},
\end{equation}
where $N_e$ is the number of electrons in the full case and $N_v$ the number of vacancies.\\
\indent For the theoretical laser pulses we use Gaussian time envelopes, and constant space profiles, so that
\begin{equation}
I(t,\mathbf{r})=I_0e^{-\frac{(t-t_0)^2}{2\sigma^2}},
\end{equation}
where $I_0$ is the peak intensity and $\sigma$ is the Gaussian's standard deviation related to the full width half maximum by ${\rm FWHM}=2\sqrt{2\ln 2}\sigma$, and $t_0$ is the time at which the laser has its peak (i.e. $I(t_0,0)=I_0$). 
Alternatively, if the dimension of the irradiated sample is comparable to the radius of the the laser focus $z_0$ a Lorentzian envelope is used along the direction $z$ in the focus plane and a Gaussian envelope with standard deviation $\omega_0$, along the pulse propagation direction $r$.
The space envelope therefore reads
\begin{equation}\label{lorenzo}
I(r,z)=I_t\frac{1}{1+(z/z_0)^2}e^{-\frac{2r^2}{\omega_0^2\left[1+(z/z_0)^2\right]}}.
\end{equation}
Note that, at photon energies of few keV, for the elements of the first row that we are considering in the present study, the photoionization happens mainly via $K$-shell photoionization ($1s$ orbitals), for $\hbar\omega=1~{\rm keV}$ in fact, the cross section $\sigma_{1s}$, given (see e.g. Ref. \cite{drake}) by
\begin{equation}
\sigma_{1s}=\frac{256\pi}{3}\alpha\left(\frac{a_0}{Z}\right)^2\left(\frac{E_{{\rm bind},1s}}{\hbar\omega}\right)^{7/2},
\end{equation}
where $Z$ is the nuclear charge of the element, $\alpha=1/137$ and $a_0=5.29\times10^{-11}$ m are the fine structure constant and the Bohr radius respectively, is of the order of $5\times10^{4}~{\rm barn}$, while $\sigma_{2s}$ and $\sigma_{2p}$ are roughly $10^{3}$ and $10~{\rm barn}$ respectively. Therefore, In the numerical calculations we assume $\sigma_{2s}=\sigma_{2p}=0$. Test simulations with real values for all the cross sections, revealed a percentage of photo-electron coming from the outer shells smaller than the 3\% at peak intensities $\geq10^{19}~{\rm W/cm^2}$, making the latter assumption rather reasonable.\\
\indent Note also that in our simulations, when an electron is released by photoionization, the direction of its velocity is extracted randomly from a normal distribution while its mother ion recoils consistently with
\begin{eqnarray}
\frac{m_{\rm ion}v^2_{1,{\rm ion}}}{2}+\hbar\omega-E_{\rm bind}&=&\frac{m_{\rm e}v^2_{2,{\rm e}}}{2}+\frac{m_{\rm ion}v^2_{2,{\rm ion}}}{2};\nonumber\\
m_{\rm ion}\mathbf{v}_{1,{\rm ion}}&=&m_{\rm e}\mathbf{v}_{2,{\rm e}}+m_{\rm ion}\mathbf{v}_{2,{\rm ion}},
\end{eqnarray}  
where $m_{\rm ion}$ and $m_{\rm e}$ are the masses of the ion and the electron and $\mathbf{v}_{1,{\rm ion}}$ and $\mathbf{v}_{2,{\rm ion}}$, $\mathbf{v}_{2,{\rm e}}$ their velocities before and after the photon absorption.
\subsection*{Auger decay}
The $K-$shell photoionization produces molecular or atomic ion states prone to decay via Auger processes (see e.g. Ref. \cite{2012PhRvA..85f3405A}) in which an external electron refill the inner vacancy while a second is emitted with energy corresponding to the energy difference between the initial (unstable) and the final configurations.\\
\indent For a given ion $i$, the transition probability from the state $a$ to the state $b$ via an Auger decay between times  $t$ and $t+\Delta t$ is computed from the energy amplitude between the two configurations $\Gamma^{ab}_i$ and reads
\begin{equation}\label{probauger}
{\mathcal P}^{{\rm auger}}_{ab,i}=\frac{\Gamma^{ab}_i}{\hbar}\frac{n_e(n_e-1)}{n (n-1)}\Delta t,
\end{equation}
where $n_e$ and $n$ are the total number of electrons in principle able to take part to the transition, and the number of electrons present in the external shells in the equivalent neutral respectively. For the species of interest in this thesis, we used the values of  $\Gamma$ given in \cite{2011JChPh.135m4314K}.\\
\indent It must be also mentioned that an analogous approach for the calculation of Auger transition rates has been recently independently developed in \cite{2012JChPh.136n4304I} and \cite{2013JChPh.138p4304I}, and used to investigate the dynamics and the fragmentation of water and methane molecules under intense X-ray lasers.
\subsection*{Recombination}
The spatial finiteness of the plasmas studied here and the short time scales considered ($<0.5{\rm fs}$) do not allow for the thermalization of ions and electrons, 
and thus any approach to treat recombination that based on the Boltzmann-Saha equation (see e.g. \cite{bellino}) is inapplicable.  
To account for the recombination of electrons with ions a simple classical procedure is carried out.\\
\indent Following the approach of Ref. \cite{2007PhRvA..76d3203G}, between every two time steps the revolution angle $\theta_{i,j}(t)$ of the $i$th electron around the nearest ion $j$ is determined. 
If such electron revolves two times around the same ion $j$ (i.e. $\theta_{i,j}(t+\Delta t)=4\pi$) it is considered localized to that ion and the electronic configuration of the latter is updated with an electron in an excited state (i.e. a ``classical recombination'' happened). 
A simpler procedure to assign an electron to the neighbouring ion would be instead 
checking whether the total energy of the $i$th electron in the reference system of the nearest ion of charge $q_j$
\begin{equation}
 \mathcal{E}_{ij}=\frac{eq_j}{|\mathbf{r}_i-\mathbf{r}_j|}+\frac{m_e}{2}\left[(v_{x,i}-v_{x,j})^2+(v_{y,i}-v_{y,j})^2+(v_{z,i}-v_{z,j})^2\right]+eU(\mathbf{r}_j)
\end{equation}
is negative, together with the fact that the two particles are at a distance $|\mathbf{r}_i-\mathbf{r}_j$ smaller than a certain fiducial scale length. In the definition above, $U(\mathbf{r}_j)$ is the cluster electrostatic potential at the ion's position. 
In both cases sketched here, the process is obviously strongly influenced by the way in which the Coulomb interaction is regularized for small separation.   
\section{The analysis of the end products}
To conclude the chapter, we give a brief description of the tools used to analyze the end products of our numerical simulations such as the routines to extract the final shapes of the charge distributions and the averaging procedures in the Monte Carlo simulations. 
\subsection*{The shape of the final charge distribution}
To extract the aspect ratio of a given non-spherical distribution of charges we use a procedure based on tensor diagonalization. Following the algorithm used in \cite{MZ97} for the end products of gravitational $N-$body simulations, we compute the second order tensor
\begin{equation}
I_{ij}\equiv \sum_{k=1}^N r_i^{(k)}r_j^{(k)}
\end{equation}
for the particles inside the sphere of radius $r_{98}$ containing the 98\% of the total number of particles in the system, where $r_i$ are the Cartesian components of the position vector in the reference frame with origin in the centre of mass.
$I_{ij}$ is related to the inertia tensor  by 
\begin{equation}
\mathcal{I}_{ij}={\rm Tr}(I_{ij})\delta_{ij}-I_{ij}.
\end{equation}
The matrix $I_{ij}$ is diagonalized iteratively, requiring that the percentage difference of the largest eigenvalue between two iterations to be smaller than $10^{-3}$. This procedure requires on average 10 iterations, and we call $I_1\geq I_2\geq I_3$ the three eigenvalues. We finally apply a rotation to the system in order to have the three eigenvectors oriented along the coordinate axes. For a heterogeneous ellipsoid of semi-axes $a,b$ and $c$, we would obtain $I_1=Aa^2$, $I_2=Ab^2$ and $I_3=Ac^2$, where $A$ is a constant depending on the density profile. Consistently, for the end-products we define $b/a=\sqrt{I_2/I_1}$ and $c/a=\sqrt{I_3/I_1}$, so that the ellipticities in the principal planes are $\epsilon_1=1-\sqrt{I_2/I_1}$ and $\epsilon_2=1-\sqrt{I_3/I_1}.$
\subsection*{The realization averages of distributions and global quantities}
As we have previously mentioned, when using the Monte Carlo approach to treat quantum phenomena, each simulation represents only a single statistical realization of a given chain of events. On the other hand, in laboratory experiments of laser-cluster interaction, one measures the yield of integrated quantities over the products of {\it several} clusters irradiated by the same laser pulse. To have a meaningful comparison between measured quantities and their simulated counterparts, or to have a reasonable picture of their scaling with the laser intensity $I_0$, the latter have to be processed with an averaging procedure. When one performs a set of $N_r$ realizations of a given numerical experiment initialized with identical initial conditions (i.e. cluster size, laser intensity etc.), but different seeds for the machine's random number generator, a time dependent quantity $S(t)$ such as for instance the number of absorbed photons $n_\gamma$ is averaged as
\begin{equation}\label{media1}
\langle S(t)\rangle=\frac{1}{N_r}\sum_{i=1}^{N_r} S_i(t)
\end{equation}  
while a quantity at fixed time $t$ that depends itself on an average over the total number of ions $N_{\rm ion}$ such as their average charge state $\langle q\rangle_{\rm ion}$ is given by
\begin{equation}\label{media2}
\langle S\rangle=\frac{1}{N_r}\sum_{i=1}^{N_r} \frac{1}{N_{\rm ion}}\sum_{j=1}^{N_{\rm ion}} S_{j,i}.
\end{equation}  
We have observed that typically, for clusters containing $\sim 1000$ atoms or molecules, one needs to average over up to 50 realizations to have reasonably smooth $\langle S\rangle$ vs $I_0$ curves.\\
\indent It must be also pointed out that in real experiments involving jets of clusters irradiated by a short and intense laser fields, it is unlikely that all the clusters of the jet are hit by the pulse at focus, if one needs for some reason to take that into account in the numerical calculations, a pulse average is required. Usually, in order to do so, following Ref. \cite{1999PhRvA..60.1771N}, one has first to chose a time and spatial profile for the theoretical laser pulse and sample them  to perform $N_r=N_t\times N_s$ independent simulations with constant intensity $I_{i,j}$ in time and space from which the sampled quantity $S_{ij}$ is extracted. In this case, a time and space averaged  quantity $\langle S\rangle_{s,t}$ is finally obtained as
\begin{equation} 
\langle S\rangle_{s,t}=\sum_{i=1}^{N_s}\left\{\ln\left(\frac{I_{i+1}}{I_i}\right)\sum_{j=1}^{i}S_{i,j}\left[\sqrt{\ln\left(\frac{I_i}{I_j}\right)}-\sqrt{\ln\left(\frac{I_i}{I_{j+1}}\right)} \right]\right\}.
\end{equation}
As an alternative, one can instead assume an effective space profile for the laser pulse, for instance that given in Eq. (\ref{lorenzo}), perform different runs where the targets are displaced with respect to the laser focus according to a normal distribution, 
and finally average the wanted quantities with Eqs. (\ref{media1}) or (\ref{media2}). In the results presented in this work we have employed only the latter method. 
\chapter{Summary and outlook}\label{sommario}
In this thesis we have presented an exploratory study of clusters exposed to short and intense x-ray pulses. We have considered the implications of the massive charging (i.e. one charge per atom/molecule) happening on a time scale of the order of few femtoseconds 
on the dynamics of ions and electrons inside the cluster, and the process of expansion (i.e. Coulomb explosion). Such extreme conditions are nowadays experimentally  accessible due to high intensity femtosecond x-ray pulses attainable in modern free-electron laser facilities. 
For the microscopic description of such conditions a numerical code has been developed, integrating atomic processes such as photoionization, Auger decay and recombination with $N-$body molecular dynamics.\\

\indent Before studying specific systems, which are of particular interest in the view of recent experiments, we have performed simulations of the explosion of charged spherical clusters under different conditions of composition and density profile, and studied their end products by means of idealized analytic models based on continuum approximation.\\ 
\indent It appeared clear that, even though the expansion processes considered here are mainly collisionless, the discrete nature of the systems still induces effects on the energy distributions which are impossible to be accounted for in the continuum model. 
In particular, for systems starting with a homogeneous charge density profile, the energy distribution shows a peaked structure close to the cutoff energy and therefore strongly departs from that which is predicted when assuming a continuum system.\\ 
\indent We have verified that this feature, discussed in Ref. \cite{mika2013}, 
persist for different particle arrangements in the initial condition, with a general trend of getting milder for more randomized particle ``displacement'' and sharper with increasing number of particles.\\
\indent  Remarkably, the differences between numerical and theoretical 
energy distributions of exploding clusters seem to be enhanced by the regularization of the Coulomb interaction. The latter, although leaving the long-range behavior basically unaltered, induces for particles close to the surface of the system an unphysical ``external field effect'' (i.e. such particles suffer an extra force pushing them outside). 
Such a fact obviously affects the analogous gravitational $N-$body codes and has not been documented in the literature.\\

\indent In view of molecular clusters or large molecules we have extended our study to two-component (i.e. light and heavy atoms) and non-spherical systems. We find that the energy distribution for the light component of two species systems significantly diverges from that predicted by analytical (continuum) models, as its dynamics is strongly influenced by the system's discreteness at early stages of the explosion.\\ 

\indent In the Coulomb explosion of homogeneous systems with ellipsoidal shapes, it turns out that the initial aspect ratio is not conserved during the expansion but instead the initially prolate systems become oblate as they expand and vice versa. 
Expressions for the energy spectra are given for the case of axisymmetric systems and limit aspect ratios are derived. Molecular dynamics simulations show remarkably good agreement with our analytic predictions.\\ 

\indent Prompted by experiments with methane clusters at the x-ray free electron laser LCLS, we have studied molecular clusters. 
To shed some light on the surprising experimental findings we made a systematic study over a broad range of physically relevant intensities (from $5\times10^{16}$ to $10^{19}{\rm W/cm^2}$) and pulse lengths between 3 and 100 fs. 
More importantly, a whole series of hydride clusters with iso-electronic configuration but different photoabsorption and Auger transition and different atomic masses was analyzed.\\
\indent Spanning such ranges of intensities and pulse lengths, we observed a remarkably non-monotonic trend of the ratio between the average kinetic energies of protons and heavy ions showing a maximum at a well determined laser intensity for fixed photon energy and pulse length. 
This effect is in apparent contrast with that which is found in the case of the pure Coulomb explosion of multi-species clusters where such a ratio is markedly monotonic, for every combination of charge ratios and percentages of the two species.\\
\indent This has been interpreted as the presence of three distinct electron-induced regimes of energy redistribution between the two species of ions. In the first case (at low intensities), the cluster is weakly charged having suffered few photoionizations and expands as a whole, as a quasi-neutral plasma. 
In the intermediate regime enough charge is created via photoionization. The electrons stripped by the (time dependent) cluster electrostatic field do not leave the system but adjust to screen its charge in the core, leaving a charged shell at the surface from which protons escape with large kinetic energies due to their lower mass.\\
\indent Finally, at large intensities the whole cluster is strongly charged and expands via pure Coulomb explosion as in idealized models neglecting the contribution of electrons. It must be pointed out that, in contrast to what is observed in clusters irradiated by long wavelength lasers,
for the conditions studied here, the charging processes happen on a time scale that is of roughly two orders of magnitude smaller than that of any appreciable (heavy) ion dynamics. The trapped electrons that do not have enough time to thermalize are, however, inducing a previously 
unobserved species segregation channel.\\ 

\indent We stress the fact that the different scenarios described above, are obtainable simply by tuning the laser intensity and, having noted that for the intermediate intensities the backbone of the cluster 
structure stays basically unaltered, may be of some relevance for x-ray based single molecule imaging.\\

\indent In conclusion, we point out that studying the dynamics of laser irradiated clusters might be relevant for other areas of physics and their technological application. 
For instance, the possibility to control (by tuning the laser intensity) the fraction of energy absorbed by the cluster which is transferred to the light ions' kinetic energy, and therefore to a certain extent, control their energy distribution, 
opens to applications in particle acceleration or intra-cluster nuclear fusion, where a narrow ion velocity spectrum is needed \cite{2001PhRvL..87c3401L}. In this case, deuterated molecular hydride clusters (i.e. hydrogen in \ce{CH4} or \ce{H2O} is substituted by its isotope deuterium) are to be used.  
The nanoplasmas produced by the interaction of intense lasers with clusters are characterized by kinetic and potential energies $K$ and $U$ such that their Coulomb coupling parameter $\Gamma$, expressed as function of its minimum inter-particle distance $a$, 
temperature $T$ and the electron charge $e$ by $\Gamma={U}/{K}={e^2}/{ak_BT}$, is of the order of unity \cite{rost2009}. This places the laser generated cluster plasmas at the edge between ideal plasmas ($\Gamma\gg 1$) \cite{bellino}, and strongly correlated plasmas ($\Gamma\ll 1$) \cite{2009JPhA...42u4002F}.
The latter are thought to exist in nature in astrophysical environments such as the cores of giant planets or white dwarfs stars. Even if their intrinsic parameters such as densities and temperatures may differ by orders of magnitude, plasmas of equal $\Gamma$ are expected to behave similarly and show analogous dynamical properties ( see e.g. \cite{rost2009}). 
Thus the study of laser generated nanoplasmas is very likely to shed some light on the physics of non-experimentally accessible plasmas. 
Nevertheless, due to their small sizes and relatively short life-times (being overall non-neutral, they expand under their self consistent Coulomb potential doubling their initila radius in roughly 10 fs), the direct investigation of cluster plasmas dynamics is still prone to a number of complications, and one relies in general on the indirect information obtained by the fragmentation products and on numerical simulations.\\

\indent From the point of view of non-equilibrium thermodynamics, finite and globally non-neutral Coulomb systems, where the electrons are confined by the space charge of the ions, share some peculiar properties with gravitational systems \cite {2005MNRAS.361.1227C}, \cite{2008JPhA...41P5501V} (or more generally with systems interacting via long-range forces, see \cite{2008AIPC..970...39C}, \cite{2010PhyA..389.4389B} and \cite{pfdc2}), such as for example 
a negative specific heat in the microcanonical ensemble \cite{1999PhyA..263..293L}, \cite{1970ZPhy..235..339T}, that is the more the system is heated, the more it cools. 
The laser induced electron-ion plasmas treated in Chap. \ref{chapLCLS}, likely fall in this regime and therefore their study could be particularly relevant to test experimentally recent results in non-equilibrium statistical mechanics and thermodynamics.   
\appendix
\chapter{Cluster production}\label{clustru}
\begin{figure} [h!t]
        \centering 
         \includegraphics[width=\textwidth]{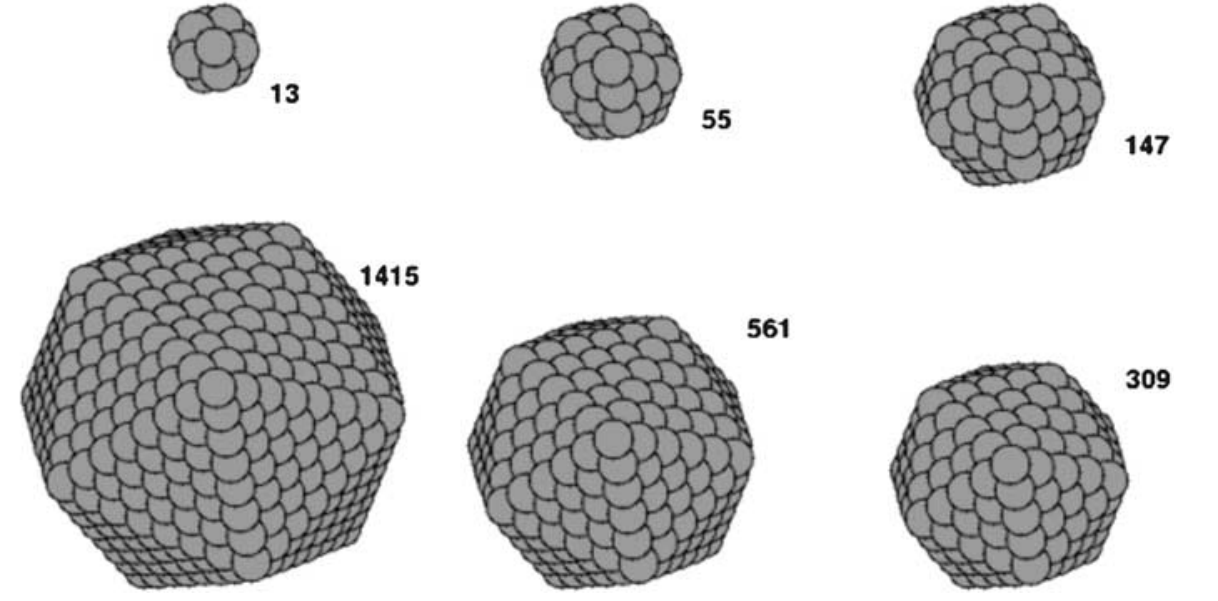}
         \caption{Rare gas clusters closed shell structure according to their (indicated) number of atoms $N_*$, so called ``magic numbers''. The figure is reproduced from Ref. \cite{rost2009}.}
\label{ico}
\end{figure}

Clusters can be classified in different ways according to their physical and structural properties, in the first instance we distinguish between atomic an molecular cluster whereas they are composed by atoms or molecules. 
The first are held together by either metallic, covalent or ionic bonds, or by London forces as in the case of rare gas clusters, while the second prevalently by van der Waals forces between 
the induced electric dipoles on molecules \cite{take} or ionic bonds. We recall that throughout this thesis we used the additional classification, speaking of {\it homogeneous clusters} when they are composed by a single species of atoms, and {\it heterogeneous clusters} 
when they are instead composed of more then one atomic species, or by hetero-nuclear molecules as in the case of molecular hydride clusters.\\
\indent The number $N_*$ of constituting particles and the type of bonds supporting their structure are responsible for the shape of the cluster. Rare gas clusters for instance (but also the molecular clusters of our interest) are characterized by the so called isocaederical structure (see Fig. \ref{ico}) 
where particles are arranged on concentric 
shells so that with increasing $N_*$, the fraction of atoms sitting at the surface $n_s$ decreases proportionally to $N_*^{-1/3}$, which implies that for small sizes the majority of the cluster's constituents is located at its surface.\\
\indent Clusters are artificially produced by letting expand into void a sonic jet of gas with pressure $P$ and temperature $T$, coming out of a conical nozzle with aperture $\alpha$ and orifice diameter $d$ of the order of 1 $\rm \mu m$ (see \cite{1992RScI...63.2374H}, \cite{jhonny} and references therein). The average number of atoms or molecules in clusters formed in this way, is given by
\begin{equation}
\langle N_*\rangle=A(H_*/10^3)^\gamma
\end{equation}
where $A$ and $\gamma$ are constants and $H_*$ is the so called (reduced) Hagena parameter \cite{1981SurSc.106..101H} given by
\begin{equation}
H_*=Pd_{\rm eq}^qT^{0.25q-2.5}K_*;\quad d_{\rm eq}=0.74d/\tan(\alpha/2),
\end{equation}
in which the constant $K_*$ depends on the species and it is related to the minimum inter-particle distance in the solid phase $r_{\rm min}$ and the sublimation enthalpy $h_s$ and reads
\begin{equation}
K_*=1/(r_{\rm min}^{q-3}T_*^{0.25q-1.5});\quad T_*=h_s/k_B.
\end{equation}
in all formulas above, the parameter $q$ is found experimentally to fall in the interval 0.5-1, see \cite{danyl} and \cite{1992RScI...63.2374H}. The distribution of sizes $\d P/\d N_*$ in a get of clusters is given by a lognormal distribution 
(see e.g. \cite{pocsik} and \cite{mendam}) and reads
\begin{equation}
\frac{\d P}{\d N_*}=\frac{1}{(N_*-N_0)\sigma\sqrt{2\pi}}\exp\left[-\frac{(\ln(N_*-N_0)-\mu)^2}{2\sigma^2}\right],
\end{equation}
where $N_0$ is the position of the ``zero'' and $\mu$ and $\sigma$ the logarithms of the geometric mean and standard deviation.
\chapter{Dynamical friction}\label{frizione}
\begin{figure} [h!t]
        \centering 		
         \includegraphics[width=0.7\textwidth]{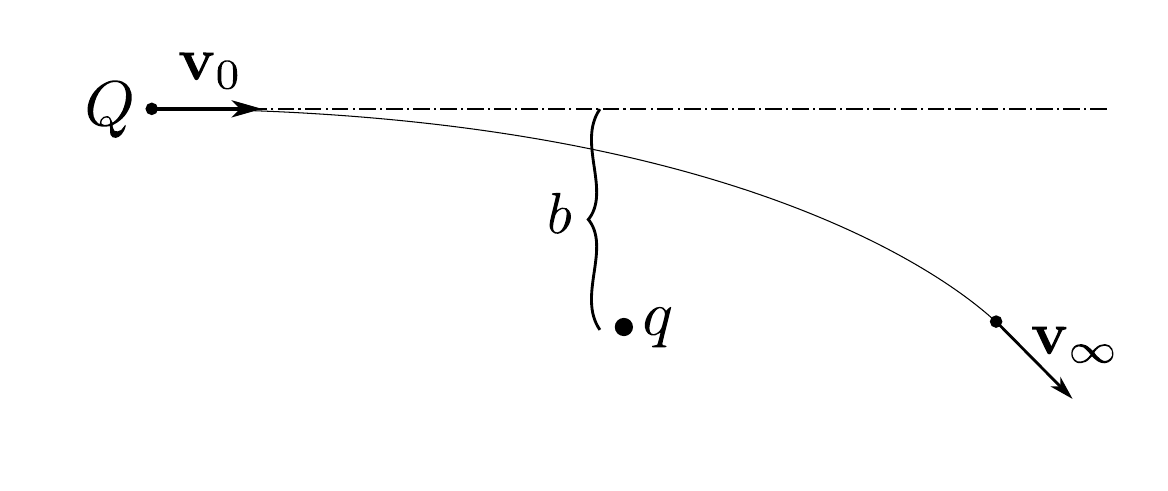}
         \caption{Unperturbed orbit (dashed dotted line) of a particle of charge $Q$ and relative orbit (solid line) when deflected by a field particle of charge $q$ for the case $Qq<0$.}
\label{figdf}
\end{figure}

Following the classical treatment of Chandrasekhar \cite{chandra43} and Spitzer \cite{spitz}, we derive here the formula of the dynamical friction, defined as the drag effect on a test particle of charge $q_t$ and mass $m_t$ moving through a background of particles with charge $q_f$ and mass $m_f$ due to long range collisions (i.e. Rutherford scattering \cite{jhonny}).\\
\indent  Let us consider first as in the original derivation (see e.g. \cite{binney08}) a binary collision in the so called {\it impulsive approximation}, for which the deflection of the trajectories due to a long-range force is approximated as due to a finite-time interaction, (see \cite{1966AIAAJ...4.1417R}). Single orbital deflections, like that sketched in Fig. \ref{figdf}, are computed independently,  and their contributions are summed vectorially over all the encounters in the field distribution assuming no correlation between consecutive collision (Markovian hypothesis). Over the relative orbit we define the relative velocity as $\mathbf{v}=\mathbf{v}_{t}-\mathbf{v}_{f}$ and the reduced mass as usual as
\begin{eqnarray}
\mu=\frac{m_fm_t}{m_f+m_t}.
\end{eqnarray}  
According to Newton's Third Law of Dynamics, the velocity change $\Delta\mathbf{v}_{t}$ along the test particle's (unbound) orbit is exactly given by\begin{eqnarray}\label{newt3}
\Delta\mathbf{v}_{t}=\frac{\mu}{m_t}\Delta\mathbf{v},
\end{eqnarray}
where $\Delta\mathbf{v}$ is the vectorial change of the {\it relative} velocity, obtained from the solution of the two body problem. We define in impulsive approximation the modulus of effective acceleration felt by the test particle as
 \begin{eqnarray}
 a=\frac{q_tq_f}{\mu b^{2}}
 \end{eqnarray} 
where $b$ is the so called {\itshape impact parameter} is defined as the smallest distance between the unperturbed trajectory of the test particle and the target, or more rigorously as 
 \begin{eqnarray}
 b=\frac{L}{\mu v_{t}(+\infty)}
 \end{eqnarray} 
where $L$ is the modulus of the total angular momentum on the hyperbolic orbit and $v_{t}(+\infty)$ is the modulus of the asymptotic velocity of the test particle. If we now take as the effective time on which the interaction takes place during the binary collision
\begin{eqnarray}
t_{\rm eff}=\frac{2b}{|\mathbf{v}|}
\end{eqnarray} 
we can express the velocity changing in the direction perpendicular to the initial relative trajectory as
\begin{eqnarray}\label{perpe}
\Delta\mathbf{v}_{\perp}\sim\frac{2q_tq_f}{b|\mathbf{v}|}.
\end{eqnarray} 
Note that, in the latter the above expression becomes an exact equality in the limit of large $b$ or relative velocity. From impulsive approximation we can state that $||\mathbf{v}||^{2}=||\mathbf{v}+\Delta\mathbf{v}_{\perp}+\mathbf{v}_{\parallel}||^{2}$; now combining  the latter with equation (\ref{newt3}) for the component $\mathbf{v}_{t\parallel}$ of the test particle's velocity parallel to the initial $\mathbf{v}_{t}$, we easily obtain at first order
\begin{eqnarray}\label{pippi}
\Delta\mathbf{v}_{t\parallel}=\frac{\mu\Delta\mathbf{v}_{\parallel}}{m_t}\sim-\frac{\mu||\Delta\mathbf{v}_{\perp}||^{2}}{2m_t||\mathbf{v}||^{2}}\mathbf{v}.
\end{eqnarray} 
By substituting equation (\ref{perpe}) on the r.h.s. of relation (\ref{pippi}), after easy algebra we finally get
\begin{eqnarray}
\Delta\mathbf{v}_{t\parallel}\simeq-\frac{2q_t^{2}q_f^{2}(m_t+m_f)}{b^{2}m_t^{2}m_f||\mathbf{v}||^{4}}\mathbf{v}.
\end{eqnarray} 
We define now the number of binary collisions with impact parameter ranging from $b$ to $b+\Delta b$ in the time interval $\Delta t$ as
\begin{eqnarray}
\Delta N=2\pi bdb||\mathbf{v}_{t}-\mathbf{v}_{f}||\Delta t nf{(v_{f})}d^{3}\mathbf{v}_{f}
\end{eqnarray} 
where $n$ is the number density of field particles and $f{(v_{f})}$ is their velocity distribution function that here we will assume to be a Maxwellian with dispersion $\sigma$, that reads
\begin{eqnarray}\label{distv}
f{(v_{f})}=\frac{e^{-v^{2}_{f}/(2\sigma^{2})}}{(2\pi)^{3/2}\sigma^{3}}.
\end{eqnarray} 
It is now easy to write down as finite difference the velocity change on a time interval $\Delta t$ for the test particle in the parallel direction after $\Delta n$ encounters in the form
\begin{eqnarray}
\frac{\Delta\mathbf{v}_{t\parallel}}{\Delta t}=-4\pi n\frac{(m_t+m_f)}{m_t^{2}m_f}q_t^{2}q_f^{2}\frac{f{(v_{f})}\mathbf{v}}{b||\mathbf{v}_{t}-\mathbf{v}_{f}||^{3}}dbd^{3}\mathbf{v}_{f}.
\end{eqnarray}
The integration over $b$ can be carried out separately since it does not involves the velocity, obtaining
\begin{eqnarray}
\int_{b_{\rm min}}^{b_{\rm max}}\frac{db}{b}=\ln(b_{\rm max})-\ln(b_{\rm min})=\ln\frac{b_{\rm max}}{b_{\rm min}}=\ln\Lambda
\end{eqnarray}
that is the well known Coulomb logarithm. It must be pointed out that, since we are working in impulsive approximation, a divergence appears for $b_{\rm min}\rightarrow 0$ (ultraviolet divergence), while an intrinsic divergence already exists for $b_{\rm min}\rightarrow +\infty$ (infrared divergence); we can get rid of them setting $b_{\rm min}$ to its minimum inter-particle separation and $b_{\rm max}$ equal to the Debye length $\lambda_{D}$ of the system when the test particle moves through particles with charge of the opposite sign, or to the system's size in the opposite case. 
Note that, in unscreened ``infinite systems"  the infrared divergence remains, due to the long range nature of Coulomb interaction. Note also that, due to this, the contribution to the deflection of the far particles is in general larger than that due to short impact parameter encounters.\\
\indent The last step is now to integrate over the velocity. In order to do so according to the classical approach, we assume that the velocity distribution for the field particles is spherically symmetric around the test particle and we take the velocity averaged Coulomb logarithm $\ln\bar\Lambda$. The integration in $d^{3}\mathbf{v}_{f}$ gives finally
\begin{eqnarray}\label{dinfri}
\frac{d\mathbf{v}_{t\parallel}}{dt}=-4\pi n\frac{(m_t+m_f)}{m_t^{2}m_f}q_t^{2}q_f^{2}\ln\bar\Lambda\frac{\Theta{(v_{t})}}{v_{t}^{3}}\mathbf{v}_{t}.
\end{eqnarray} 
where the so called fractional volume function $\Theta{(\mathbf{v}_{t})}$ is given assuming $f(v_f)$ from Eq. (\ref{distv}) by
\begin{eqnarray}
\Theta{(\mathbf{v}_{t})}=4\pi\int_{0}^{v_{t}}f{(v_{f})}v_{f}^{2}dv_{f}={\rm Erf}(v_t/\sqrt{2}\sigma)-\frac{2v_t\exp(-v_t^2/2\sigma^2)}{\sigma\sqrt{2\pi}},
\end{eqnarray}
and the term
\begin{equation}
\omega_{\rm coll}=4\pi n\frac{(m_t+m_f)}{m_t^{2}m_f}q_t^{2}q_f^{2}\frac{\ln\bar\Lambda}{v_{t}^3}
\end {equation}
is sometimes referred to as collision frequency.\\
\indent Equation (\ref{dinfri}) gives us the effective deceleration suffered by the test particle crossing the sea of identical particles of charge $q_f$ and mass $m_f$. It can be immediately noticed that, for different charge states of the same species, (same mass different charges), 
the particles with higher charge states suffer the largest deceleration. This can be read for example considering an initially coherent beam of different ions of the same atomic species shot through a gas of identical particles, 
in the fact that the highest charged ions are more likely spread radially around the beam's initial direction of propagation.\\
\indent Recently, this formalism has been extended in the gravitational case to systems characterized by a mass spectrum $\Psi(m_f)$ \cite{2010AIPC.1242..117C}, so that their total number density in the definition of $\Delta N$ is given by
\begin{equation}
n=\int_0^{\infty}\Psi(m_f)\d m_f,
\end{equation}
and formally each component has its fractional volume function
\begin{eqnarray}
\Theta{(\mathbf{v}_{t},m_f)}=4\pi\int_{0}^{v_{t}}f{(v_{f},m_f)}v_{f}^{2}dv_{f}.
\end{eqnarray}
In our case of charged particles, two case are physically relevant, namely the discrete mass spectrum and discrete charge spectrum given by
\begin{equation}
\Psi(m_f)=n_1\delta(m_f-m_1)+n_2\delta(m_f-m_2);\quad \Psi(q_f)=n_1\delta(q_f-m_1)+n_2\delta(q_f-m_2),
\end{equation}
corresponding respectively to systems where all particles have the same charge $q_f$ and different masses, or same mass $m_f$ and different charges.
Here we have considered for simplicity only two species with number densities $n_1$ and $n_2$ and masses (charges) $m_1$ and $m_2$ ($q_1$ and $q_2$). Obviously this can be generalized to systems with an arbitrarily number of species, and to the more complicated case for both mass and charge spectra.
Following \cite{2010AIPC.1242..117C}, by integrating over all the different species $m_f$, Eq. (\ref{dinfri}) becomes
\begin{eqnarray}\label{dinfrimulti}
\frac{d\mathbf{v}_{t\parallel}}{dt}=-4\pi n\frac{(m_t+\langle m_f\rangle)}{m_t^{2}\langle m_f\rangle}q_t^{2}q_f^{2}\langle\ln\bar\Lambda\rangle\frac{\tilde\Theta{(v_{t})}}{v_{t}^{3}}\mathbf{v}_{t}.
\end{eqnarray} 
where   
\begin{equation}
n\langle m_f\rangle=\int_0^{\infty}m_f\Psi(m_f)\d m_f=n_1m_1+n_2m_2,
\end{equation}
the term $\tilde\Theta(v_t)$ is the total velocity volume factor depending on the choice of $f(v_f,m_f)$ and $\langle\ln\bar\Lambda\rangle$ is the mass averaged Coulomb logarithm depending instead on the choice of $n_1$ and $n_2$. Analogous considerations and steps can be carried out 
for the case of a charge spectrum $\Psi(q_f)$.
\chapter{Some remarks on kinetic theory}\label{kt}
Another theoretical treatment of the Coulomb explosion of spherical nanoplasmas based this time on Kinetic Theory \cite{landau10}, accounting for different initial density profiles and easily generalizable to multi-component systems, is that developed in Refs \cite{kovalev05}, \cite{2007PhPl...14e3103K} and \cite{2008PlPhR..34..920N}, see also \cite{tesimika}. 
This approach makes use of the concept of {\it one-particle phase space distribution function} $f(\mathbf{r},\mathbf{v},t)$, defined as the (differential) fraction of system at time $t$ in given point ($\mathbf{r},\mathbf{v}$) of the (six-dimensional) one particle phase space \cite{bellino}, in a way that once the time is fixed, the number density $n(\mathbf{r})$ is obtained as
\begin{equation}\label{CMB1}
n(\mathbf{r})=\int_{\Omega}f(\mathbf{r},\mathbf{v})\d^3\mathbf{v},
\end{equation} 
and the velocity field as
\begin{equation}\label{CMB2}
v(\mathbf{r})=\frac{1}{n(\mathbf{r})}\int_{\Omega}vf(\mathbf{r},\mathbf{v})\d^3\mathbf{r},
\end{equation} 
where the integrals are extended the domain $\Omega$ in phase space that is ``accessible" for the system. 
Instead of following in time the trajectories of particles or fluid elements, here one studies the time evolution of $f$ through the so-called Collisionless Boltzmann Equation, also known among Plasma physicists as Vlasov Equation\footnote{Precisely speaking, one should refer to Eq. (\ref{vlasov}) coupled to the Maxwell equations as {\it Vlasov equations}, see e.g. \cite{stix} se also \cite{1982A&A...114..211H} and references therein for an extended discussion.}  
\begin{equation}\label{vlasov}
\frac{{\rm D}f}{{\rm D}t}\equiv\frac{\partial f}{\partial t} + \mathbf{v} \cdot \frac{\partial f}{\partial\mathbf{r}}+\nabla\Phi\cdot\frac{\partial f}{\partial\mathbf{v}}=0,
\end{equation}
where the self consistent electrostatic potential $\Phi$ is related to $f$ through $n(\mathbf{r})$ by the Poisson equation now written as
\begin{equation}\label{poisson1}
\Delta\Phi(\mathbf{r})=4\pi q\int_{\Omega}f(\mathbf{r},\mathbf{v}){\rm d}^3\mathbf{v}.
\end{equation}
When the system is in an equilibrium state or in an asymptotic state (i.e. $\partial f/\partial t=0$), the phase space distribution $f$ can be written as function of energy and angular momentum through the phase space coordinates $\mathbf{r}$ and $\mathbf{v}$. 
If $f$ is expressible as a function of the energy $\mathcal{E}$ only, it is related to the differential energy distribution $n(\mathcal{E})$ introduced in Chapter \ref{chapmultice} by the relation
\begin{equation}
f(\mathcal{E})=n(\mathcal{E})\cdot g^{-1}(\mathcal{E}),
\end{equation}
where 
\begin{equation}
g(\mathcal{E})=4\pi\int_{\Omega|_\mathbf{v}}\sqrt{2(\mathcal{E}-\Phi(\mathbf{r}))}\d\mathbf{r},
\end{equation}
is the phase space volume with energy in the range $(\mathcal{E},\mathcal(E)+\d\mathcal{E})$.\\
\indent In our case of spherically symmetric systems, where the density is assumed to be angularly isotropic, if the distribution function at $t=0$ can be factorized as 
\begin{equation}
f=f_{\rm rad}(r,v_{\rm rad},0)f_{\rm tan}(\mathbf{v}^2_{\rm tan},0),
\end{equation}
it evolves under (\ref{vlasov}) so that its tangential part $f_{\rm tan}$ is stationary. Therefore in this case it is sufficient to study only the evolution of the radial part of the distribution function $f_{\rm rad}(r,v_{\rm rad},t)$. For simplicity hereafter we will drop the suffix $_{\rm rad}$.\\
\indent Adding an initial condition, in our case $f_0(r,v,t=0)=\delta(v)n_0(r)$, to the system of Eqs. (\ref{vlasov}) and (\ref{poisson1}) one has a Cauchy problem whose explicit solution in integral form for $f$ is, see  \cite{kovalev05} for the derivation,
\begin{equation}\label{kovasolution}
f(r,v,t)=\frac{1}{r^2}\int_0^{\infty} n_0(\eta)\delta\left[r-R(\eta,t)\right]\delta\left[v-U(\eta,t)\right]\eta^2\d\eta,
\end{equation}
where we the functions $R$ and $U$ are defined as solutions of 
\begin{eqnarray}\label{kovalevmodel}
\frac{\partial R(\eta,t)}{\partial t}=U(\eta,t)\nonumber\\
\frac{\partial U(\eta,t)}{\partial t}=\frac{q\eta^2\nabla\Phi(\eta,t)}{mR^2(\eta,t)}\nonumber\\
R(\eta,t=0)=\eta;U(\eta,t=0)=0.
\end{eqnarray}
Following the approach of \cite{2007PhPl...14e3103K}, it is possible to obtain an (explicit) approximate expression for $f$ at any time as a function of the initial number density profile $n_0$. By substituting in (\ref{kovalevmodel}) $r^2\nabla\Phi(r,t)$ with its value for $t=0$ and using the factorization of the $\delta$ distribution $\delta\left[g(x)\right]=\sum_k\delta(x-x_k)/|\partial g/\partial x_k|$, Equation (\ref{kovasolution}) is rewritten as
\begin{equation}
f(r,v,t)=\sum_k\frac{(2s_k-1)^2n_0(\eta_k)}{s_k^4|\partial R(\eta,t)/\partial \eta|_{\eta=\eta_k}}\delta(v-U(\eta_k,t)),
\end{equation}
where the $s_k$ are implicit solutions of
\begin{equation}
t\sqrt{\frac{2q\eta^2\nabla\Phi(\eta)}{m\eta^3}}=\frac{s(s-1)}{2s-1}+\frac{1}{2}\ln(2s-1)
\end{equation}
for specific times $t$ and coordinate $\eta$, $R(\eta,t)=\eta s^2(\eta,t)/(2s(\eta,t)-1)$ and $U(\eta,t)=(s(\eta,t)-1)\sqrt{2q\eta^2\nabla\Phi(\eta)/(m\eta)}/s(\eta,t)$ and the summation in $k$ runs over all the solutions of
\begin{equation}
\frac{s^2\eta}{2s-1}-r=0.
\end{equation}
Finally, the density and velocity profiles at time $t$ are the obtained from $f(r,v,t)$ solving the integrals \ref{CMB1} and \ref{CMB2}.
Knowing $v(r)$, from the usual expression
\begin{equation}
n(K)=\frac{4\pi r^2 n(r,t)}{{\d K}/{\d r}},
\end{equation}
it is possible to compute the time dependent kinetic energy distribution $n(K)$ which is given in Ref. \cite{kovalev05} as
\begin{equation}\label{distrokovalev}
n(K)=\frac{2\pi}{\sqrt{K}}\int_{-\infty}^{+\infty}f_*(v)\d v\times\sum_l\frac{\eta_l^2 n(\eta_l)}{|\partial U(\eta,v,t)/\partial \eta|_{\eta=\eta_l,U=\pm\sqrt{K}}},
\end{equation}
where $f_*(v)$ is the part in velocity of the distribution function and the summation runs over the $l$ solutions of
\begin{equation} 
U(\eta,v,t)=\pm\sqrt{K}.
\end{equation}
In the limit of $t\rightarrow\infty$, (implying also that $a(\eta,t)\rightarrow\infty$) when all the energy is converted in kinetic energy so that $n(\mathcal{E})_{\infty}=n(K)_{\infty}$, Equation (\ref{distrokovalev}) is rewritten according to \cite{tesimika} as
\begin{equation}
n(\mathcal{E})=2\pi\sum_l\frac{\eta_l^4n(\eta_l)}{|q\nabla\Phi(\eta_l)/m-\eta_l^3\omega(\eta_l)|},
\end{equation} 
where this time the sum in $l$ is over the solutions of
\begin{equation}
\frac{2q\nabla\Phi(\eta_l)}{m\eta_l}-\mathcal{E}=0
\end{equation}
and the radial dependent plasma frequency is defined by
\begin{equation}
\omega(r)=\sqrt{\frac{4\pi q^2n(r)}{m}}.
\end{equation}
It must be pointed out that the model introduced here allows one to take into account multi-stream motion and particle overtaking as well as to treat regimes where the hydrodynamical model is inapplicable, see \cite{2005PlPhR..31..178B}.\\
\indent The approach discussed above can be easily extended to homogeneous two component systems with densities $n_1$ and $n_2$, and unit charges $q_1$ and $q_2$, with ``kinematic parameters'' $a$ and $b$ defined by Eq. (\ref{qsum}), and described by two distribution functions $f_1$ and $f_2$ (see e.g. ref. \cite{kov}, \cite{2007PhPl...14e3103K}, \cite{2009LPB....27..321P} and \cite{2010PhPl...17b3110A}). 
Following the approach of \cite{kov}, let us suppose that $q_1n_2\ll q_2n_2$ (corresponding to low values of $b$). In this way, the contribution of the ``more mobile'' component 1 to the cluster's radial electrostatic field $E(r,t)$ is negligible, and thus the radial component of the electric field can be written as
\begin{align}\label{elettrico}
E(r,t)=
\begin{cases}
r(1-b^2)^3/3;\quad {\rm for}\quad r<(1-b^2)^{-1}\\
r^{-2}/3;\quad {\rm for}\quad r\geq(1-b^2)^{-1},
\end{cases}
\end{align}
where $t=\sqrt{6}\int_0^u\d u^\prime/(1-u^{\prime 2})^2$. According to \cite{2007PhPl...14j3110K} and \cite{2007PhPl...14e3103K}, the radial part of $f_1$ has the same implicit form given by Eq. (\ref{kovasolution}), where this time in the definition of $R$ and $U$ as solutions
of Eq. (\ref{kovalevmodel}), enters the electric field given by (\ref{elettrico}). By integrating numerically Equation (\ref{distrokovalev}) in the limit of large times, one finds \cite{2009LPB....27..321P}, that the asymptotic $n_1(\mathcal{E})$ for the fast component 
has a power law tail at low energies and an universal singular cutoff at maximal energy $\mathcal{E}_{\rm max}$ proportional to $(\mathcal{E}_{\rm max}-\mathcal{E})^{-1/2}$. The cut off energy can be estimated as function of the clusters parameters as
\begin{equation}
\mathcal{E}_{\rm max}\simeq\frac{3}{2}\frac{Q_{1}}{R}(a-1/3)\zeta,
\end{equation}
where $Q_1$ is the total charge of the fast component and $\zeta$ is the ratio of the unit masses of the slow and fast component.\\

\indent As a final remark, it should be pointed out that the numerically obtained $n_1(\mathcal{E})$ of Sect. \ref{multisystem} reproduce the power-law decay at low energies 
but do not show sharp cutoffs at $\mathcal{E}_{\rm max}$, having instead a seemingly universal decay (when energies are normalized with respect to their maximum value). The reason of that is the fact that for the values of $a$ and $b$ used in our simulations, the dynamics of the slower ions influences that of the fast component and, in addition, 
effects due to the system's particulate nature are not entirely negligible.

\newpage

\section*{Versicherung}
\noindent
Hiermit versichere ich, dass ich die vorliegende Arbeit ohne unzul\"assige
Hilfe Dritter und ohne Benutzung anderer als der angegebenen Hilfsmittel
angefertigt habe; die aus fremden Quellen direkt oder indirekt \"ubernommenen
Gedanken sind als solche kenntlich gemacht. Die Arbeit wurde bisher weder im
Inland noch im Ausland in gleicher oder \"ahnlicher Form einer anderen
Pr\"ufungsbeh\"orde vorgelegt.\\
Die Arbeit wurde am Max-Planck-Institut f\"ur Physik komplexer Systeme in der
Abteilung \glqq Endliche Systeme\grqq \ angefertigt und von
Prof. Dr. Jan-Michael Rost und Prof. Dr. Ulf Saalmann betreut.\\
Ich erkenne die Promotionsordnung der Fakult\"at Mathematik und
Naturwissenschaften der Technischen Universit\"at Dresden
von 23.02.2011~an.\\
\\
\\
\\

\quad\quad\quad\quad\quad\quad\quad\quad\quad\quad\quad\quad\quad\quad\quad\quad Pierfrancesco Di Cintio
\end{document}